\documentclass[12pt]{article}
\pdfoutput=1

\usepackage{jheppub}
\makeatletter
\def\@fpheader{\relax}
\makeatother

\usepackage{amsthm,amsmath,amssymb,mathtools}
\usepackage{mathrsfs}
\usepackage{bm}
\usepackage{float}
\usepackage{caption}
\usepackage{subcaption}
\usepackage[most]{tcolorbox}
\usepackage[mathscr]{euscript}
\usepackage{dsfont}
\usepackage[mathscr]{euscript}
\usepackage{mathrsfs}

\def\mA{\mathcal{A}}

\renewcommand\mod{\, \text{mod}\ }

\title{$\boldsymbol{SL(3,\mathbb{Z})}$ Modularity and New Cardy Limits of the $\boldsymbol{\mathcal{N}=4}$ Superconformal Index}

\author[a]{Vishnu Jejjala}
\author[b]{\!\!, Yang Lei}
\author[a]{\!\!, Sam van Leuven}
\author[c]{\!\!, Wei Li}

\affiliation[\,a]{Mandelstam Institute for Theoretical Physics, School of Physics, NITheP, and CoE-MaSS, \\ University of the Witwatersrand, Johannesburg, South Africa}
\affiliation[\,b]{Kavli Institute for Theoretical Sciences (KITS),  \\ University of Chinese Academy of Sciences, 100190 Beijing, P.R.~China}
\affiliation[\,c]{Institute of Theoretical Physics, \\ Chinese Academy of Sciences, 100190 Beijing, P.R.~China}

\emailAdd{vishnu@neo.phys.wits.ac.za}
\emailAdd{leiyang@ucas.edu.cn}
\emailAdd{svleuven@xs4all.nl}
\emailAdd{weili@mail.itp.ac.cn}

\abstract{
	The entropy of $1/16$-th BPS AdS$_5$ black holes can be microscopically accounted for by the superconformal index of the $\mathcal{N}=4$ super-Yang--Mills theory.
	One way to compute this is through a Cardy-like limit of a formula for the index obtained in~\cite{Goldstein:2020yvj} using the ``$S$-transformation'' of the elliptic $\Gamma$ function.
	In this paper, we derive more general $SL(3,\mathbb{Z})$ modular properties
	of the elliptic $\Gamma$ function.
	We then use these properties to obtain a three integer parameter family of generalized Cardy-like limits of the $\mathcal{N}=4$ superconformal index. 
	From these limits, we obtain entropy formulae that have a similar form as that of the original AdS$_5$ black hole, up to an overall rescaling of the entropy.
	We interpret this both on the field theory and the gravitational side.
	Finally, we comment on how our work suggests a generalization of the Farey tail to four dimensions.
}

\date{}

\begin{document} 
	
	\parskip=10pt
	\newcommand\smallspace{\vspace{.1in}}
	\newcommand\negspace[1]{\vspace{-#1in}}
	\newcommand\etp[1]{\enlargethispage{#1\baselineskip}}
	
	\maketitle 
	
	\section{Introduction and summary}\label{sec:intro}
	
	Explaining the microscopic origin of the Bekenstein--Hawking entropy of black holes remains one of the outstanding puzzles of quantum gravity.
	The first significant breakthrough on this problem was the Strominger--Vafa enumeration of degenerate supersymmetric vacua of a D-brane system associated to the five-dimensional extremal Reissner--Nordstr\"om solution~\cite{Strominger:1996sh}.
	While this counting successfully reproduces the entropy, we do not know what the microstates are in the strong coupling regime of gravity.
	With the development of the AdS/CFT correspondence~\cite{Maldacena:1997re,Gubser:1998bc,Witten:1998qj}, one hoped to do better.
	The $1/2$-, $1/4$-, and $1/8$-th BPS solutions in AdS$_5$ are incipient black holes with zero horizon area in supergravity~\cite{Myers:2001aq}.
	The first supersymmetric black hole with AdS$_5\times S^5$ asymptopia and a finite sized horizon preserves two out of thirty-two supercharges~\cite{Gutowski:2004ez, Gutowski:2004yv,Chong:2005da,Chong:2005hr,Kunduri:2006ek}.
	Its entropy scales like $N^2$.
	Until recent years, it was expected that this entropy was not captured by the superconformal index of the $\mathcal{N}=4$ $SU(N)$ super-Yang--Mills gauge theory, which at large-$N$ exhibits only an $\mathcal{O}(1)$ growth of states~\cite{Kinney:2005ej}.
	However, it turns out that when real chemical potentials are replaced with complex chemical potentials, subject to a linear constraint, the superconformal index does exhibit the expected $\mathcal{O}(N^2)$ asymptotic scaling~\cite{Cabo-Bizet:2018ehj,Choi:2018hmj,Benini:2018ywd}.
	At large-$N$, the black hole entropy is exactly reproduced from a Legendre transformation of the index and corresponds to the extremization of an entropy functional \`a la the attractor mechanism~\cite{Hosseini:2017mds}.
	Further developments on the $\mathcal{N}=4$ index and its relation to AdS$_5$ black holes include~\cite{Choi:2018vbz,Honda:2019cio,ArabiArdehali:2019tdm,Cabo-Bizet:2019eaf,ArabiArdehali:2019orz,Murthy:2020rbd,Agarwal:2020zwm,GonzalezLezcano:2020yeb,Copetti:2020dil,Cabo-Bizet:2020ewf,Amariti:2020jyx,Lezcano:2021qbj,Benini:2021ano,Choi:2021lbk,ArabiArdehali:2021nsx}.
	These methods have also been extended to the analysis of superconformal field theories with $\mathcal{N}=1$ supersymmetry~\cite{Kim:2019yrz,Cabo-Bizet:2019osg,Amariti:2019mgp,Lezcano:2019pae,Lanir:2019abx,Cabo-Bizet:2020nkr,Benini:2020gjh,Amariti:2021ubd,Cassani:2021fyv}.
	Connections with AdS$_3$/CFT$_2$, specifically the two-dimensional Cardy formula and near horizon limits of the AdS$_5$ black holes, were explored in~\cite{Goldstein:2019gpz,Nian:2020qsk,David:2020ems}.
	Finally, there have also been explorations beyond the BPS limit~\cite{Larsen:2019oll}.

	In previous work~\cite{Goldstein:2020yvj}, we computed the superconformal index of the $\mathcal{N}=4$ theory through a residue calculation that leads to a Higgs branch localization type formula for the index~\cite{Peelaers:2014ima,Yoshida:2014qwa,Nieri:2015yia}.
	The resulting formula has the schematic form:
	\begin{equation}\label{eq:schem-form-index}
		I=\sum Z_{\text{P}} Z_{\text{v}}Z_{\text{av}} \,.
	\end{equation}
	Here, the sum is over residues; $Z_{\text{P}}$ captures the perturbative part of a localization computation, and $Z_{\text{v}}$ and $Z_{\text{av}}$ capture the non-perturbative vortex and anti-vortex contributions.
	It turns out that $Z_{\text{P}}$ consists of elliptic $\Gamma$ functions whereas $Z_{\text{v}}$ and $Z_{\text{av}}$ consist of $q$-$\theta$ functions.
	We collect the definition and important properties of these functions in Appendix~\ref{app:defs}.
	
	It is well known that the $\theta(z;\tau)$ function has modular properties under the Jacobi group $SL(2,\mathbb{Z})\ltimes \mathbb{Z}^2$, which can be used to compute its ``high temperature'' $\tau\to 0$ limit explicitly.
	Indeed, this is precisely analogous to Cardy's derivation of his eponymous formula~\cite{Cardy:1986ie}.
	It is perhaps less well known that the elliptic $\Gamma$ function has modular properties under $SL(3,\mathbb{Z})\ltimes \mathbb{Z}^{3}$~\cite{Felder_2000}.
	Using these modular properties, we were able to calculate a Cardy-like limit of the superconformal index.
	The resulting function matches precisely with the free energy of~\cite{Hosseini:2017mds}, which they observed to yield the AdS$_5$ black hole entropy upon a Legendre transformation.
	
	The Cardy-like limit considered in our previous work~\cite{Goldstein:2020yvj} is evaluated using a specific modular property of the elliptic $\Gamma$ function,  which is analogous to the modular property of ordinary modular forms under the $S$-transformation in $SL(2,\mathbb{Z})$.
	In $SL(2,\mathbb{Z})$, the most general modular transformation has the form:
	\begin{equation}
		\tau\to\frac{a\tau+b}{c\tau+d}\,, \qquad ad-bc=1 \,.
	\end{equation}
	The behavior of a modular form under such a transformation can be used to evaluate what we will call a generalized Cardy limit of the modular form.
	This more general limit is obtained by sending $\tau\to -\frac{d}{c}$.
	Such limits of the partition function can be interpreted on the gravitational side.
	In particular, close to the point $\tau= -\frac{d}{c}$ the Euclidean gravitational path integral is dominated by a specific member of the so-called $SL(2,\mathbb{Z})$ family of BTZ black holes~\cite{Maldacena:1998bw,Dijkgraaf:2000fq,Manschot:2007ha}.
	Only the Euclidean BTZ with $(c,d)=(1,0)$ can be continued to Lorentzian signature, where it becomes the usual BTZ black hole.
	
	It is natural to ask if more general elements in $SL(3,\mathbb{Z})$ yield modular properties for the elliptic $\Gamma$ function that facilitate similar generalized Cardy limits, and subsequently what the gravitational interpretation of these limits is.
	This is the question that we address in the present paper.
	Such generalized Cardy limits were first studied in~\cite{Cabo-Bizet:2019eaf} and more recently in~\cite{Cabo-Bizet:2020ewf,ArabiArdehali:2021nsx}.

	\paragraph{Summary of main results:}
	
	In this work, we systematically explore $SL(3,\mathbb{Z})$ modular properties of the elliptic $\Gamma$ function, which we then  apply to a Higgs branch localization type formula of the $\mathcal{N}=4$ superconformal index derived in~\cite{Goldstein:2020yvj}, to study its generalized Cardy limits.
	In particular, we derive a three integer parameter family of modular properties for the elliptic $\Gamma$ function, of which the fully explicit cases are given by:
	\begin{equation}\label{eq:summary-mod-prop-gamma}
		\Gamma(z;\tau,\sigma)=e^{-i\pi Q'_{\mathbf{m}}\left(m z;\tau,\sigma\right)}\frac{\Gamma\left(\tfrac{z}{m\sigma+1-m n_2};\tfrac{\tau-\sigma+n_2-n_1}{m\sigma+1-m n_2},\tfrac{\sigma-n_2}{m\sigma+1-m n_2}\right)}{\Gamma\left(\tfrac{z+\tau-\sigma+n_2-n_1}{m\tau+1-m n_1};\tfrac{\tau-\sigma+n_2-n_1}{m\tau+1-m n_1},\tfrac{\tau-n_1}{m\tau+1-m n_1}\right)} \,, 
	\end{equation}
	which is a generalization of the modular property first described in~\cite{Felder_2000} and used in the present context in~\cite{Gadde:2020bov,Goldstein:2020yvj}.
	The function $Q'_{\mathbf{m}}$ is essentially the $Q$ polynomial appearing in the original modular property, but with an overall $\frac{1}{m}$ rescaling plus a change of arguments and a constant shift that both depend on three free integers $\mathbf{m}=(m,n_1,n_2)$.
	This new modular property is precisely constructed such that one can study certain generalized Cardy limits of the elliptic $\Gamma$ function, in which:
	\begin{equation}
		\tau\to n_1-\tfrac{1}{m}\,,\quad \sigma\to n_2-\tfrac{1}{m} \,.
	\end{equation}
	In this limit, the right hand side of the modular property~\eqref{eq:summary-mod-prop-gamma} essentially reduces to the phase prefactor, giving a simple expression for the limit.
	
	The expression for the superconformal index derived in~\cite{Goldstein:2020yvj} contains a part which consists of a product of elliptic $\Gamma$ functions.
	To leading order in the Cardy limit, this can be argued to be the only relevant part of the index.
	Simply applying~\eqref{eq:summary-mod-prop-gamma} to each elliptic $\Gamma$ function and taking the generalized Cardy limit yields a function closely related to the free energy discussed in~\cite{Hosseini:2017mds}, which was first derived in field theory in~\cite{Cabo-Bizet:2018ehj,Choi:2018hmj,Benini:2018ywd}.
	Explicitly, it is given by:
	\begin{equation}\label{eq:summary-free-energy}
		\lim_{\text{gen Cardy}} \log I_N =-i\pi\frac{N^2}{m}\frac{[m\phi_1][m\phi_2]([m\phi_1]+[m\phi_2]+1)}{(m\tau+1-mn_1)(m\sigma+1-mn_2)}\,.
	\end{equation}
	This formula was derived using different methods in~\cite{Cabo-Bizet:2019eaf,Cabo-Bizet:2020ewf,ArabiArdehali:2021nsx}.
	Upon Legendre transformation, this leads to an entropy formula that is almost equivalent to the entropy formula for supersymmetric AdS$_5$ black holes~\cite{Kunduri:2006ek,Benini:2018ywd,Choi:2018hmj,Cabo-Bizet:2018ehj}:
	\begin{equation}
		S=\frac{2\pi}{m}\sqrt{Q_1 Q_2+Q_2 Q_3+Q_1 Q_3-\tfrac{1}{2}N^2(J_1+J_2)}\,.
	\end{equation}
	However, it differs from the ordinary AdS$_5$ black hole entropy formula  by a factor $\frac{1}{m}$.
	In~\cite{AharonyTalk}, this entropy is connected to certain orbifolds of the original supersymmetric AdS$_5$ black hole.  
	
	We interpret this reduction by $\frac{1}{m}$ in the field theory by observing that our expression for the index in the generalized Cardy limit implies a certain reduction of the original Hilbert space of the theory, such that the reduced Hilbert space only consists of states with charges that are multiples of $m$.
	On the gravitational side, we observe that the generalized Cardy limit effectively quotients the Hopf fiber of the $S^3$ by a factor $m$.
	Such a quotient reduces the area element by a factor $\frac{1}{m}$, which through the Bekenstein--Hawking formula leads to the required reduction in the entropy.
	
	Finally, we  discuss how the modular properties of the full $\mathcal{N}=4$ index suggest a formula for the gravitational path integral in terms of a sum over certain elements in the modular $SL(3,\mathbb{Z})$ group.
	In particular, we discuss similarties and differences with the Farey tail expansion of the elliptic genus of~\cite{Dijkgraaf:2000fq,Manschot:2007ha} in two dimensions.

	\paragraph{Plan of paper:}
	
	The remainder of this paper is organized as follows.
	In Section~\ref{ssec:review-index}, we review the Higgs branch localization type formula for the index derived in our previous work~\cite{Goldstein:2020yvj} and the associated Cardy-like limit using the $S$-transformation of the elliptic $\Gamma$ function.
	In Section~\ref{sec:mod-prop-gamma}, we generalize the modular property of the elliptic $\Gamma$ function studied in~\cite{Felder_2000,Gadde:2020bov} to more general $SL(3,\mathbb{Z})$ elements and study the corresponding generalized Cardy limits.
	This section is more technical in nature.
	In Section~\ref{sec:revis-index}, we turn to the physical application of the results derived in Section~\ref{sec:mod-prop-gamma}.
	In particular, we apply the result of Section~\ref{sec:mod-prop-gamma} to study the generalized Cardy limit of the $\mathcal{N}=4$ superconformal index; based on this we conjecture an expression for the index which resembles a sum over geometries; and finally, we compute the entropy in these limits and discuss its interpretation from both the field theory and the gravity side.
	Section~\ref{sec:disc} contains a discussion on the relation between the sum over geometries and modularity of the index, and a selection of open problems.
	Appendix~\ref{app:defs} collects relevant properties of $q$-$\theta$ function and the elliptic $\Gamma$ function;
	Appendix~\ref{app:more-gen-order-3-elms} studies more general order three elements of $SL(3,\mathbb{Z})$ than the one focused on in the main text; and finally, Appendix~\ref{app:deriv-Q} provides a detailed derivation of the $Q$ polynomials associated to the various modular properties.

	\paragraph{Note added:}
	During the completion of this work, the preprint~\cite{ArabiArdehali:2021nsx} appeared which studies similar topics using complementary techniques.

	\section{Superconformal index of the \texorpdfstring{$\mathcal{N}=4$}{N=4} SYM theory}\label{ssec:review-index}

	The superconformal index of the $\mathcal{N}=4$ $SU(N)$ super-Yang--Mills theory is defined as the weighted trace over the Hilbert space $\mathcal{H}$ of the theory quantized on $S^3$.
	A recent review is~\cite{Gadde:2020yah}.
	Explicitly, the index is defined as~\cite{Kinney:2005ej}:
	\begin{equation}\label{eq:trace-defn-index}
		I_N=\mathrm{tr}_{\mathcal{H}}(-1)^F p^{J_1}q^{J_2}(p q)^{\frac{r_1-r_3}{2}}f_1^{r_2+r_3}f_2^{r_3}e^{-\beta \lbrace \mathcal{Q},\mathcal{Q}^\dagger\rbrace} \,.
	\end{equation}
	The charges $J_{1,2}$ parameterize the angular momenta along $S^3$ corresponding to the $SO(2)\times SO(2)\subset SO(4)$ Cartan generators.
	We can write them in terms of Cartan generators $j_{1,2}$ of $SU(2)_1\times SU(2)_2$ as $J_{1,2}=j_1\pm j_2$.
	These charges are half integer valued.
	The $r_i$ correspond to $R$-charges associated to the Cartan of $SU(4)$.
	The $SO(6)$ Cartan generators $Q_a$ used in~\cite{Benini:2018ywd} are related to the $r_i$ via:
	\begin{equation}
		Q_1=\frac{r_1+2r_2+r_3}{2} \,, \qquad Q_2=\frac{r_1+r_3}{2} \,,\qquad Q_3=\frac{r_1-r_3}{2} \,,
	\end{equation}
	which are also half integer valued.
	The supercharge $\mathcal{Q}$ has charges: 
	\begin{equation}
		Q_1=Q_2=Q_3=-J_1=-J_2=\frac{1}{2} \,.
	\end{equation}
	Due to the insertion of $(-1)^F$, with $F=2Q_3$ the fermion number operator, the index localizes on $\mathcal{H}_{\text{BPS}}$, the 1/16-th BPS Hilbert space, corresponding to the vanishing locus of the operator:
	\begin{equation}
		\lbrace \mathcal{Q},\mathcal{Q}^\dagger\rbrace=E-J_1-J_2-Q_1-Q_2-Q_3 \,.
	\end{equation}
	The fugacities can be expressed in terms of chemical potentials as:
	\begin{equation}
		p=e^{2\pi i\sigma} \,,\quad q=e^{2\pi i \tau} \,, \quad f_1=e^{2\pi i\phi_1} \,,\quad f_2=e^{2\pi i \phi_2} \,.
	\end{equation}
	Our parameterization is equivalent to the one used in~\cite{Benini:2018ywd} upon identifying $f_i=y_i$.
	As emphasized in~\cite{Benini:2018ywd}, since the charges $2J_{1,2}$ and $2Q_{1,2,3}$ of any state in $\mathcal{H}$ are all equal $\mod 2$, the expression for the index is manifestly periodic under $\tau,\sigma\to\tau,\sigma+1$ and $\phi_{1,2}\to \phi_{1,2}+1$.
	
	Since the index is independent of continuous deformations of the theory which preserve $\mathcal{Q}$, one may compute it at weak coupling~\cite{Kinney:2005ej}.
	In this case, the trace can  be explicitly performed, and the resulting expression is given by~\cite{Kinney:2005ej,Romelsberger:2005eg,Dolan:2008qi}:
	\begin{equation}\label{eq:defn-suN-index}
		I_N=\frac{\kappa_N}{N!} \prod^{N-1}_{k=1}\oint_{\left|x_k\right|=1}\frac{dx_k}{2\pi i x_k}\prod_{1\leq i\neq j\leq N}\frac{\prod^3_{a=1}\Gamma(x_{i j}f_a)}{\Gamma(x_{i j})} \,.
	\end{equation}
	Here, we have defined $f_3=p q f_1^{-1}f_{2}^{-1}$, $x_N=(x_1\cdots x_{N-1})^{-1}$, $x_{ij}=x_ix_j^{-1}$, and we use shorthand notation for the elliptic $\Gamma$ function $\Gamma(x)\equiv\Gamma(z;\tau,\sigma)$, defined in Appendix~\ref{app:defs}, with $x_i=e^{2\pi i z_i}$.
	The integral over the gauge fugacities $x_i$ ensures the projection onto gauge invariant states.
	Furthermore, $\kappa_N$ consists of the Cartan factors of both the chiral multiplets and the vector multiplet and is given by:
	\begin{equation}
		\kappa_N=(p;p)^{N-1}_{\infty}(q;q)^{N-1}_{\infty}\left(\Gamma(f_1)\Gamma(f_2)\Gamma(f_3)\right)^{N-1} \,, \label{eq:qpoc}
	\end{equation}
	where the $q$-Pochhammer symbol is also defined in Appendix~\ref{app:defs}.
	
	Various approaches to evaluating~\eqref{eq:defn-suN-index} exist in the literature. 
	These can be divided into two main categories: approximation of the integrand in some limit (e.g., large-$N$ limit or limits of the chemical potentials) after which the integral can be evaluated through a saddle point approximation~\cite{Cabo-Bizet:2018ehj,Choi:2018hmj,Choi:2018vbz,Honda:2019cio,ArabiArdehali:2019tdm,Kim:2019yrz,Cabo-Bizet:2019osg,Amariti:2019mgp,Larsen:2019oll,Cabo-Bizet:2019eaf,Goldstein:2019gpz,ArabiArdehali:2019orz,Cabo-Bizet:2020nkr,Murthy:2020rbd,Agarwal:2020zwm,Benini:2020gjh,GonzalezLezcano:2020yeb,Copetti:2020dil,Amariti:2020jyx,Choi:2021lbk,ArabiArdehali:2021nsx,Cassani:2021fyv}. 
	On the other hand, an exact evaluation of the contour integral is also possible, only after which one takes a large-$N$ or chemical potential limit~\cite{Benini:2018ywd,Lanir:2019abx,Goldstein:2020yvj,Cabo-Bizet:2020ewf}. 
	The method we employed is of the latter type, i.e., we first evaluate the contour integral explicitly by picking up residues from the various poles of the integrand, and subsequently take a limit of the chemical potentials.
	
	In our previous work~\cite{Goldstein:2020yvj}, we obtain the following expression for the index:
	\begin{align}\label{eq:suN-index-final}
		\begin{split}
			I_N^{\prime}&=\frac{(\Gamma(f_1)\Gamma(f_2)\Gamma(f_3))^{N-1}}{N!\Gamma(1)^{N-1}}\sum^\prime_{(a_i)}\prod^{N-1}_{i< j}\frac{\prod^3_{b=1}\Gamma((f_{a_i}f_{a_j}^{-1})^{\pm}f_b)}{\Gamma((f_{a_i}f_{a_j}^{-1})^{\pm})}\prod^{N-1}_{i=1}\frac{\prod^3_{b=1}\Gamma(f_{a_i}^{\pm}f_b)}{\Gamma(f_{a_i}^{\pm})}\\
			&\times \sum^\prime_{(k_i),(l_i)\geq (0)} Z^{(a_i),(k_i)}_{\mathrm{V}}(\phi_{a},\sigma;\tau) Z^{(a_i),(l_i)}_{\mathrm{V}}(\phi_{a},\tau;\sigma) 
			\,,
		\end{split}
	\end{align}
	where we use the notation $\Gamma\left(x^\pm\right)=\Gamma\left(x\right)\Gamma\left(x^{-1}\right)$.
	We will make a couple of comments about this expression, while referring the reader to~\cite{Goldstein:2020yvj} for full details.
	\begin{itemize}
		\item The expression takes the form of~\eqref{eq:schem-form-index} as mentioned in Section~\ref{sec:intro}.
		In particular, the residue sum is realized as the sum over the $(N-1)$ tuples:
		\begin{equation}
			(a_i)\equiv (a_1,\ldots,a_{N-1}) \,,\quad (k_i)\equiv (k_1,\ldots,k_{N-1}) \,,\quad (l_i)\equiv (l_1,\ldots,l_{N-1}) \,,
		\end{equation}
		where $a_i=1,2,3$ and $k_i,l_i\geq 0$.
		One does not sum over every such tuple, which is indicated by the primes on the respective sums.
		The precise summation domains depend on the values of the chemical potentials, of which we have not found a simple description in the generic case. 
		Furthermore, the part depending on the elliptic $\Gamma$ functions represents $Z_{\text{P}}$~\eqref{eq:schem-form-index}, while the non-perturbative contributions are encoded in functions $Z_{\mathrm{V}}$, which are expressed purely in terms of $\theta$-functions.
		\item The precise form of the vortex partition functions of the numerator $Z_{\mathrm{V}}$ depends on the sign of $k_i-k_j$ and $l_i-l_j$.
		For example, if both are positive or both negative for all $i<j$, then the vortex partition function is given by:
		\begin{align}\label{eq:vortex-part-suN}
			\begin{split}
				Z^{(a_i),(k_i)}_{\mathrm{V}}&(\phi_{a},\sigma;\tau)=\prod^{N-1}_{i< j}\frac{\prod^{k_i-k_j}_{m=1}\theta_q(f_{a_i}^{-1}f_{a_j}p^{-m})}{\prod^{k_i-k_j-1}_{m=0}\theta_q(f_{a_i}f_{a_j}^{-1}p^m)}\prod^{N-1}_{i=1}\frac{\prod^{k_i}_{m=1}\theta_q(f_{a_i}^{-1}p^{-m})}{\prod^{k_i-1}_{m=0}\theta_q(f_{a_i}p^m)}\\
				&\times\prod^3_{b=1}\Bigg(\prod^{N-1}_{i< j}\frac{\prod^{k_i-k_j-1}_{m=0}\theta_q(f_{a_i}f_{a_j}^{-1}f_bp^m)}{\prod^{k_i-k_j}_{m=1}\theta_q(f_{a_i}^{-1}f_{a_j}f_bp^{-m})}
				\prod^{N-1}_{i=1}\frac{\prod^{k_i-1}_{m=0}\theta_q(f_{a_i}f_bp^m)}{\prod^{k_i}_{m=1}\theta_q(f_{a_i}^{-1}f_bp^{-m})}\Bigg) \,.
			\end{split}
		\end{align}
		Here, we use the shorthand notation $\theta_q(x)\equiv\theta(z;\tau)$, which is defined in Appendix~\ref{app:defs}.
		\item The attentive reader will note that the expression~\eqref{eq:suN-index-final} is not quite equal to the final expression (2.47) in~\cite{Goldstein:2020yvj}.
		In particular, in that paper we canceled the elliptic $\Gamma$ functions in the denominator of~\eqref{eq:suN-index-final} against the $m=0$ terms in the denominator of the first line of~\eqref{eq:vortex-part-suN}.
		The reason for doing this is that whenever $a_i=a_j$, the factors $\frac{1}{\Gamma((f_{a_i}f_{a_j}^{-1})^\pm)}$ and $\frac{1}{\Gamma(f_{a_i}^\pm)}$ have a zero.
		These zeros cancel against the poles coming from the vortex partition function, making the whole non-vanishing and well-defined.
		The reason to not cancel these zeros and poles facilitates the analysis in Sections~\ref{ssec:anom-pol-index} and~\ref{ssec:mod-Z}.
		\item The factor $\Gamma(1)^{N-1}$ in the denominator is included to cancel the same factor arising from the product over $b$ of the last product on the first line.
		This latter factor is not part of the expression of the index, and therefore should be cancelled. 
		The reason that it is included is for notational convenience. 
		\item The prime on the index indicates that only a certain class of poles has been taken into account in the residue sum.
		This class is singled out by the fact that it contributes to the residue sum for the most generic values of the chemical potentials.
		However, for more special values of the chemical potentials, there may also be other classes contributing.
		The associated residue can be obtained from residues included in the sum by a simple transformation on the fugacities, described in more detail in~\cite{Goldstein:2020yvj}.
		\item Up to subtleties with the precise summation domain, related to the previous point, the expression holds for arbitrary \emph{complexified} chemical potentials $\tau$, $\sigma$ and $\phi_a$, as long as one keeps the $\phi_a$ strictly unequal (modulo integer shifts).\footnote{
			However, as explained in Appendix~B of~\cite{Goldstein:2020yvj}, it is possible to adapt the computation of the index such that one can also obtain the expression at the unrefined point $f_1=f_2=f_3$. In particular, it turns out that for purposes of the Cardy-like limit, the unrefined limit can be safely taken in the expression for the Cardy limit obtained from the refined expression.}
	\end{itemize}
	Given these comments, we note that the expression~\eqref{eq:suN-index-final} is incomplete and not fully explicit.
	Despite this, it can be used to evaluate an exact ``Cardy-like'' limit of the superconformal index, as we will now explain.
	
	Firstly, recall that the usual Cardy limit in two-dimensional CFT refers to a high temperature limit $\beta\to 0$ of the torus partition function.
	The superconformal index, due to the insertion of $(-1)^F$, does not depend on temperature.
	However, a Cardy-like limit can still be defined by taking $\tau$ and $\sigma$, the chemical potentials which couple to the angular momenta, both to $0^{+i}$, while keeping $\phi_a$, the chemical potentials coupling to the $R$-charges $Q_{1,2,3}$, fixed~\cite{Choi:2018hmj,ArabiArdehali:2019tdm}.\footnote{
		Notice that in our analysis of the Cardy limit, $\frac{\tau}{\sigma}\notin \mathbb{R}$ in general, in contrast with, for example,~\cite{Choi:2018hmj,Benini:2018ywd,ArabiArdehali:2019tdm,Benini:2020gjh}.}
	In the microcanonical ensemble, this corresponds to a large charge limit where the charges scale as $Q_a\sim \mu^2$ and $J_i\sim \mu^3$ for $\mu\to \infty$~\cite{Choi:2018hmj}.
	
	In the Cardy-like limit, there is an important simplification of~\eqref{eq:suN-index-final}.
	Using modular properties of both $\theta$ functions, of which the $Z_{\mathrm{V}}$ consist, and of the elliptic $\Gamma$ function, it can be shown that only the perturbative part of the partition function contributes at leading order.
	Moreover, the modular property of the elliptic $\Gamma$ allows an explicit computation of this limit, which turns out to be a relatively simple function.
	This eliminates complications associated to the vortex partition functions.
	Another advantage of the modular property is that the resulting Cardy-like limit of the $\Gamma$ function holds for finite imaginary part of the $\phi_a$ chemical potentials.
	In particular, it can be viewed as a justification for the analytic continuation of earlier works~\cite{Choi:2018hmj,ArabiArdehali:2019tdm}, which relied on taking real values of the chemical potentials.
	
	At this stage, one still has to deal with the summation of various residues labelled by $(a_i)$.
	It turns out that all residues contribute at leading order in the Cardy limit, but for generic values of the $\phi_a$ they depend on the tuples $(a_i)$, making a resummation necessary to obtain a useful expression for the limit of the index.
	However, an additional simplification occurs in a special region in parameter space, close to the unrefined point $f_1=f_2=f_3$, and at large-$N$.
	Here, it can be shown that each residue included in the expression for $I_N'$ contributes universally.
	In particular, this means that one does not need to worry about interfering phases between various residues as, for instance, in the Bethe Ansatz scenario~\cite{Benini:2018ywd}. 
	Furthermore, in this region one can also show that residues not included in the expression for $I_N'$ do not contribute since the associated poles will lie outside the integration contour.
	Taken together, this means that~\eqref{eq:suN-index-final} can be used to derive an exact expression for the full superconformal index $I_N$ in the Cardy-like limit.
	
	It turns out that the resulting limit of the index can be expressed in terms of bracketed chemical potentials, as also observed in the large-$N$ limit in~\cite{Benini:2018ywd}.
	The precise definition of the brackets may be found in~\cite{Goldstein:2020yvj}, or as the special case $m=n_1=n_2=1$ of the bracket defined in Section~\ref{ssec:gen-cardy-gamma}.
	For now, it suffices to note that the brackets are periodic with period 1: $[z+1]=[z]$.
	There exist two possible limits of the index, depending on the values of $[\phi_1]$ and $[\phi_2]$.
	In one case, the resulting limit is given by:
	\begin{align}\label{eq:Qprime-gen-res}
		\begin{split}
			\log I_N &=-i\pi N^2 \frac{[\phi_1][\phi_2][\phi_3]}{\tau\sigma}+\mathcal{O}\left(\frac{([\phi_{a}]-[\phi_{b}])^2}{\tau\sigma}\right)+\mathcal{O}(\tau^{-1})+\mathcal{O}(\sigma^{-1}) \,,
		\end{split}
	\end{align}
	where $[\phi_3]=-[\phi_1]-[\phi_2]-1$, and the expression holds at large-$N$ and close to the unrefined point:
	\begin{equation}
		[\phi_1]=[\phi_2]=[\phi_3]=-\tfrac{1}{3}\,.
	\end{equation}
	Furthermore, the pair $(a,b)$ takes the values $(1,2)$, $(1,3)$, and $(2,3)$.
	Crucially, the integer $-1$ in $[\phi_3]$ emerges from a careful limit of the modular property.
	
	In the second case, we can compute the limit close to another unrefined point, when $[\phi_3]=-[\phi_1]-[\phi_2]-2$, which is given by:
	\begin{equation}
		[\phi_1]=[\phi_2]=[\phi_3]=-\tfrac{2}{3}\,.
	\end{equation}
	In this case, the Cardy-like limit of the index yields:
	\begin{align}\label{eq:Qprime-gen-res-twin}
		\begin{split}
			\log I_N&=-i\pi N^2 \frac{[\phi_1]^\prime[\phi_2]^\prime[\phi_3]^\prime}{\tau\sigma}+\mathcal{O}\left(\frac{([\phi_{a}]'-[\phi_{b}]')^2}{\tau\sigma}\right)+\mathcal{O}(\tau^{-1})+\mathcal{O}(\sigma^{-1}) \,.
		\end{split}
	\end{align}
	where we defined:
	\begin{equation}
		[\phi_a]^\prime=[\phi_a]+1 \,.
	\end{equation}
	and $[\phi_3]'=-[\phi_1]'-[\phi_2]'+1$.
	Upon Legendre transformation, these expressions give rise to the Bekenstein--Hawking entropy for AdS$_5$ black holes with angular momenta $J_{1,2}$ and electric charges $Q_{1,2,3}$~\cite{Gutowski:2004ez, Gutowski:2004yv,Chong:2005da,Chong:2005hr,Kunduri:2006ek}.
	This agrees with the results of~\cite{Cabo-Bizet:2018ehj,Choi:2018hmj,Benini:2018ywd}.
	
	Furthermore, notice that these expressions are periodic under $\phi_{1,2}\to \phi_{1,2}+1$, which reflects the same periodicity of the index~\eqref{eq:trace-defn-index} discussed above.
	However, the expressions do not reflect the periodicity under $\tau,\sigma\to \tau,\sigma+1$.
	We will see that the generalized Cardy limit of the index results in a very similar expression as above, from which we can understand how the periodicity in $\tau$ and $\sigma$ emerges.

	\section{Modular properties of the elliptic \texorpdfstring{$\Gamma$}{Gamma} function}\label{sec:mod-prop-gamma}
	
	In this section, we will derive a new modular property for the elliptic $\Gamma$ function.
	We subsequently employ this property to compute a generalized Cardy limit of the $\Gamma$ function.
	Before getting there, we take the first three subsections to review some crucial preliminaries.
	
	The main inspiration for this section is the recent work on modularity of supersymmetric partition functions~\cite{Gadde:2020bov}, which itself is inspired by earlier mathematical work on the elliptic $\Gamma$ function~\cite{Felder_2000}.
	In the first two subsections, we will review relevant parts of~\cite{Gadde:2020bov}, such as the main modular property of four-dimensional supersymmetric partition functions and aspects of the associated modular group $SL(3,\mathbb{Z})\ltimes \mathbb{Z}^3$.
	In the subsequent section, we will specialize to the partition function of a free chiral multiplet and introduce the formalism of~\cite{Felder_2000}, which will help us to find a new modular property of the elliptic $\Gamma$ function.
	After these preliminaries, we will turn to the derivation of a new modular property and the associated generalized Cardy limit.

	\subsection{Modularity of supersymmetric partition functions}\label{ssec:review-Gadde}
	
	Consider an $\mathcal{N}=1$ SCFT defined on a manifold that can be viewed as the gluing of two solid three-tori along their boundaries.
	The boundaries may be identified up to the action of a large symmetry.
	If the theory has a global symmetry group of rank $r$, the full group of large symmetries on the boundary torus is given by $G=SL(3,\mathbb{Z})\ltimes \mathbb{Z}^{3r}$.
	The group $SL(3,\mathbb{Z})$ acts on the moduli of the torus whereas the $\mathbb{Z}^{3r}$ part shifts the global symmetry chemical potentials by periods of the torus.
	The manifold glued with a large symmetry $g\in G$ will be denoted by $M_g$.
	(A more precise definition for the gluing will be given in Section~\ref{ssec:sl3}.)
	Manifolds thus constructed include $S^2\times T^2$, where the gluing group element $g$ is the identity, and $L(r,s)\times S^1$ where $L(r,s)$ is a lens space.
	
	The special case $S^3\times S^1$ can be obtained when the gluing group element $g$ is an $S$-transformation inside a certain $SL(2,\mathbb{Z})\subset SL(3,\mathbb{Z})$.
	In this paper, we will be mainly concerned with the superconformal index, i.e., the geometry $S^3\times S^1$, although we make some comments about lens spaces in Section~\ref{ssec:lens}.
	
	We will be interested in the (normalized) supersymmetric partition functions defined on the manifold $M_g$.
	We use the short hand notation:
	\begin{equation}
		\hat{Z}^{a}_g\left(\boldsymbol{\rho}\right)\equiv \hat{Z}^{a}\left[M_g\right]\left(\boldsymbol{\rho}\right) \,,
	\end{equation}
	where $\boldsymbol{\rho}$ indicates the full set of chemical potentials, i.e., the complex structure moduli of $M_g$ and the chemical potentials for the global symmetry:
	\begin{equation}\label{eq:Mg-moduli}
		\boldsymbol{\rho}\equiv (z_1,\ldots,z_r;\tau,\sigma) \,.
	\end{equation}
	In addition, the hat denotes a normalized partition function, defined as:
	\begin{equation}
		\hat{Z}^{a}\left[M_g\right]\left(\boldsymbol{\rho}\right)\equiv\frac{Z^{a}\left[M_g\right]\left(\boldsymbol{\rho}\right)}{Z^{a}\left[M_1\right]\left(g^{-1}\boldsymbol{\rho}\right)} \,,
	\end{equation}
	where $Z^{a}\left[M_g\right]$ is the supersymmetric partition function on $M_g$ in a given Higgs branch vacuum $|a\rangle$ of the mass deformed theory and $M_1 = S^2\times T^2$ is the manifold obtained by gluing with the identity element.\footnote{
		See e.g.,~\cite{Peelaers:2014ima,Gadde:2020bov} and references therein for more details on Higgs branch localization and the type of twist used to preserve supersymmetry on $M_g$.}
	The ordinary partition function obtained through Higgs branch localization would be given by: $Z\left[M_g\right]=\sum_a Z^{a}\left[M_g\right]$, where $a$ runs over all the Higgs branch vacua of the mass deformed theory, which are typically finite in number.
	In the case of the index of the $\mathcal{N}=4$ theory, as reviewed in Section~\ref{ssec:review-index}, $a=(a_i)$ in the notation introduced there.
	
	The action $g \boldsymbol{\rho}$ of a group element $g\in G$ on $\boldsymbol{\rho}$ will be described in more detail in the following section.
	For now, we note that $SL(3,\mathbb{Z})$ acts projectively on $\tau$ and $\sigma$, while the $\mathbb{Z}^3$ factor is generated by shifts of $z_i$ with cycles $1$, $\tau$ and $\sigma$. 
	
	It is proposed in~\cite{Gadde:2020bov} that $\hat{Z}^{a}_g$ can be regarded as an element of the first group cohomology $H^1(G,N/M)$.\footnote{\label{fn:FV}
		This is partially based on the mathematical work~\cite{Felder_2000}, which in physical language establishes the claim in the special case when $\hat{Z}_g$ is a partition function of the free chiral multiplet.}
	Here, $N$ and $M$ represent, respectively, the spaces of meromorphic and holomorphic, nowhere vanishing, functions on the associated fugacity space.
	The space $M$ can be represented by complex phases $e^{i\phi}$, with $\phi$ a holomorphic function of the chemical potentials.
	We refer the reader who is interested in more mathematical details to Section~7 of~\cite{Felder_2000}. (See also~\cite{Gadde:2020bov}.)
	Instead, we will focus on a concrete implication of this statement, namely that $\hat{Z}^{a}_g$ obeys the following property:
	\begin{equation}\label{eq:sl3-reln-part-funct}
		\begin{aligned}
			\hat{Z}^{a}_{g_1 \cdot g_2}\left(\boldsymbol{\rho}\right)&=e^{i\phi_{g_1,g_2}\left(\boldsymbol{\rho}\right)}\hat{Z}^{a}_{g_1}\left(\boldsymbol{\rho}\right)\hat{Z}^{a}_{g_2}\left(g^{-1}_1\boldsymbol{\rho}\right)\\
			&=\hat{Z}^{a}_{g_1}\left(\boldsymbol{\rho}\right)\hat{Z}^{a}_{g_2}\left(g^{-1}_1\boldsymbol{\rho}\right)\quad \mod M \,.
		\end{aligned}
	\end{equation}
	Here, $e^{i\phi_{g_1,g_2}(\boldsymbol{\rho})}$ can be understood as an element in $H^2(G,M)$ and satisfies the property:
	\begin{equation}
		e^{i\phi_{g_1\cdot g_2,g_3}\left(\boldsymbol{\rho}\right)}e^{i\phi_{g_1, g_2}\left(\boldsymbol{\rho}\right)}=
		e^{i\phi_{g_1,g_2\cdot g_3}\left(\boldsymbol{\rho}\right)}e^{i\phi_{g_2,g_3}\left(g_1^{-1}\boldsymbol{\rho}\right)}.
	\end{equation}
	Also, in the last equality of~\eqref{eq:sl3-reln-part-funct} we work $\mod M$ and may therefore cancel the phase.
	Finally, note that~\eqref{eq:sl3-reln-part-funct} allows one to compute the normalized partition function on any manifold $M_{g}$ as soon as one knows $\hat{Z}^{a}_{g_i}$ and $e^{i\phi_{g_1,g_2}\left(\boldsymbol{\rho}\right)}$ for $g_{i}$ the generators of $G$.\footnote{
		The phases $e^{i\phi_{g_1,g_2}\left(\boldsymbol{\rho}\right)}$ contain all the important information associated with the generalized Cardy limit we will consider.
		It turns out that knowing $e^{i\phi_{g_1,g_2}\left(\boldsymbol{\rho}\right)}$ for basic relations in $G$ suffices to compute the phase for any relation in $G$. 
		We will come back to this in Section~\ref{ssec:partn-fn-sl3} and Appendix~\ref{app:deriv-Q} in more detail.}
	As we will see below, modulo the phase factor, all \emph{normalized} partition functions can be expressed in terms of a single $\hat{Z}_g$ for $g=S_{23}\in G$, whose associated manifold is $S^3\times S^1$.
	This implies that the normalized partition function on any manifold $M_g$ can be expressed as a product of normalized superconformal indices.
	
	One may compare the property~\eqref{eq:sl3-reln-part-funct} to the more standard property of the superconformal index of two-dimensional SCFTs (a.k.a.\ the elliptic genus~\cite{Kawai:1993jk}) under the Jacobi group $J=SL(2,\mathbb{Z})\ltimes \mathbb{Z}^2$:
	\begin{equation}\label{eq:sl2-reln-part-funct}
		Z\left(z;\tau\right)=e^{i\phi_{g}\left(z;\tau\right)}Z\left(g^{-1}(z;\tau)\right), \quad g\in J \,.
	\end{equation}
	Notice that in this case, the partition function is not labelled by a group element.
	Indeed, the two-dimensional index can be thought of as an element in $H^0(J,N/M)$, whereas $e^{i\phi_{g}\left(z;\tau\right)}$ sits in $H^1(J,M)$.
	A crucial aspect of~\eqref{eq:sl2-reln-part-funct} is that it provides a useful relation between the torus partition function with parameter $\tau$ and the torus partition function on any modular image of $\tau$.
	The same is not immediately true for~\eqref{eq:sl3-reln-part-funct}, since, as it stands, it only allows one to compute a partition function on some manifold $M_{g_1g_2}$ in terms of a product of partition functions on $M_{g_1}$ and $M_{g_2}$. (We will discuss the interpretation of this equation in more detail in Section~\ref{ssec:mod-norm-Z}.)
	
	One way to obtain interesting relations between partition functions on a single manifold $M_g$ is when $g$ has finite order, i.e., $g^n=1$ for some $n$.
	This is because in this case the property~\eqref{eq:sl3-reln-part-funct} implies:
	\begin{equation}\label{eq:mod-prop-order-n}
		\begin{aligned}
			1&=\hat{Z}^{a}_{g^n}(\boldsymbol{\rho})\quad \mod M\\
			&=\hat{Z}^{a}_{g}\left(\boldsymbol{\rho}\right)\,\hat{Z}^{a}_{g}\left(g^{-1} \boldsymbol{\rho}\right)\cdots\hat{Z}^{a}_{g}\left(g^{-n+1} \boldsymbol{\rho}\right) \quad\mod M \,.
		\end{aligned}
	\end{equation}
	Together with the associated phase, this type of equations will play a central role in the rest of this paper.
	Notice that the modular property described in~\cite{Gadde:2020bov} is a special case of this relation for the element $Y\in SL(3,\mathbb{Z})$ which cyclically permutes the periods of the torus.
	
	Over the next two subsections, we will review some background on the group $SL(3,\mathbb{Z})\ltimes \mathbb{Z}^{3}$ and the work~\cite{Felder_2000}.
	Thereafter, we will describe an interesting set of finite order elements in $SL(3,\mathbb{Z})$, motivated by~\eqref{eq:mod-prop-order-n}, and derive the associated modular properties for the elliptic $\Gamma$ function.
	In the last subsection, we will detail how this new modular property can be used to find a generalized Cardy limit of $\Gamma(z;\tau,\sigma)$.

	\subsection{The group \texorpdfstring{$SL(3,\mathbb{Z})\ltimes \mathbb{Z}^3$}{SL(3,Z)xZ3}}\label{ssec:sl3}
	
	In this section, we will closely follow the presentation of Gadde~\cite{Gadde:2020bov} on $G=SL(3,\mathbb{Z})\ltimes \mathbb{Z}^3$, adapted to the contents of Felder and Varchenko~\cite{Felder_2000}.
	
	The action of $G$ is defined on the complex structure moduli $\hat{\tau}$ and $\hat{\sigma}$ of a three torus $T^3$ and a chemical potential $\hat{z}$, associated to a line bundle over $T^3$.\footnote{\label{fn:tilde-moduli}We use the hatted moduli to indicate moduli of the three torus, while the unhatted moduli denote the complex structure moduli of the full manifold $M_g$ as in~\eqref{eq:Mg-moduli}. Generally, these moduli will be related by some $G$ transformation. We will describe the explicit map only in Section~\ref{sec:disc}, where it will play an important role.}
	The action is most conveniently understood by thinking of a rectangular torus and using affine coordinates $\hat{Z}$ and $\hat{x}_i$:
	\begin{equation}\label{eq:def-xi}
		(\hat{z};\hat{\tau},\hat{\sigma})\simeq \left(\frac{\hat{Z}}{\hat{x}_1};\frac{\hat{x}_2}{\hat{x}_1},\frac{\hat{x}_3}{\hat{x}_1}\right) \,.
	\end{equation}
	The $SL(3,\mathbb{Z})$ acts on the vector $\hat{x}_i$ by (left) matrix multiplication.
	The $\mathbb{Z}^3$ factor is generated by shifts of $\hat{Z}$ with $\hat{x}_i$.
	In the four-dimensional geometry, i.e., the solid torus $D_2\times T^2$, we think of $\hat{x}_{1,3}$ as the cycle lengths of the non-contractible $T^2$, whereas $\hat{x}_2$ is taken to be the length of $\partial D_2$.\footnote{
		Note that our convention differs from~\cite{Gadde:2020bov} by $\hat{\tau}\leftrightarrow\hat{\sigma}$.}
	
	The group $SL(3,\mathbb{Z})$ can be generated by $3\times 3$ elementary matrices $\{T_{ij}\}$ which obey certain relations.
	These matrices $T_{ij}$ with $1\leq i\neq j\leq 3$ can be represented as $3\times 3$ matrices that differ from the identity matrix by an element $1$ at the position $ij$, e.g.,
	\begin{equation}
		T_{13}=\begin{pmatrix}
			1&\ 0\ &1\\
			0&\ 1\ &0\\
			0&\ 0\ &1
		\end{pmatrix} \,, \qquad
		T_{21}=\begin{pmatrix}
			1&\ 0\ &0\\
			1&\ 1\ &0\\
			0&\ 0\ &1
		\end{pmatrix} \,.
	\end{equation}
	Note that $T_{ij}$ and $T_{ji}$ comprise the $T$ and $STS^{-1}$ matrices of the  three $SL(2,\mathbb{Z})$ subgroups of $SL(3,\mathbb{Z})$.
	These matrices obey the $SL(3,\mathbb{Z})$ relations:
	\begin{equation}\label{eq:sl3-relns}
		T_{ij}T_{kl}=T_{kl}T_{ij} \quad (i\neq l , j\neq k) \,,\quad T_{ij}T_{jk}=T_{ik}T_{jk}T_{ij} \,, \quad (T_{13}T^{-1}_{31}T_{13})^4=1 \,.
	\end{equation}
	The last relation holds for any string of the form $T_{ij}T^{-1}_{ji}T_{ij}$, as can be checked explicitly through conjugation of the equation by $T_{kl}T^{-1}_{lk}T_{kl}$ for appropriate $k$ and $l$, while making use of the first two relations.
	
	As one may have guessed, the three $S$ matrices inside $SL(3,\mathbb{Z})$ can be written in terms of the $T_{ij}$ as:
	\begin{equation}
		S_{12}=T_{12}T_{21}^{-1}T_{12} \,,\quad S_{23}=T_{23}T_{32}^{-1}T_{23} \,,\quad S_{13}=T_{13}T_{31}^{-1}T_{13} \,.
	\end{equation}
	Explicitly, these are given by:
	\begin{equation}
		S_{12}=\begin{pmatrix}
			0&\ 1\ &0\\
			-1&\ 0\ &0\\
			0&\ 0\ &1
		\end{pmatrix} \,, \qquad 
		S_{23}=\begin{pmatrix}
			1&\ 0\ &0\\
			0&\ 0\ &1\\
			0&\ -1\ &0
		\end{pmatrix} \,, \qquad 
		S_{13}=\begin{pmatrix}
			0&\ 0\ &1\\
			0&\ 1\ &0\\
			-1&\ 0\ &0
		\end{pmatrix} \,.
	\end{equation}
	Notice that both $S_{12}$ and $S_{23}$ exchange the contractible circle with a non-contractible circle when viewed as acting on the solid torus.
	Since one can view $S^3$ as a torus fibration over an interval, where the $(1,0)$ cycle shrinks on one end and the $(0,1)$ cycle on the other end, we see that the manifolds $M_{S_{12}}$ and $M_{S_{23}}$ have the topology of $S^3\times S^1$.
	
	\begin{figure}[t]
		\centering
		\includegraphics[scale=0.4]{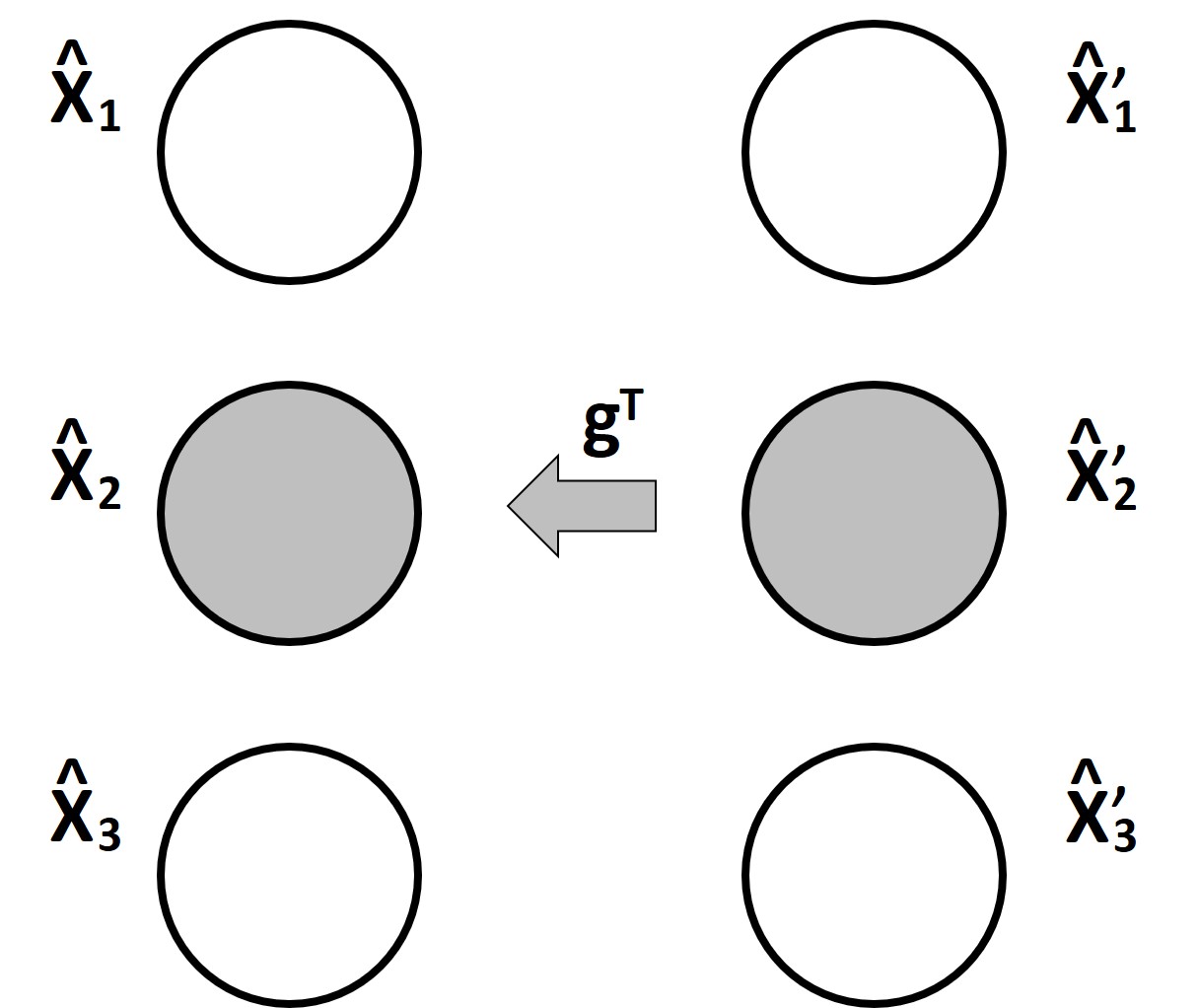}
		\caption{This figure is directly taken from~\cite{Gadde:2020bov}, but includes a small clarification as indicated in footnote~\ref{fn8}. It shows the construction of the manifold $M_g$ for $g\in G$. The left and right part represent two solid three-tori with cycle lengths $\hat{\textbf{x}}$ and $\hat{\textbf{x}}\,'=\left(g^{-1}\right)^T\hat{\textbf{x}}$ respectively. The shaded circle indicates the contractible circle. The image of the cycles $\hat{\textbf{x}}\,'$ under $g^{T}$ is glued to the cycles $\hat{\textbf{x}}$, where we suppress orientation reversal on $\hat{\textbf{x}}\,'$.}
		\label{fig:Mg}
	\end{figure}
	
	More generally, the manifold $M_g$ is constructed as indicated in Figure~\ref{fig:Mg}.
	In particular, one considers two solid three-tori with parameters $\hat{\textbf{x}}$ and $\left(g^{-1}\right)^T\hat{\textbf{x}}$, where $T$ indicates the transpose.
	Then, one identifies the image of the cycles $\left(g^{-1}\right)^T\hat{\textbf{x}}$ under $g^{T}$ to the cycles $\hat{\textbf{x}}$.\footnote{
		Alternatively, as is common in the math literature, one can view the gluing as the right multiplication with $g$ on the row vector $\hat{\textbf{x}}^Tg^{-1}$.\label{fn8}}
	
	Finally, let us refer to the generators of the $\mathbb{Z}^3$ factor of $G$ by $t_{1,2,3}$.
	The total group $G$ is then generated by $\lbrace T_{ij},t_3\rbrace$, since $t_1$ and $t_2$ may be obtained from $t_3$ by conjugation with the appropriate $S_{ij}$ element.
	
	In the following, an important role will be played by the elements of  $G$ whose action cannot be extended to the solid torus $D_2\times T^2$.
	The generators that belong to this class are the elements $T_{12}$, $T_{32}$ and $t_2$.
	This is because their action, respectively, shifts $\hat{x}_1$, $\hat{x}_3$ and $\hat{Z}$ by $\hat{x}_2$, the period of the cycle of $T^3$ which contracts in the solid torus.
	Notice that for the same reason, the actions of $S_{12}$ and $S_{23}$ cannot be extended to the solid torus either.
	All other generators can be extended and generate the group $H=SL(2,\mathbb{Z})\ltimes (\mathbb{Z}^2)^2$.
	The $SL(2,\mathbb{Z})$ is generated by $T_{13}$ and $T_{31}$.
	In addition, $t_{1,3}$, $T_{21}$ and $T_{23}$ make up the additional $(\mathbb{Z}^2)^2$.

	\subsection{Partition functions of the free chiral multiplet and \texorpdfstring{$SL(3,\mathbb{Z})$}{SL(3,Z)}}\label{ssec:partn-fn-sl3}
	
	In this section, we will review Section 7 of~\cite{Felder_2000}.
	We focus on establishing the relevant equations while omitting many of the mathematical details.
	
	As we have commented in footnote~\ref{fn:FV}, the paper~\cite{Felder_2000} establishes the proposal that the normalized partition function of a free chiral multiplet is an element of the first group cohomology of $G$, i.e., $\hat{Z}^{a}_g\in H^1\left(G,N/M\right)$.
	In this case, since the free chiral has a single vacuum we will drop the superscript $a$ and write $\hat{Z}_g\equiv \hat{Z}_g^{a}$.
	It evaluates on the generators of $G$ outside $H$ in terms of the following representatives:
	\begin{align}\label{eq:hatZ-on-gens-1}
		\begin{split}
			\hat{Z}_{T_{12}}\left(\textbf{x}\right)&=\Gamma\left(\tfrac{Z-x_2}{x_3};\tfrac{x_1-x_2}{x_3},-\tfrac{x_1}{x_3}\right)^{-1}\quad\mod M \,,\\
			\hat{Z}_{T_{32}}\left(\textbf{x}\right)&=\Gamma\left(\tfrac{Z}{x_1};\tfrac{x_2-x_3}{x_1},\tfrac{x_3}{x_1}\right)\quad \mod M \,,\\
			\hat{Z}_{t_2}\left(\textbf{x}\right)&=\theta\left(\tfrac{Z-x_2}{x_1};\tfrac{x_3}{x_1}\right)\quad \mod M \,,
		\end{split}
	\end{align}
	while for the generators in $H$ it evaluates as:
	\begin{align}\label{eq:hatZ-on-gens-2}
		\begin{split}
			\hat{Z}_{T_{ij}}\left(\textbf{x}\right)&=1\quad \mod M \,, \quad \text{for} \quad j\neq 2 \,,\\
			\hat{Z}_{t_i}\left(\textbf{x}\right)&=1\quad \mod M \,, \quad \text{for} \quad i\neq 2 \,.
		\end{split}
	\end{align}
	Here, we defined the unhatted affine moduli $\textbf{x}\equiv (Z;x_1,x_2,x_3)$ of $M_g$, in terms of which we define its complex structure moduli as $\boldsymbol{\rho}=(z;\tau,\sigma)\equiv (\frac{Z}{x_1}; \frac{x_2}{x_1},\frac{x_3}{x_1})$.
	This should be distinguished from the hatted moduli of the $T^3$ defined in~\eqref{eq:def-xi}.
	We will think of $x_i$ as complex variables.
	The action of $G$ on $\textbf{x}$ is identical to the action of $G$ on $\hat{\textbf{x}}$, described in Section~\ref{ssec:sl3}. 
	In the following, we will be using $\textbf{x}$ instead of $\boldsymbol{\rho}$ whenever an action of $G$ is involved, since this action is easier to describe.
	
	These equations, together with the property~\eqref{eq:sl3-reln-part-funct}, allow one to determine $\hat{Z}_g$ for any $g\in G$ up to a phase.
	Notice that $\hat{Z}_g$ is non-trivial only for the generators $g\in G$ which cannot be extended to the solid torus.\footnote{
		See the discussion at the end of the previous section.}
	It is not difficult to check, using the gluing prescription of the previous section, that $M_{T_{12}}$ and $M_{T_{32}}$ have the topology of the lens space $L(1,1)\times S^1$ which is nothing but $S^3\times S^1$.
	On the other hand, the generators $T_{ij}$ for $j\neq 2$ lead to $M_{T_{ij}}$ with $S^2\times T^2$ topology, on which one may have anticipated that normalized partition function should be trivial (up to a phase).
	
	Because of the above, we will introduce the following (abuse of) terminology.
	We will refer to an element $g$ inside $H$ as \emph{cohomologically trivial}.
	Instead, if the decomposition of an element in $G$ contains at least one $T_{12}$, $T_{32}$ or $t_2$, we will call such an element \emph{cohomologically non-trivial}.
	We will always have in mind that the presentation of an element $g\in G$ is irreducible in the sense that the number of cohomologically non-trivial generators it consists of cannot be reduced by making use of the $SL(3,\mathbb{Z})$ relations~\eqref{eq:sl3-relns}.
	Finally, notice that the property~\eqref{eq:sl3-reln-part-funct} implies that $\hat{Z}_g$ for $g\in G$ containing $k_1$ $T_{12}$ and $k_2$ $T_{32}$ factors will be evaluated in general to a partition function which consists of $k_1+k_2$ elliptic $\Gamma$ functions.
	
	As mentioned in the previous section, the elements $S_{12}$ and $S_{23}$ lead to a manifold $M_g$ with $S^3\times S^1$ topology.
	We will now show that $\hat{Z}_{S_{23}}=\hat{Z}_{T_{23}T_{32}^{-1}T_{23}}$ computes the usual (normalized) superconformal index of a free chiral multiplet.
	To this end, we first compute $\hat{Z}_{T_{23}T_{32}^{-1}T_{23}}$ using the definitions~\eqref{eq:hatZ-on-gens-1} and~\eqref{eq:hatZ-on-gens-1}, and the property~\eqref{eq:sl3-reln-part-funct}:
	\begin{align}\label{eq:sci-from-z-t32}
		\begin{split}
			\hat{Z}_{T_{23}T_{32}^{-1}T_{23}}\left(\textbf{x}\right)&=\frac{1}{\hat{Z}_{T_{32}}\left(T_{32}T_{23}^{-1}\textbf{x}\right)}\quad \mod M\\ &=\Gamma\left(\tfrac{Z+x_3}{x_1};\tfrac{x_2}{x_1},\tfrac{x_3}{x_1}\right)\quad \mod M \,,
		\end{split}
	\end{align}
	where in the last line we simplified using properties of the elliptic $\Gamma$ function collected in Appendix~\ref{app:defs}, while in the first line we used the fact that:
	\begin{equation}\label{eq:Zhat-inv}
		\begin{aligned}
			1&=\hat{Z}_{g^{-1}\cdot g}\left(\textbf{x}\right)\quad \mod M\\
			&=\hat{Z}_{g^{-1}}\left(\textbf{x}\right)\hat{Z}_{g}\left(g\,\textbf{x}\right)\quad \mod M \,.
		\end{aligned}
	\end{equation}
	We can check that this is the correct result for the normalized superconformal index of the anomaly-free chiral multiplet by comparing with explicit expressions available in the literature~\cite{Closset:2013sxa}.
	Let us denote by $Z_{S_{23}}$ the superconformal index and by $Z_1$ the partition function associated to the theory on $S^2\times T^2$.
	These are given by:
	\begin{equation}
		Z_{S_{23}}\left(\boldsymbol{\rho}\right)=\Gamma(z;\tau,\sigma) \,, \qquad Z_{1}\left(\boldsymbol{\rho}\right)=\frac{1}{\theta(z;\sigma)} \,.
	\end{equation}
	The normalized partition function is now given by:
	\begin{equation}\label{eq:sci-free-chiral}
		\hat{Z}_{S_{23}}\left(\boldsymbol{\rho}\right)=\frac{Z_{S_{23}}\left(\boldsymbol{\rho}\right)}{Z_1\left(S_{23}^{-1}\boldsymbol{\rho}\right)}=\Gamma(z+\sigma;\tau,\sigma) \,,
	\end{equation}
	which is precisely what we found in~\eqref{eq:sci-from-z-t32} if we use the identifications spelled out below~\eqref{eq:hatZ-on-gens-2}.
	
	Even though the equivalence of the above computations should only hold $\mod M$, we see that it holds exactly.
	This can be understood by using
	the ``gauge fixed'' formalism of~\cite{Felder_2000}, which allows us to compute $\hat{Z}_g$ exactly, i.e., not $\mod M$.
	This formalism will be crucial for the computation of the (generalized) Cardy limit of the elliptic $\Gamma$ function.
	In this formalism, one takes the partition functions $\hat{Z}_{T_{ij}}$ equal to the representatives in~\eqref{eq:hatZ-on-gens-1} and~\eqref{eq:hatZ-on-gens-2}.
	In particular, this means that: 
	\begin{equation}
		\hat{Z}_{h}(\mathbf{x})=1 \quad \textrm{for}\quad h\in H\,.
	\end{equation}
	In addition, using properties of the elliptic $\Gamma$ function one may notice that for these representatives, the relation~\eqref{eq:Zhat-inv} holds without working $\mod M$:
	\begin{equation}
		\hat{Z}_{g^{-1}}\left(\textbf{x}\right)=\hat{Z}_{g}\left(g\,\textbf{x}\right)^{-1} \,.
	\end{equation}
	Furthermore, specific phases are associated to the evaluation of $\hat{Z}_{(\cdot)}$ on the basic $SL(3,\mathbb{Z})$ relations~\eqref{eq:sl3-relns}.
	The meaning of this is summarized by the equations:\footnote{We will not make use of the $\mathbb{Z}^3$ factor of $G$, generated by the $t_i$. Therefore, we do not include relations in the group associated to the $t_i$. The interested reader may find these in Section 7 of~\cite{Felder_2000}.}
	\begin{align}\label{eq:Felder-relations}
		\begin{split}
			&\hat{Z}_{T_{ij}}\left(\textbf{x}\right)\hat{Z}_{T_{kl}}\left(T_{ij}^{-1}\textbf{x}\right)=\phi^{k,l}_{i,j}\left(\textbf{x}\right)\hat{Z}_{T_{kl}}\left(\textbf{x}\right)\hat{Z}_{T_{ij}}\left(T_{kl}^{-1}\textbf{x}\right),\qquad i\neq l \,,\quad j\neq k \,,\\
			&\hat{Z}_{T_{ij}}\left(\textbf{x}\right)\hat{Z}_{T_{jk}}\left(T_{ij}^{-1}\textbf{x}\right)=\phi^{j,k}_{i,j}\left(\textbf{x}\right)\hat{Z}_{T_{ik}}\left(\textbf{x}\right)\hat{Z}_{T_{jk}}\left(T_{ik}^{-1}\textbf{x}\right)\hat{Z}_{T_{ij}}\left(T_{jk}^{-1}T_{ik}^{-1}\textbf{x}\right)  \,,\\
			&\hat{Z}_{S_{13}}\left(\textbf{x}\right)\hat{Z}_{S_{13}}\left(S_{13}^{-1}\textbf{x}\right)\hat{Z}_{S_{13}}\left(S_{13}^{-2}\textbf{x}\right)\hat{Z}_{S_{13}}\left(S_{13}^{-3}\textbf{x}\right)=1 \,.
		\end{split}
	\end{align}
	These equations reflect non-trivial Berry curvature on the parameter space~\cite{Gadde:2020bov}.
	Indeed, the first equation can be viewed as describing how the partition functions changes under the non-trivial loop in parameter space:
	\begin{equation}
		\textbf{x}\to T_{kl}^{-1}\,\textbf{x}\to T_{ij}^{-1}T_{kl}^{-1}\,\textbf{x}\to T_{kl}T_{ij}^{-1}T_{kl}^{-1}\,\textbf{x}\to T_{ij}T_{kl}T_{ij}^{-1}T_{kl}^{-1}\,\textbf{x}=\textbf{x},
	\end{equation}
	and similarly for the other equations.
	Furthermore, $\phi^{k,l}_{i,j}(\textbf{x})\equiv e^{i\pi L^{k,l}_{i,j}(\textbf{x})}$ and $\phi^{j,k}_{i,j}(\textbf{x}) \equiv e^{i\pi L^{j,k}_{i,j}(\textbf{x})}$ with:
	\begin{align}\label{eq:L-phases}
		\begin{split}
			L^{3,2}_{1,2}(\textbf{x})=-L^{1,2}_{3,2}(\textbf{x})&=Q\left(\tfrac{Z-x_1}{x_1};\tfrac{x_2-x_3}{x_1},\tfrac{x_3-x_1}{x_1}\right)+Q\left(\tfrac{Z-x_1+x_3}{x_1-x_3};\tfrac{x_3}{x_1-x_3},\tfrac{x_2-x_1}{x_1-x_3}\right) \,,\\
			L^{1,2}_{3,1}(\textbf{x})&=Q\left(\tfrac{Z-x_1}{x_1};\tfrac{x_2-x_3}{x_1},\tfrac{x_3-x_1}{x_1}\right) \,,\\
			L^{3,2}_{1,3}(\textbf{x})&=-Q\left(\tfrac{Z-x_1+x_3}{x_1-x_3};\tfrac{x_2-x_1}{x_1-x_3},\tfrac{x_3}{x_1-x_3}\right) \,,
		\end{split}
	\end{align}
	where $Q$ is defined by: 
	\begin{align}\label{eq:defn-Q-FV}
		\begin{split}
			Q(z;\tau,\sigma)&=\frac{z^3}{3\tau\sigma}-\frac{\tau+\sigma-1}{2\tau\sigma}z^2+\frac{\tau^2+\sigma^2+3\tau\sigma-3\tau-3\sigma+1}{6\tau\sigma}z\\
			&+\frac{1}{12}(\tau+\sigma-1)(\tau^{-1}+\sigma^{-1}-1) \,.
		\end{split}
	\end{align}
	and all other $L$ vanish.
	
	Consider now some arbitrary relation $r=1$ in the group $SL(3,\mathbb{Z})$, where the relation is a reduced expression $r=e_1\cdots e_n$ in terms of generators $e_k\in \lbrace T_{ij}^{\pm 1}\rbrace$, where by reduced is meant that any adjacent $e_{i+1}=e_i^{-1}$ are canceled.
	Evaluating the partition function on this relation~\cite{Felder_2000}:
	\begin{equation}
		\hat{Z}_{e_1}\left(\textbf{x}\right)\prod^{n-1}_{i=1} \hat{Z}_{e_{i+1}}\left(e_i^{-1}\cdots e_1^{-1} \textbf{x}\right) =e^{i\pi f_{r}\left(\textbf{x}\right)} \,,
	\end{equation}
	one can compute the associated phase $f_{r}\left(\textbf{x}\right)$ by shuffling around the $\hat{Z}_{e_i}$ using the relations~\eqref{eq:Felder-relations} and simplifying whenever one encounters two adjacent factors expressed as $\hat{Z}_{T_{ij}}(\textbf{x})Z_{T_{ij}^{-1}}(T_{ij}^{-1}\textbf{x})=1$ .
	Thus, $f_{r}\left(\textbf{x}\right)$ is computed as an accumulation of the non-trivial phases appearing in~\eqref{eq:Felder-relations}.
	This brings us back to an understanding of why~\eqref{eq:sci-from-z-t32} reproduces the exact superconformal index $\hat{Z}_{S_{23}}$.
	There is no phase acquired in the computation because no basic relation has been used to derive it.
	More details and non-trivial examples of phase computations are collected in Appendix~\ref{app:deriv-Q}.

	\subsection{Finite order elements of \texorpdfstring{$SL(3,\mathbb{Z})$}{SL(3,Z)}}
	\label{ssec:finite-order-elms}
	
	As mentioned at the end of Section~\ref{ssec:review-Gadde}, useful modular properties of $\hat{Z}_g$ can be derived from finite order elements in $SL(3,\mathbb{Z})$.
	For example, evaluating~\eqref{eq:mod-prop-order-n} for a free chiral multiplet will give rise to a modular property of the elliptic $\Gamma$ function involving a multiple of $n$ $\Gamma$ functions.
	It is our goal to find a finite order element that allows us to compute a generalized Cardy limit of the form $\tau\to -\frac{n_1}{m_1}$ and $\sigma\to -\frac{n_2}{m_2}$, which could for example be achieved by a relation of the form:
	\begin{equation}\label{eq:wish-mod-prop}
		\Gamma(z;\tau,\sigma)=e^{i\pi P(z;\tau,\sigma)}\Gamma\left(\tfrac{z+\ldots}{m_1\tau+n_1};\tfrac{\ldots}{m_1\tau+n_1},\tfrac{\ldots}{m_1\tau+n_1}\right)\Gamma\left(\tfrac{z+\ldots}{m_2\sigma+n_2};\tfrac{\ldots}{m_2\sigma+n_2},\tfrac{\ldots}{m_2\sigma+n_2}\right)\,,
	\end{equation}
	where we have not explicitly written some linear function of $\tau$ and $\sigma$ in the numerators.
	Such a relation would be derived from some order three elements.
	The point of this property is that the $\Gamma$ functions on the right hand side could trivialize in the generalized Cardy limit, giving a simple expression for the limit of $\Gamma(z;\tau,\sigma)$ in terms of the phase $e^{i\pi P(z;\tau,\sigma)}$.
	In this section, we will therefore describe in some detail a class of finite order elements in $SL(3,\mathbb{Z})$.
	
	We start by recalling that the characteristic polynomial of an element of order $n$ in $SL(3,\mathbb{C})$ should divide $\lambda^n-1$.
	Let the corresponding eigenvalues be denoted by the $n$\textsuperscript{th} roots of unity $\zeta_{1,2,3}$.
	For the element to be in $SL(3,\mathbb{Z})$, the determinant and trace constraint require that $\zeta_1\zeta_2\zeta_3=1$ and $\zeta_1+\zeta_2+\zeta_3\in \mathbb{Z}$.
	It is then not difficult to convince oneself that the finite order elements in $SL(3,\mathbb{Z})$ are of order $1$, $2$, $3$, $4$, and $6$.
	We will now argue that only specific order three elements lead to modular properties which could result in a simple expression for a generalized Cardy limit of the elliptic $\Gamma$ function.
	
	Clearly, order one and two elements lead to trivial modular properties.
	Furthermore, the (cohomologically non-trivial) order four and order six elements lead to modular properties involving (at least) four or six elliptic $\Gamma$ functions.
	This follows from~\eqref{eq:mod-prop-order-n} when $\hat{Z}_g$ is taken as the partition function of a free chiral multiplet.
	Although this could potentially lead to interesting relations, they generally will not lead to an interesting Cardy limit.
	The reason for this is best illustrated by returning briefly to the ordinary modular property we used in~\cite{Goldstein:2020yvj}:
	\begin{equation}\label{eq:ord-mod-prop-Gamma}
		\Gamma(z;\tau,\sigma)=e^{-i\pi Q(z;\tau,\sigma)}\frac{\Gamma\left(\frac{z}{\tau};\frac{\sigma}{\tau},-\frac{1}{\tau}\right)}{\Gamma\left(\frac{z-\tau}{\sigma};-\frac{\tau}{\sigma},-\frac{1}{\sigma}\right)} \,,
	\end{equation}
	where the ingredients are defined in Appendix~\ref{app:defs}.
	This property follows from~\eqref{eq:mod-prop-order-n} for the order three element  $Y=S_{23}^{-1}S_{13}$ discussed in~\cite{Gadde:2020bov}, as we review in Appendix~\ref{app:Y-phase}.
	In the Cardy limit $\tau,\sigma\to 0$, the two $\Gamma$ functions on the right hand side trivialize (when restricted to some domain in $z$, as discussed in detail in~\cite{Goldstein:2020yvj}).
	Therefore, this modular property provides a \emph{useful} Cardy-like limit, in the sense that this limit of the elliptic $\Gamma$ function is replaced by the simple function $e^{-i\pi Q(z;\tau,\sigma)}$.
	
	Now, if one were given an order four or order six element, the associated modular property would result in a similar rewriting of $\Gamma(z;\tau,\sigma)$, but now with (at least) three or five elliptic $\Gamma$ functions on the right hand side.
	Generically, the denominators of their arguments do not share the same vanishing locus in $(\tau,\sigma)$ space, implying that not all elliptic $\Gamma$ functions trivialize in the same Cardy-like limit.
	Therefore, such modular properties appear not useful in the sense explained above.
	
	We pause here to stress that once given a useful modular property, such as~\eqref{eq:ord-mod-prop-Gamma}, it can also be applied to partition functions on manifolds $M_g$ for which $g$ is not necessarily of finite order.
	An interesting example is the partition function of a chiral multiplet on a lens space, which was first acquired in~\cite{Benini:2011nc} (see also~\cite{Razamat:2013jxa,Razamat:2013opa,Gadde:2020bov}):
	\begin{equation}\label{eq:partition-lens}
		Z_{L(m,1)\times S^1 } = \Gamma(z+\sigma;\tau,\sigma) \Gamma(z;m \tau-\sigma,\tau) \,, \quad m\in\mathbb{Z} \,.
	\end{equation}
	Here, $L(m,1)$ can be thought of as $M_g$ for $g=S_{23}T_{23}^{-m}S_{23}$.
	Even though this element is not of finite order, it is clear that applying~\eqref{eq:ord-mod-prop-Gamma} to both $\Gamma$ functions, one obtains a modular property for the lens space partition function which has a useful Cardy limit.
	
	For the reasons explained above, we will from now on only be interested in order three elements.
	It turns out that in $SL(3,\mathbb{Z})$, there are two conjugacy classes of order three elements~\cite{nmj/1118798212}.
	In terms of generators, simple representatives are given by:
	\begin{equation}
		X_k=T_{23}T_{13}^{-k}S_{23}\,,\quad k=0,1\,.
	\end{equation}
	We note here that the order three element $Y$, which cyclically permutes the cycles of the torus, sits in the conjugacy class of $X_1$, and can be obtained explicitly through conjugation by $T_{21}T_{31}$.
	Let us denote by $X_k^g$ the element obtained from $X_k$ by conjugation with a general element $g\in G$:
	\begin{equation}
		X^g_k\equiv g\, X_k \, g^{-1}\,.
	\end{equation}
	We now use~\eqref{eq:sl3-reln-part-funct} to evaluate the partition function on the relation $\left(X_k^g\right)^3=1$:
	\begin{align}\label{eq:gen-mod-prop-conjugate-ord-3-element}
		\begin{split}
			1&=\hat{Z}_{\left(X_k^g\right)^3}\left(\boldsymbol{\rho}\right)\;\mod M\\
			&=\hat{Z}_{g}\left(\boldsymbol{\rho}\right)\hat{Z}_{X_k}\left(g^{-1}\boldsymbol{\rho}\right)\hat{Z}_{X_k}\left(X_k^{-1}g^{-1}\boldsymbol{\rho}\right)\hat{Z}_{X_k}\left(X_k^{-2}g^{-1}\boldsymbol{\rho}\right)\hat{Z}_{g^{-1}}\left(g^{-1}\boldsymbol{\rho}\right)\;\mod M\\
			&=\hat{Z}_{X_k}\left(g^{-1}\boldsymbol{\rho}\right)\hat{Z}_{X_k}\left(X_k^{-1}g^{-1}\boldsymbol{\rho}\right)\hat{Z}_{X_k}\left(X_k^{-2}g^{-1}\boldsymbol{\rho}\right)\;\mod M\,,
		\end{split}
	\end{align}
	where in the last line we used~\eqref{eq:Zhat-inv}.
	Since $X_k$ only contains one factor outside $H$, namely $S_{23}$, the partition function of a free chiral multiplet evaluates to a single elliptic $\Gamma$ function:
	\begin{equation}
		\hat{Z}_{X_k}\left(\boldsymbol{\rho}\right)=\hat{Z}_{S_{23}}\left(T_{13}^{k}T_{23}^{-1}\boldsymbol{\rho}\right)=\Gamma\left(\tfrac{z+\sigma}{k\sigma+1};\tfrac{\tau-\sigma}{k\sigma+1},\tfrac{\sigma}{k\sigma+1}\right)
		\,.
	\end{equation}
	In particular, if we define $\boldsymbol{\rho}'\equiv g^{-1}\boldsymbol{\rho}$, the modular property evaluates to:
	\begin{align}\label{eq:gen-conj-mod-prop}
		\begin{split}
			1&=\Gamma\left(\tfrac{z+\sigma'}{k\sigma'+1};\tfrac{\tau'-\sigma'}{k\sigma'+1},\tfrac{\sigma'}{k\sigma'+1}\right)\Gamma\left(\tfrac{z+\tau-\sigma'}{k\tau'+1};\tfrac{-\tau'}{k\tau'+1},\tfrac{\tau'-\sigma'}{k\tau'+1}\right)\Gamma\left(z+\tau';\sigma',-\tau'\right)\;\mod M\,,
		\end{split}
	\end{align}
	where we recall that $k=0,1$.
	Here, we see that any order three element written as $gX_kg^{-1}$ for $k=0$ does not lead to a useful modular property, since the arguments of the three $\Gamma$ functions are clearly not of the desired form written in~\eqref{eq:wish-mod-prop}.
	For $k=1$, the situation is slightly better in that~\eqref{eq:gen-conj-mod-prop} would now allow us to compute the limit $\tau,\sigma\to -1$ of the elliptic $\Gamma(z;\tau,\sigma)$ function, since the $\Gamma$ function has the following property (see Appendix~\ref{app:defs}):
	\begin{equation}
		\Gamma(z;\tau,\sigma)=\frac{1}{\Gamma(z-\tau;\sigma,-\tau)}\,.
	\end{equation}
	However, clearly this still falls short of our desired expression~\eqref{eq:wish-mod-prop}.
	
	It turns out that this can be improved upon, for example, by commuting the $S_{23}$ factor of $X_k$ in the conjugate element $X_k^g$ to the right.
	Let us write this representation $X_k^g$ as:
	\begin{equation}
		X_k^g=gT_{23}T_{13}^{-k}\tilde{g}^{-1}S_{23}\,,
	\end{equation}
	where we defined $\tilde{g}^{-1}=S_{23}g^{-1}S_{23}^{-1}$.
	In this case, one finds:
	\begin{align}
		\begin{split}
			1&=\hat{Z}_{\left(gX_kg^{-1}\right)^3}\left(\boldsymbol{\rho}\right)\;\mod M\\
			&=\hat{Z}_{gT_{23}T_{13}^{-k}\tilde{g}^{-1}}\left(\boldsymbol{\rho}\right)\hat{Z}_{S_{23}}\left(S_{23}\left(X_k^g\right)^{-1}\boldsymbol{\rho}\right)\hat{Z}_{gT_{23}T_{13}^{-k}\tilde{g}^{-1}}\left(\left(X_k^g\right)^{-1}\boldsymbol{\rho}\right)\\
			&\times\hat{Z}_{S_{23}}\left(S_{23}\left(X_k^g\right)^{-2}\boldsymbol{\rho}\right)\hat{Z}_{gT_{23}T_{13}^{-k}\tilde{g}^{-1}}\left(\left(X_k^g\right)^{-2}\boldsymbol{\rho}\right)\hat{Z}_{S_{23}}\left(S_{23}\boldsymbol{\rho}\right)\;\mod M\,.
		\end{split}
	\end{align}
	In this equation, there are three partition functions labeled by $S_{23}$.
	It is not difficult to check that for generic $g$ these would lead to three $\Gamma$ functions of the form expressed in~\eqref{eq:wish-mod-prop}, although there will be some correlation between the generalized Cardy limit of $\tau$ and $\sigma$, i.e., for a given limit of $\tau\to -\frac{n_1}{m_1}$, the limit of $\sigma\to-\frac{n_2}{m_2}$ is not fully independent.
	However, the issue with this equation is that if $gT_{23}T_{13}^{-k}\tilde{g}^{-1}\notin H$, there will be additional elliptic $\Gamma$ functions in the modular property.
	These will generically spoil the existence of a useful Cardy limit for similar reasons as in the case of the order four and six elements discussed above.
	Therefore, to obtain a useful modular property, we need to constrain $g$ such that
	\begin{equation}
		gT_{23}T_{13}^{-k}\tilde{g}^{-1}\in H\,.
	\end{equation}
	An example of such $g$ is given by:
	\begin{equation}
		g=T_{21}^{n_1}T_{31}^{n_2}T_{12}^{m}T_{13}^{m},
	\end{equation}
	for arbitrary integers $m$ and $n_{1,2}$.
	Slightly more generally, we can also write  
	\begin{equation}
		g=T_{21}^{n_1}T_{31}^{n_2}T_{12}^{m}T_{13}^{m+k}T_{32}T_{23}^{-1},
	\end{equation}
	where the integers are again arbitrary and $k=0,1$.
	
	More systematically, we can equivalently approach this problem by considering elements of the form $A=h\cdot S_{23}$ for $h\in H$ such that the element has order three.
	The matrix $A$ will have the general form:
	\begin{equation}
		A=\begin{pmatrix}
			A_{11} & A_{12} & 0\\
			A_{21} & A_{22} & A_{23} \\
			A_{31} & A_{32}& 0
		\end{pmatrix} \,,
	\end{equation}
	where the zeros reflect the fact that $h$ does not contain $T_{12}$ and $T_{32}$ matrices.
	The matrix $A$ has order three when the following equations are satisfied:
	\begin{align}
		\begin{split}
			\det \left(A- (\omega_3)^\ell \, I \right)&=0 \,,
		\end{split}
	\end{align}
	where $\omega_3 = e^{2\pi i/3}$ and $\ell=0,1,2$ gives the three cubic roots of unity, and $I$ is the identity matrix.
	In addition, the entries of $A$ are constrained to be integral.
	An intermediate step yields constraints for $A_{2j}$ and $A_{31}$, such that $A$ and $A^2$ now read:
	\begin{align}\label{eq:matrixA}
		\begin{split}
			A&=\begin{pmatrix}
				n\ &\ m\ &\ 0\\
				-\frac{n^2+p}{m}\ &\ -n\ &\ 1\\
				\frac{1+n\, p}{m}\ &\ p\ &\ 0
			\end{pmatrix} \,, \qquad 
			A^2=\begin{pmatrix}
				-p\ &\ 0\ &\ m\\
				\frac{1+n\, p}{m}\ &\ 0 \ &\ -n\\
				\frac{n-p^2}{m}\ &\ 1 \ &\  p
			\end{pmatrix},
		\end{split}
	\end{align}
	with $m,n,p\in\mathbb{Z}$.
	To obtain an integral matrix, we solve for $p=-n^2-k\,m$ with $k\in \mathbb{Z}$.
	Then, the matrix is integral for $m\,|\, 1-n^3$, i.e.:
	\begin{equation}\label{eq:a2-a1cubic-def-d}
		(1-n)(n^2+n+1)=d\, m\,, \quad d\in \mathbb{Z} \,.
	\end{equation}
	Given a solution to this equation, we may already anticipate a result that will become more clear in Section~\ref{ssec:gen-cardy-gamma}.
	Namely, the modular property associated to $A^3=1$ will yield a useful generalized Cardy limit for:
	\begin{equation}\label{eq:ant-most-gen-cardy-lim}
		\tau\to -\tfrac{n}{m} \,,\qquad \sigma\to -\tfrac{n^2}{m}-k \,.
	\end{equation}
	This is because the first rows of $A$ and $A^2$ determine the denominators of the arguments of the elliptic $\Gamma$ functions on the right hand side of the modular property.
	Therefore, we see that the solution space to~\eqref{eq:a2-a1cubic-def-d} determines the possible useful Cardy limits of the elliptic $\Gamma$ function \emph{which can be obtained through order three elements}.
	Notice in particular that the solution space of~\eqref{eq:ant-most-gen-cardy-lim} does not contain arbitrary coprime $n$ and $m$, in contrast to Cardy limits that can be taken in two-dimensional CFTs.
	In Section~\ref{ssec:future}, we will suggest other types of relations in $SL(3,\mathbb{Z})$ which may lead to modular properties that allow arbitrary rational limits of $\tau$ and $\sigma$.
	
	In the following, we will focus on the simplest family of solutions to~\eqref{eq:a2-a1cubic-def-d} for which we can find a manifestly integral matrix:
	\begin{equation}\label{eq:simplest-soln-a1-a2}
		1-n=n_1 \, m \,,  \quad n_1\in \mathbb{Z} \,.
	\end{equation}
	It will turn out that this order three element already yields a modular property which essentially captures all of the non-trivial physics corresponding to the generalized Cardy limit.
	In addition, for this element we are able to rigorously determine the modular property, including the phase in Appendix~\ref{app:Xabc-phase}.
	A more detailed classification of the total space of solutions to~\eqref{eq:a2-a1cubic-def-d} will be given in Appendix~\ref{app:more-gen-order-3-elms}, and a partial derivation of the phase associated to the more general modular property will be given in Appendix~\ref{app:most-gen-mod-prop}.
	
	Before plugging the solution~\eqref{eq:simplest-soln-a1-a2} into the matrix $A$, we find it convenient to replace $k$ by: 
	\begin{equation}
		k\equiv n_1\, (2-m\, n_1)-n_2 \,,
	\end{equation}
	for $n_2\in \mathbb{Z}$ arbitrary.
	We will write $\mathbf{m}=(m,n_1,n_2)$ and refer to the resulting matrix as $X_{\mathbf{m}}$, which is given by:
	\begin{align}\label{eq:Xabc-matrix}
		\begin{split}
			X_{\mathbf{m}}&=\begin{pmatrix}
				\ 1-m\, n_1\ &\ m\ &\ 0\ \\
				\ (2-m\, n_1)n_1-n_2\ &\ m\, n_1-1\ &\ 1\ \\
				\ (1-m\, n_1)n_2+n_1\ &\ m\, n_2-1\ &\ 0\ 
			\end{pmatrix} \,, \\
			X^2_{\mathbf{m}}&=\begin{pmatrix}
				\ 1-m\, n_2\ &\ 0\ &\ m\ \\
				\ (1-m\, n_1)n_2+n_1\ &\ 0\ &\ m\, n_1-1\ \\
				\ (2-m\, n_2)n_2-n_1\ &\ 1\ &\ m\, n_2-1\ 
			\end{pmatrix} \,.
		\end{split}
	\end{align}
	It is easy to check that $X_{\mathbf{m}}^3=I$.
	Notice that the order three element $Y=S_{23}^{-1}S_{13}$, which permutes the periods of the torus cyclically and has been the focus of~\cite{Gadde:2020bov,Goldstein:2020yvj}, corresponds to the special case $m=n_1=n_2=1$; more precisely: $X_{(1,1,1)}=Y^{-1}$.
	We also anticipate, similarly to~\eqref{eq:ant-most-gen-cardy-lim}, that the associated generalized Cardy limit will be:
	\begin{equation}\label{eq:ant-Xabc-gen-cardy-lim}
		\tau\to n_1-\tfrac{1}{m} \,,\qquad \sigma\to n_2-\tfrac{1}{m} \,.
	\end{equation}
	We will close this section by noting that the element $X_{\mathbf{m}}$ can be decomposed into generators as follows:
	\begin{equation}\label{eq:Xabc-generators}
		X_{\mathbf{m}}=T_{23}T_{21}^{n_1-n_2}T_{31}^{n_2}T_{13}^{-m}T_{31}^{n_1}T_{21}^{-n_2}S_{23} \,.
	\end{equation}
	Notice that this has the expected form of $X_{\mathbf{m}}=h\cdot S_{23}$ with $h\in H$.
	This decomposition is crucial for the computation of the phase in the modular property, which we perform in Appendix~\ref{app:Xabc-phase}.
	We also notice that conjugation by $T_{21}$ and $T_{31}$ shifts $n_{1,2}$:
	\begin{align}
		\begin{split}
			T_{21}X_{(m,n_1,n_2)}T_{21}^{-1}&=X_{(m,n_1+1,n_2)}\,,\\
			T_{31}X_{(m,n_1,n_2)}T_{31}^{-1}&=X_{(m,n_1,n_2+1)}\,.
		\end{split}
	\end{align}
	Finally, one may check, using the gluing prescription described in Section~\ref{ssec:sl3}, that the topology of $M_{X_{\mathbf{m}}}$ is $S^3\times S^1$.
	This is due to the fact that the decomposition of $X_{\mathbf{m}}$ into generators only contains a single factor of $T_{23}$ and $S_{23}$.
	A lens space $L(m,1)$ for $m>1$ is associated instead to the element $S_{23}T_{23}^{-m}S_{23}$.

	\subsection{New modular property}
	\label{ssec:new-mod-prop}

	In this section, we will derive a new modular property for the elliptic $\Gamma$ function associated to the order three element $X_{\mathbf{m}}$ found in the previous subsection.
	Even though our derivation only holds for a free chiral multiplet, the result is still relevant for more general $\mathcal{N}=1$ SCFTs, including the $\mathcal{N}=4$ theory.
	This is because the perturbative part $Z_{\text{P}}$ of their partition functions can be written in terms of elliptic $\Gamma$ functions, as reviewed in Section~\ref{ssec:review-index} for the $\mathcal{N}=4$ theory.
	This turns out to be the only relevant part of the partition function to leading order in Cardy-like limits, as we will see in more detail in Section~\ref{ssec:gen-cardy-index}.
	Therefore, modular properties for the chiral multiplet partition function, consisting of a single elliptic $\Gamma$ function, can be applied to (at least) any gauge theory partition function to find the leading order behavior in the Cardy-like limit. 
	
	The new modular property is obtained by evaluating $\hat{Z}_g$ on the relation $X_{\mathbf{m}}^3=1$.
	We will work in the gauge fixed formalism, mentioned at the end of Section~\ref{ssec:partn-fn-sl3}.
	The explicit computation of the phase associated to $X_{\mathbf{m}}^3=1$ can be found in Appendix~\ref{app:Xabc-phase}.
	Here, we will just quote the result:
	\begin{align}\label{eq:deriv-mod-prop-Xabc}
		\begin{split}
			\hat{Z}_{X_{\mathbf{m}}}\left(\boldsymbol{\rho}\right)\,\hat{Z}_{X_{\mathbf{m}}}\left(X_{\mathbf{m}}^{-1}\boldsymbol{\rho}\right)\,\hat{Z}_{X_{\mathbf{m}}}\left(X_{\mathbf{m}}^{-2}\boldsymbol{\rho}\right)=e^{i\pi Q_{\mathbf{m}}\left(mz;\tau,\sigma\right)} \,,
		\end{split}
	\end{align}
	where in the first equality we used~\eqref{eq:sl3-reln-part-funct}.
	The function $Q_{\mathbf{m}}\left(mz;\tau,\sigma\right)$ is defined as:
	\begin{equation}\label{eq:defn-Qabc}
		Q_{\mathbf{m}}\left(mz;\tau,\sigma\right)=\tfrac{1}{m}Q\left(m z-1;m\tau+1-mn_1,m\sigma+1-mn_2\right)+\tfrac{(m+1)(m+3)}{4m} \,,
	\end{equation}
	where $Q$ is the function appearing in the original modular property~\cite{Goldstein:2020yvj,Felder_2000} and is defined in Appendix~\ref{app:defs}.
	As mentioned, this property only holds for $\hat{Z}_g$ being the normalized partition function of a free chiral multiplet.
	However, following~\cite{Gadde:2020bov} it is natural to conjecture that~\eqref{eq:deriv-mod-prop-Xabc} holds for a general $\mathcal{N}=1$ SCFT if one replaces the $Q$ function in the definition of $Q_{\mathbf{m}}$ with the anomaly polynomial of the theory.
	More precisely, this proposal reads:
	\begin{equation}\label{eq:mod-prop-gen-theory}
		\hat{Z}^{a}_{X_{\mathbf{m}}}\left(\boldsymbol{\rho}\right)\,\hat{Z}^{a}_{X_{\mathbf{m}}}\left(X_{\mathbf{m}}^{-1}\boldsymbol{\rho}\right)\,\hat{Z}^{a}_{X_{\mathbf{m}}}\left(X_{\mathbf{m}}^{-2}\boldsymbol{\rho}\right)=e^{i\pi P_{\mathbf{m}}\left(\boldsymbol{\rho}\right)} \,,
	\end{equation}
	where we remind the reader that $a$ labels the Higgs branch vacua of the mass deformed theory.
	Here, $P_{\mathbf{m}}\left(\boldsymbol{\rho}\right)$ given in terms of the anomaly polynomial $P(z_i;\tau,\sigma)$ as follows:
	\begin{equation}
		P_{\mathbf{m}}\left(\boldsymbol{\rho}\right)=\tfrac{1}{3m}P(mz_i;m\tau+1-mn_1,m\sigma+1-mn_2)+\mathcal{C}\tfrac{(m+1)(m+3)}{4m} \,.
	\end{equation}
	Here, we refer for the definition of $P(mz_i;\tau,\sigma)$ to Appendix~\ref{app:defs}.
	Moreover, $\mathcal{C}$ is the weighted sum of the number of elliptic $\Gamma$ functions in the partition function, for which $\Gamma$ functions in the numerator are counted with a plus sign and $\Gamma$ functions in the denominator with a minus sign.
	We will verify this proposal explicitly for the $\mathcal{N}=4$ theory in Section~\ref{ssec:anom-pol-index}.
	
	Let us now evaluate the modular property when $\hat{Z}_g$ is the partition function of a chiral multiplet, i.e., when it is evaluated on the generators described in Section~\ref{ssec:partn-fn-sl3}.
	First, we use~\eqref{eq:sl3-reln-part-funct} and the expression for $X_{\mathbf{m}}$ in terms of generators~\eqref{eq:Xabc-generators} to find:
	\begin{align}\label{eq:deriv-mod-prop-Xabc-2}
		\begin{split}
			\hat{Z}_{S_{23}}\left(S_{23}\boldsymbol{\rho}\right)=\frac{e^{i\pi Q_{\mathbf{m}}\left(mz;\tau,\sigma\right)}}{\hat{Z}_{S_{23}}\left(S_{23}X_{\mathbf{m}}^{-1}\boldsymbol{\rho}\right)\,\hat{Z}_{S_{23}}\left(S_{23}X_{\mathbf{m}}^{-2}\boldsymbol{\rho}\right)} \,,
		\end{split}
	\end{align}
	where we used the fact that $\hat{Z}_h=1$ for $h\in H$ in the gauge described in Section~\ref{ssec:partn-fn-sl3}.
	Notice that even though the right hand side of this equation may seem to depend on the element $X_{\mathbf{m}}$, the left hand side shows that it does not.
	
	Using the expression for $\hat{Z}_{S_{23}}$ given in~\eqref{eq:sci-free-chiral} and properties of the elliptic $\Gamma$ function collected in Appendix~\ref{app:defs}, we can rewrite this as:
	\begin{tcolorbox}[ams equation, colback=yellow!10!white]
		\label{eq:mod-prop-gamma-a}
		\Gamma(z;\tau,\sigma)=e^{-i\pi Q'_{\mathbf{m}}\left(mz;\tau,\sigma\right)}\frac{\Gamma\left(\tfrac{z}{m\sigma+1-mn_2};\tfrac{\tau-\sigma+n_2-n_1}{m\sigma+1-mn_2},\tfrac{\sigma-n_2}{m\sigma+1-mn_2}\right)}{\Gamma\left(\tfrac{z+\tau-\sigma+n_2-n_1}{m\tau+1-mn_1};\tfrac{\tau-\sigma+n_2-n_1}{m\tau+1-mn_1},\tfrac{\tau-n_1}{m\tau+1-mn_1}\right)} \,,
	\end{tcolorbox}
	\noindent where
	\begin{tcolorbox}[ams equation, colback=yellow!10!white ]
		\label{eq:abc-Q-poly}
		Q'_{\mathbf{m}}(mz;\tau,\sigma)=\tfrac{1}{m}Q\left(mz;m\tau+1-mn_1,m\sigma+1-mn_2\right)+\tfrac{m^2-1}{12m} \,.
	\end{tcolorbox}
	\noindent Note that $Q'_{\mathbf{m}}$ is related to $Q_{\mathbf{m}}$ by a shift of the $z$ argument.
	The property~\eqref{eq:mod-prop-gamma-a} will be the main modular property that we will use to compute the generalized Cardy limit of the $\mathcal{N}=4$ superconformal index.
	We will discuss its interpretation as a generalization of holomorphic blocks for the chiral multiplet~\cite{Nieri:2015yia} in Section~\ref{ssec:mod-Z}.
	
	The rewriting in~\eqref{eq:mod-prop-gamma-a} ensures that the product expression for all three $\Gamma$ functions, defined in Appendix~\ref{app:defs}, is convergent for $\mathrm{Im}(\tau)>0$, $\mathrm{Im}(\sigma)>0$ and:
	\begin{equation}\label{eq:constraint-mod-prop}
		\mathrm{Im}\left(\tfrac{\tau-\sigma+n_2-n_1}{m\sigma+1-mn_2}\right)>0 \,.
	\end{equation}
	This is because all the other arguments are $SL(2,\mathbb{Z})$ transformations of the above:
	\begin{align}
		\begin{split}
			\tfrac{\tau-n_1}{m\tau+1-mn_1}&=ST^{-m}S^{-1}T^{-n_1}\cdot\tau \,,\\
			\tfrac{\sigma-n_2}{m\sigma+1-mn_2}&=ST^{-m}S^{-1}T^{-n_2}\cdot\sigma \,,\\
			\tfrac{\tau-\sigma+n_2-n_1}{m\tau+1-mn_1}&=ST^{-m}S^{-1}\cdot \tfrac{\tau-\sigma+n_2-n_1}{m\sigma+1-mn_2} \,.
		\end{split}
	\end{align}
	We note that this way of writing the modular property is the most direct generalization of the property used in~\cite{Goldstein:2020yvj}, which we wrote in~\eqref{eq:ord-mod-prop-Gamma} and which also appears in Theorem 4.1 of~\cite{Felder_2000}.
	
	We end this section by analyzing the $\tau=\sigma$ limit of the modular property.
	This limit was analyzed for the ordinary modular property in Theorem 5.2 of~\cite{Felder_2000}, and played an important role in~\cite{Benini:2018ywd}.
	For us, the limit is subtle when $n_1=n_2\equiv n$, since in this case the second arguments of the $\Gamma$ functions on the right hand side of~\eqref{eq:mod-prop-gamma-a} hit a zero.
	This means that the $\Gamma$ functions themselves diverge.
	As in Theorem 5.2 of~\cite{Felder_2000}, this divergence cancels.
	Indeed, it should cancel given that the left hand side of the modular property is well-defined for $\tau=\sigma$.
	Since the derivation of the $\tau=\sigma$ limit is completely analogous to the proof of Theorem 5.2 given in~\cite{Felder_2000}, we will only state the result.
	We have:
	\begin{equation}\label{eq:spec-mod-prop}
		\Gamma(z;\tau,\tau)=\frac{e^{-i\pi Q'_{\mathbf{m}}\left(mz;\tau,\tau\right)}}{\theta\left(\frac{z}{m\tau+1-mn};\frac{\tau -n}{m\tau+1-mn}\right)}\prod^{\infty}_{k=0}\left(\frac{\psi^{(m,k+1)}\left(\frac{-z-\frac{(k+1)}{m}}{m\tau+1-mn}\right)}{\psi^{(m,k)}\left(\frac{z-\frac{k}{m}}{m\tau+1-mn}\right)}\right)^m\,.
	\end{equation}
	Here, the function $\psi^{(m,k)}(t)$ is defined for $\mathrm{Im}(t)>0$ as:
	\begin{equation}
		\psi^{(m,k)}(t)=\exp\left[-t\log\left(1-e^{2\pi i \left(t+\frac{k}{m}\right)}\right)-\tfrac{1}{2\pi i}\mathrm{Li}_2\left(e^{2\pi i \left(t+\frac{k}{m}\right)}\right)\right]\, ,
	\end{equation}
	where the branch of the logaritm is taken such that $\log\left(1-x\right)=-\sum^{\infty}_{j=1}\frac{x^{j}}{j}$ and $\mathrm{Li}_2\left(x\right)\equiv \sum_{j=1}^{\infty}\frac{x^{j}}{j^2}$.
	The function clearly only depends on $k\mod m$, and in the following we will take $0\leq k<m$.\footnote{Note that in~\eqref{eq:spec-mod-prop} $k$ appears explicitly in the argument of $\psi^{(m,k)}(t)$ as well. This means that those functions do depend on the exact value of $k$.}
	Let us make some further comments about this function:
	\begin{itemize}
		\item Taking $m=n=1$ gives rise to the original theorem of~\cite{Felder_2000}:
		\begin{equation}\label{eq:felder-thm-52}
			\Gamma(z;\tau,\tau)=\frac{e^{-i\pi Q\left(z;\tau,\tau\right)}}{\theta\left(\frac{z}{\tau};-\frac{1}{\tau}\right)}\prod^{\infty}_{k=0}\frac{\psi\left(\frac{-(k+1)-z}{\tau}\right)}{\psi\left(\frac{z-k}{\tau}\right)}\,,
		\end{equation}
		where we used that $\psi^{(1,k)}(t)=\psi(t)$ coincides with the definition from~\cite{Felder_2000}:
		\begin{equation}
			\psi(t)=\exp\left[-t\log\left(1-e^{2\pi i t}\right)-\tfrac{1}{2\pi i}\mathrm{Li}_2\left(e^{2\pi i t}\right)\right]\, ,
		\end{equation}
		The reason for the overall minus signs in the arguments of $\psi(t)$ in~\eqref{eq:felder-thm-52} is due to the fact that we have defined $\psi(t)$ for $\mathrm{Im}(t)>0$ as opposed to $\mathrm{Im}(t)<0$.
		\item The function $\psi^{(m,k)}(t)$ shares many properties with $\psi(t)$ (see~\cite{Felder_2000}).
		In particular, we can write it as:
		\begin{equation}
			\psi^{(m,k)}(t)=\exp\left(-2\pi i  \int^{i\infty}_{t}\frac{s\ ds}{e^{-2\pi i \left(s+\frac{k}{m}\right)}-1}\right)\,.
		\end{equation}
		This way of writing $\psi^{(m,k)}(t)$ makes clear that it can only have singularities at $t\in\mathbb{Z}-\frac{k}{m}$. 
		To study the analytic structure further, we first note it satisfies the functional equation:
		\begin{equation}\label{eq:func-eqn-psi}
			\psi^{(m,k)}(t+1)=\left(1-e^{2\pi i \left(t+\frac{k}{m}\right)}\right)^{-1}\psi^{(m,k)}(t)\,.
		\end{equation}
		We note in passing that this equation implies that even the explicit appearance of $k$ in the argument of $\psi^{(m,k)}(t)$ can be reduced to a dependence on $k\mod m$, at the cost of introducing additional factors as in~\eqref{eq:func-eqn-psi}.
		Let us now take $t=-\frac{k}{m}$.
		For this value, we have that the $m^{\text{th}}$ power of $\psi^{(m,k)}(t)$, which is what appears in~\eqref{eq:spec-mod-prop}, has an order $k$ zero:
		\begin{equation}
			\left(\psi^{(m,k)}\left(-\tfrac{k}{m}\right)\right)^m=e^{\frac{i\pi m}{12}}\left(1-e^{2\pi i \left(t+\frac{k}{m}\right)}\right)^{k}\,
		\end{equation}
		where we used that $\mathrm{Li}_2(1)=\frac{\pi^2}{6}$.
		Together with~\eqref{eq:func-eqn-psi}, it now follows that for $t=-\frac{k}{m}+j$, $j\in \mathbb{Z}_{>0}$, $\left(\psi^{(m,k)}\left(t\right)\right)^m$ has a pole of order $jm-k$.
		On the other hand, for $t=-\frac{k}{m}-j$, $j\in \mathbb{Z}_{>0}$, it has zeros of order $mj+k$.
		\item Finally, the estimate for $\mathrm{Im}(t)\to \infty$ of $\psi^{(m,k)}\left(t\right)$ is the same as it is for $\psi(t)$, and reads~\cite{Felder_2000}:
		\begin{equation}
			\psi^{(m,k)}\left(t\right)=1+\mathcal{O}\left(\left|\mathrm{Im}(t)\right|e^{-2\pi\left|\mathrm{Im}(t)\right|}\right)\,,
		\end{equation}
		as $\mathrm{Im}(t)\to \infty$.
	\end{itemize}

	\subsection{Generalized Cardy limit associated to the new modular property}\label{ssec:gen-cardy-gamma}
	
	In this section, we will define and compute the generalized Cardy limit of the elliptic $\Gamma$ function.
	We will first compute this limit by simply making use of the summation formula for $\Gamma(z;\tau,\sigma)$.
	After that, we will employ the modular property~\eqref{eq:mod-prop-gamma-a} to compute the limit, and comment on how this leads to a more general expression for the generalized Cardy limit than obtained in the first approach.
	
	Recall that the ordinary Cardy-like limit takes the chemical potentials $\tau$ and $\sigma$, which couple to the angular momenta $J_1$ and $J_2$, respectively (see Section~\ref{ssec:review-index}), to $0^{+i}$ such that the ratio $\frac{\tau}{\sigma}\notin \mathbb{R}$~\cite{Goldstein:2020yvj}.
	The generalized Cardy limit, anticipated in~\eqref{eq:ant-Xabc-gen-cardy-lim}, is defined similarly, except that it keeps the real parts of $\tau$ and $\sigma$ finite.
	More precisely:\footnote{
		One may naively think the integer shifts of $\tau$ and $\sigma$ are not important to keep, since the index~\eqref{eq:trace-defn-index} is periodic under integer shifts of $\tau$ and $\sigma$.
		However, we will see that $n_{1,2}$ play an important role in understanding how this periodicity is reproduced in generalized Cardy limits.}
	\begin{align}
		\label{eq:gen-cardy-limit}
		\begin{split}
			\tau \to n_1-\frac{1}{m} \,, \quad \textrm{with}\quad &\alpha_\tau \equiv\frac{m\,\mathrm{Im}(\tau)}{m\,\mathrm{Re}(\tau)+1-mn_1}\quad \text{fixed} \,,\\ 
			\sigma \to n_2-\frac{1}{m} \,, \quad\textrm{with}\quad& \alpha_\sigma\equiv\frac{m\,\mathrm{Im}(\sigma)}{m\,\mathrm{Re}(\sigma)+1-mn_2} \quad \text{fixed} \,,\\
			\textrm{and}\quad &   \alpha_\tau\neq\alpha_\sigma \,.
		\end{split}
	\end{align}
	The constraint on $\alpha_\tau$ and $\alpha_\sigma$ is the generalization of $\frac{\tau}{\sigma}\notin \mathbb{R}$ in the ordinary case, and follows from~\eqref{eq:constraint-mod-prop}.
	To study the limits for $\alpha_\tau=\alpha_\sigma$, at least when $n_1=n_2$, one can make use of~\eqref{eq:spec-mod-prop}.
	
	Let us now take this limit of $\Gamma(z;\tau,\sigma)$, making use of its summation formula (see Appendix~\ref{app:defs} for details on notation):
	\begin{equation}
		\log \Gamma(z;\tau,\sigma)=\sum^{\infty}_{l=1}\frac{1}{l}\frac{x^l-(x^{-1}pq)^l}{(1-p^l)(1-q^l)} \,.
	\end{equation}
	The generalized Cardy limit of this function diverges, and to leading order one finds:\footnote{
		A more precise estimate, including subleading corrections, was recently given in~\cite{ArabiArdehali:2021nsx}. See also~\cite{Ardehali:2015bla} for earlier work.}
	\begin{align}\label{eq:pre-naive-cardy-lim-Gamma}
		\begin{split}
			\lim_{\text{gen Cardy}}\log \Gamma(z;\tau,\sigma)&=-\frac{1}{4\pi^2 \left(m\tau+1-mn_1\right)\left(m\sigma+1-mn_2\right)}\sum^{\infty}_{l'=1}\frac{x^{ml'}-x^{-ml'}}{(l')^3}\\
			& =\frac{\pi i B_{3}(-mz)}{3\left(m\tau+1-mn_1\right)\left(m\sigma+1-mn_2\right)} \,.
		\end{split}
	\end{align}
	The second equality only holds for $-\frac{1}{m}\leq z\leq 0$, where we used the Fourier series of the Bernoulli polynomial $B_3(z)$:
	\begin{equation}
		B_3(z)=z^3-\tfrac{3}{2}z^2+\tfrac{1}{2}z \,.
	\end{equation}
	Let us point out that the first line of~\eqref{eq:pre-naive-cardy-lim-Gamma} implies that $\Gamma(z;\tau,\sigma)$ develops a finer periodicity under $z\to z+\frac{1}{m}$.
	In particular, if we define the fractional part of $z$ to be
	\begin{equation}
		[z]=z+n \quad \text{such that} \quad 0<z+n<1 \,,
	\end{equation}
	we can express the limit for any $z\in \mathbb{R}\setminus \frac{1}{m}\mathbb{Z}$ as:
	\begin{equation}\label{eq:naive-gen-cardy-lim-Gamma}
		\lim_{\text{gen Cardy}}\log \Gamma(z;\tau,\sigma)=\frac{\pi i B_{3}\left([-mz]\right)}{3\left(m\tau+1-mn_1\right)\left(m\sigma+1-mn_2\right)} \,.
	\end{equation}
	This finer periodicity exhibited in the generalized Cardy limit will play an important role later on.
	
	Now we turn to a derivation of the generalized Cardy limit using the modular property.
	We will see that the resulting limit generalizes~\eqref{eq:naive-gen-cardy-lim-Gamma} in two ways: it contains subleading corrections and analytically extends the expression to finite imaginary parts of $z$.
	For convenience, we repeat the modular property here:
	\begin{align}\label{eq:mod-prop-gamma-a-2}
		\begin{split}
			\Gamma(z;\tau,\sigma)&=e^{-i\pi Q'_{\mathbf{m}}\left(mz;\tau,\sigma\right)}\frac{\Gamma\left(\tfrac{z}{m\sigma+1-mn_2};\tfrac{\tau-\sigma+n_2-n_1}{m\sigma+1-mn_2},\tfrac{\sigma-n_2}{m\sigma+1-mn_2}\right)}{\Gamma\left(\tfrac{z+\tau-\sigma+n_2-n_1}{m\tau+1-mn_1};\tfrac{\tau-\sigma+n_2-n_1}{m\tau+1-mn_1},\tfrac{\tau-n_1}{m\tau+1-mn_1}\right)} \,,
		\end{split}
	\end{align}
	where $Q'_{\mathbf{m}}$ is defined in~\eqref{eq:abc-Q-poly}.
	In the limit~\eqref{eq:gen-cardy-limit}, again using the summation formula for $\Gamma(z;\tau,\sigma)$, we find that the numerator on the right hand side of~\eqref{eq:mod-prop-gamma-a-2} becomes:
	\begin{align}\label{eq:cardy-lim-Gamma-factor-1}
		\begin{split}
			\Gamma\left(\tfrac{z}{m\sigma+1-mn_2};\tfrac{\tau-\sigma+n_2-n_1}{m\sigma+1-mn_2},\tfrac{\sigma-n_2}{m\sigma+1-mn_2}\right)=&\exp\left(\sum^{\infty}_{l=1}\frac{1}{l}\frac{e^{2\pi il \frac{z}{m\sigma+1-mn_2}}-e^{2\pi il \frac{\tau-n_1-z}{m\sigma+1-mn_2}}}{(1-e^{2\pi il\frac{\tau-\sigma+n_2-n_1}{m\sigma+1-mn_2}})(1-e^{2\pi il\frac{\sigma-n_2}{m\sigma+1-mn_2}})}\right)\\
			\longrightarrow{}&\exp\left(\sum^{\infty}_{l=1}\frac{1}{l}\frac{e^{2\pi il \frac{z}{m\sigma+1-mn_2}}-e^{2\pi il \frac{-z-\frac{1}{m}}{m\sigma+1-mn_2}}}{1-e^{2\pi il\frac{\tau-\sigma+n_2-n_1}{m\sigma+1-mn_2}}}\right) \,.
		\end{split}
	\end{align}
	
	\begin{figure}[t]
		\centering
		\begin{subfigure}[t]{0.3\textwidth}
			\centering
			\includegraphics[width=\textwidth]{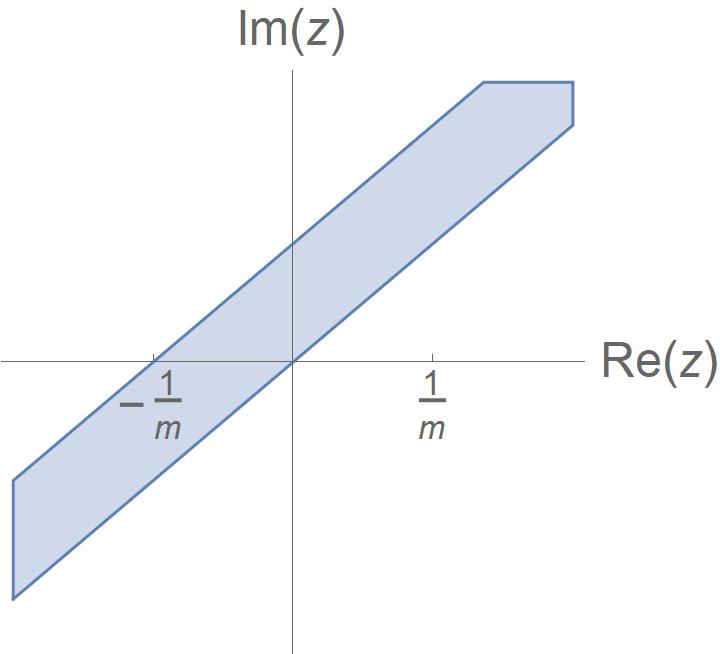}
			\caption{$m>0, \; \alpha_\sigma >0$}
			\label{fig:agg}
		\end{subfigure}
		\hspace{2cm}
		\begin{subfigure}[t]{0.3\textwidth}
			\centering
			\includegraphics[width=\textwidth]{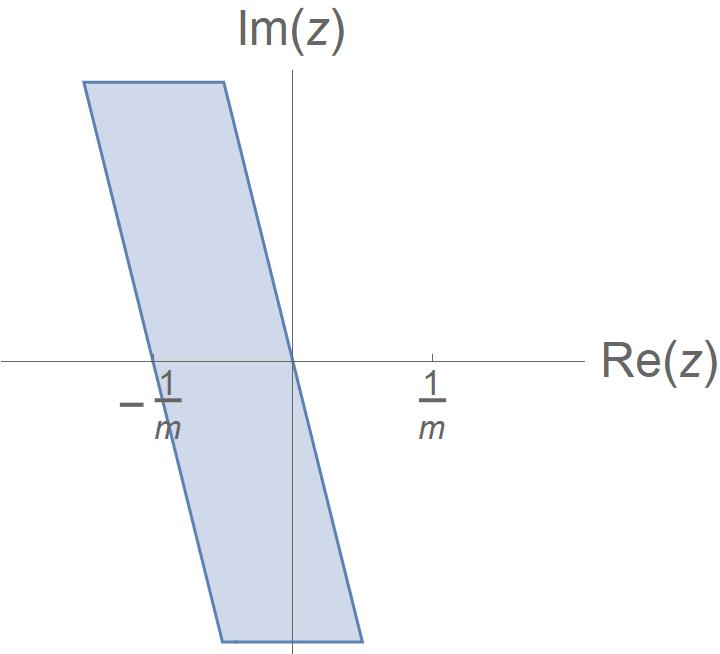}
			\caption{$m>0, \; \alpha_\sigma<0$}
			\label{fig:agl}
		\end{subfigure}
		\par
		\begin{subfigure}[t]{0.3\textwidth}
			\centering
			\includegraphics[width=\textwidth]{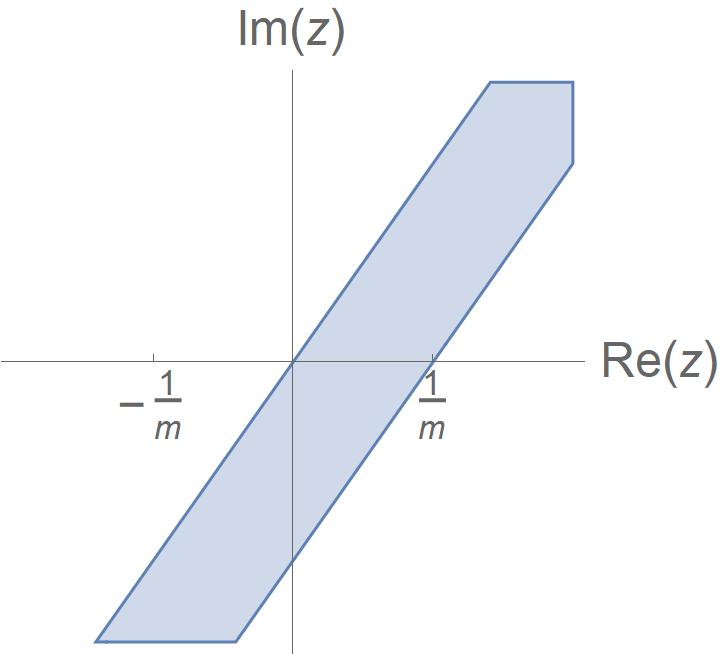}
			\caption{$m<0, \; \alpha_\sigma<0$}
			\label{fig:alg}
		\end{subfigure}
		\hspace{2cm}
		\begin{subfigure}[t]{0.3\textwidth}
			\centering
			\includegraphics[width=\textwidth]{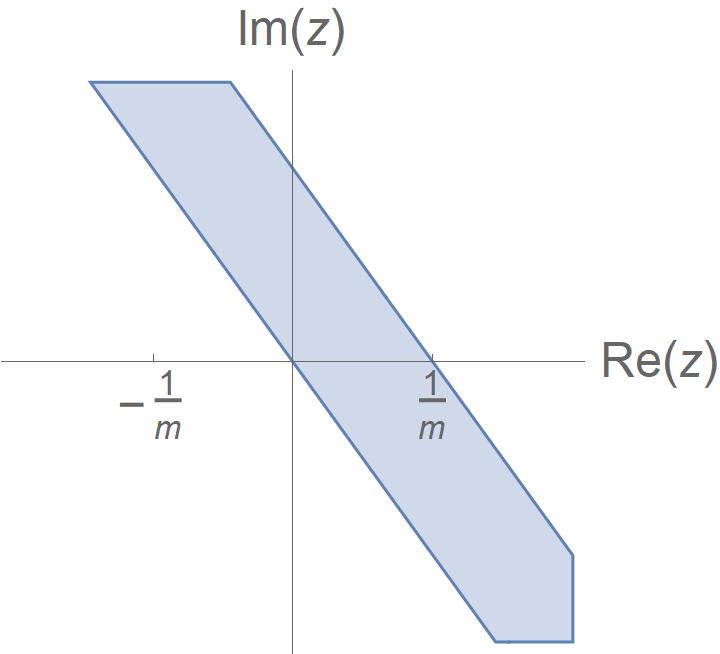}
			\caption{$m<0, \; \alpha_\sigma>0$}
			\label{fig:all}
		\end{subfigure}
		\caption{The four distinct types of strips.}
		\label{fig:strips}
	\end{figure}
	
	\noindent A similar formula can be obtained for the $\Gamma$ function in the denominator on the right hand side of~\eqref{eq:mod-prop-gamma-a-2}.
	These two limits converge if $z$ lies inside the respective domains:
	\begin{align}\label{eq:tau-sigma-domains}
		\begin{split}
			&-\frac{1}{m}\,\mathrm{Im}\left(\frac{1}{m\sigma+1-mn_2}\right)>\mathrm{Im}\left(\frac{z}{m\sigma+1-mn_2}\right)>0 \,,\\
			&-\frac{1}{m}\,\mathrm{Im}\left(\frac{1}{m\tau+1-mn_1}\right)>\mathrm{Im}\left(\frac{z}{m\tau+1-mn_1}\right)>0 \,.
		\end{split}
	\end{align}
	In addition, we have to require that $\tilde{\tau}=a\tilde{\sigma}^n$ for $a\notin \mathbb{R}$ as $\tilde{\sigma}\to 0$,\footnote{We thank Sameer Murthy for discussions on this point.} where $\tilde{\tau}$ and $\tilde{\sigma}$ are defined by:
	\begin{equation}\label{eq:tilded-tau-sigma-defn}
		\tilde{\tau}\equiv m
		\tau+1-m n_1\,,\quad \tilde{\sigma}\equiv m \sigma+1-m n_2\,.
	\end{equation}
	In this case, both $\Gamma$ functions on the right hand side of the modular property trivialize.
	Depending on the sign of $m$ and $\alpha_{\tau,\sigma}$, the respective domains can represent four types of strips with slope $\alpha_{\tau,\sigma}$, as illustrated in Figure~\ref{fig:strips}.
	
	\begin{figure}[t]
		\centering
		\begin{subfigure}[t]{0.3\textwidth}
			\centering
			\includegraphics[width=\textwidth]{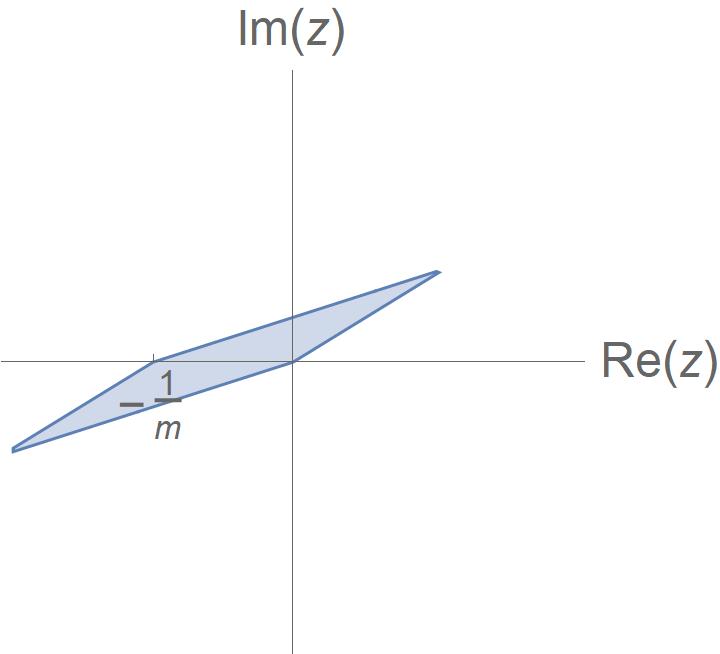}
			\caption{$n=-2$}
			\label{fig:b-2}
		\end{subfigure}
		\hfill
		\begin{subfigure}[t]{0.3\textwidth}
			\centering
			\includegraphics[width=\textwidth]{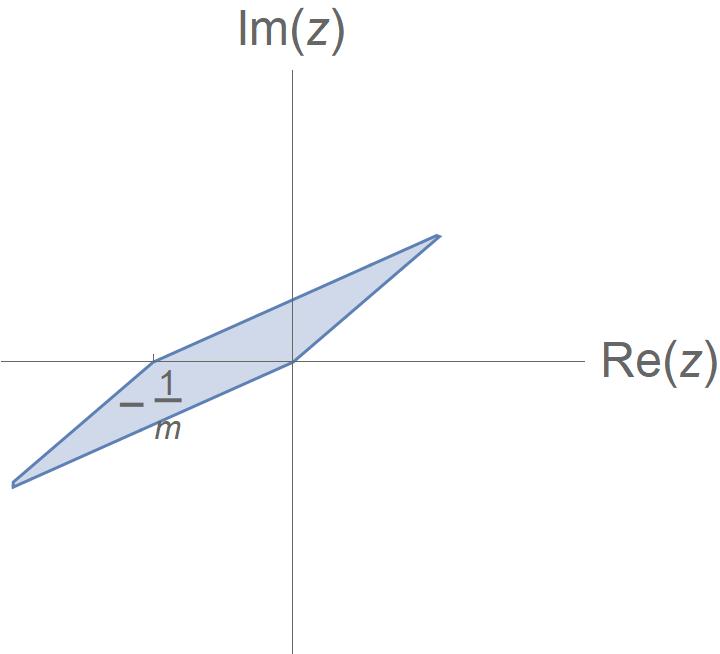}
			\caption{$n=-1$}
			\label{fig:b-1}
		\end{subfigure}
		\hfill
		\begin{subfigure}[t]{0.3\textwidth}
			\centering
			\includegraphics[width=\textwidth]{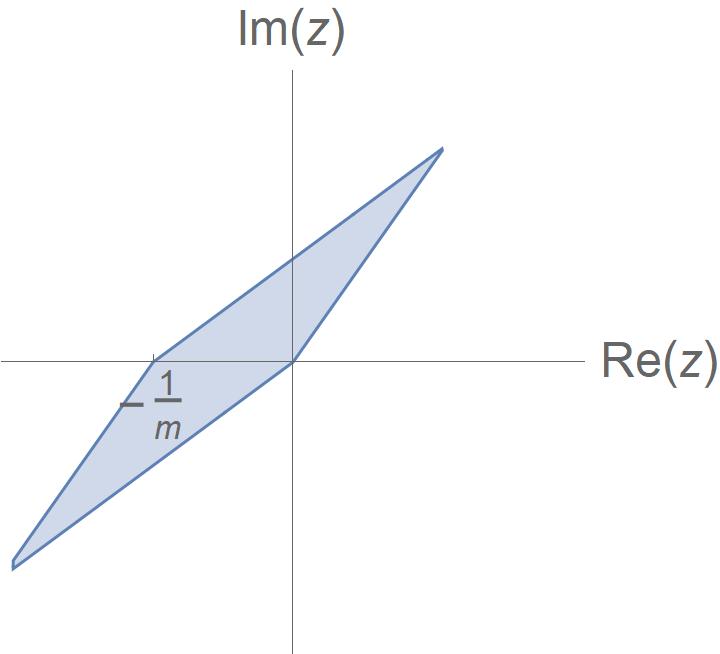}
			\caption{$n=0$}
			\label{fig:b0}
		\end{subfigure}
		\par
		\begin{subfigure}[t]{0.3\textwidth}
			\centering
			\includegraphics[width=\textwidth]{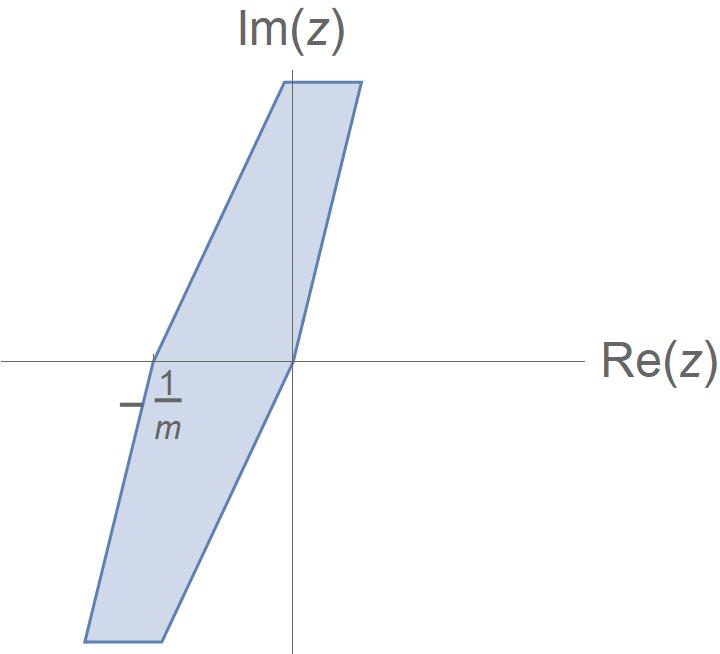}
			\caption{$n=1$}
			\label{fig:b1}
		\end{subfigure}
		\hfill
		\begin{subfigure}[t]{0.3\textwidth}
			\centering
			\includegraphics[width=\textwidth]{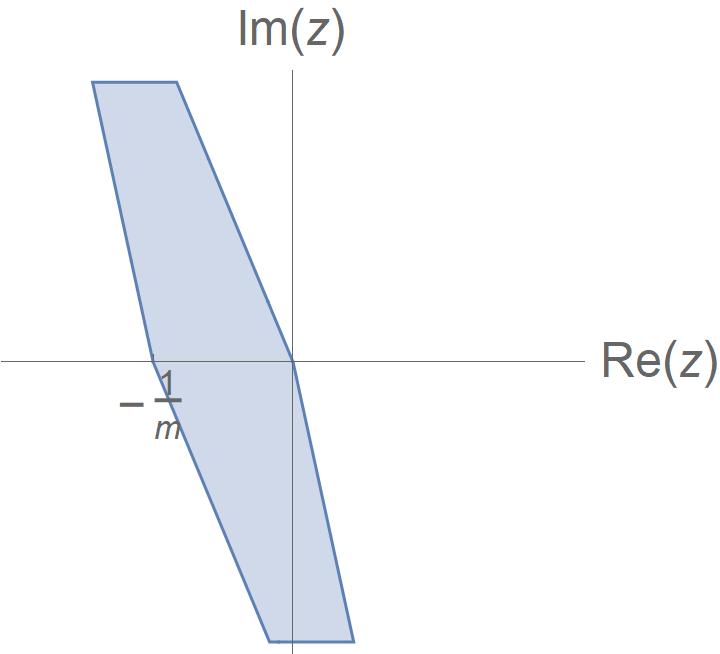}
			\caption{$n=2$}
			\label{fig:b2}
		\end{subfigure}
		\hfill
		\begin{subfigure}[t]{0.3\textwidth}
			\centering
			\includegraphics[width=\textwidth]{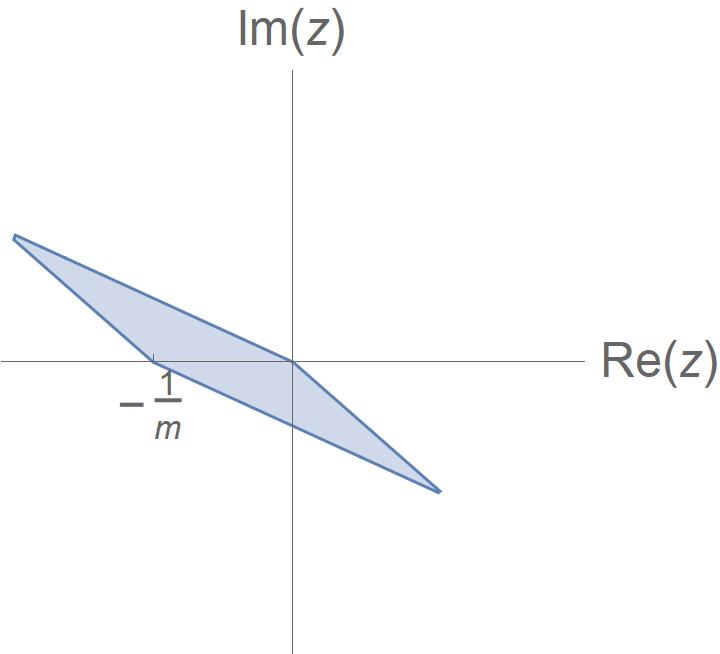}
			\caption{$n=4$}
			\label{fig:b4}
		\end{subfigure}
		\caption{Diamond domains for $m>0$ and various values of $n_1=n_2=n$. Values of $\tau$ and $\sigma$ are chosen such that they are consistent with the constraint~\eqref{eq:constraint-mod-prop}. For very negative $n$, the domain aligns with the real axis. As one increases $n$, the domain rotates anti-clockwise until it again realigns with the real axis.}
		\label{fig:diamonds}
	\end{figure}

	The full regime of convergence consists of the intersection of the two types of strips, which represents a diamond shaped domain.
	In the following we will denote the diamond shaped domain associated to the modular property for $X_{\mathbf{m}}$ by $D^{(0)}_{1/m}$.
	The upper index indicates where the right boundary intersects the real axis, while the lower index indicates the width of the domain along the real axis.
	Some examples for the diamond domains for $m>0$ and $n_1=n_2$ are given in Figure~\ref{fig:diamonds}.
	Recall that $m=n_1=n_2=1$, in Figure~\ref{fig:b1}, corresponds to the diamond domain of the original modular property analyzed in~\cite{Goldstein:2020yvj}.
	
	We notice that the strips will never fully overlap due to the constraint~\eqref{eq:constraint-mod-prop}, which as we noted above implies $\alpha_\tau\neq\alpha_\sigma$.
	This implies that the diamond domains will always be of finite extent in the $z$-plane.\footnote{Again, one can analyze the case $\alpha_\tau=\alpha_\sigma$ for $n_1=n_2$ by using~\eqref{eq:spec-mod-prop}. The essence of the analysis is similar to the case $\alpha_\tau\neq\alpha_\sigma$, and we refrain from giving the explicit details.}
	As we already alluded to in~\cite{Goldstein:2020yvj}, Figure~\ref{fig:diamonds} illustrates how those values of $z$ that fall outside the diamond domain of the modular property for some $\mathbf{m}$, may fall inside the domain for another modular property $\mathbf{m}'$.
	
	The diamond domain implies an upper bound on the allowed imaginary part of $z$.
	For concreteness, we take $\alpha_{\tau}>\alpha_\sigma>0$.
	The top and bottom of the diamond lie respectively at:
	\begin{equation}\label{eq:top-bottom-diamond}
		\mathrm{Im}(z)=\pm \frac{\alpha_\tau\alpha_\sigma}{m(\alpha_\tau-\alpha_\sigma)} \,.
	\end{equation}
	We notice that for large values of $m$ the diamonds becomes not only small in width but also in height.
	
	The conclusion of the above discussion is that the generalized Cardy limit of the modular property~\eqref{eq:mod-prop-gamma-a} yields the following expression for $\Gamma(z;\tau,\sigma)$:
	\begin{equation}\label{eq:gen-cardy-Gamma}
		\lim_{\text{gen Cardy}} \Gamma(z;\tau,\sigma)=e^{-\frac{i\pi}{m}\left[Q\left(mz;m\tau+1-mn_1,m\sigma+1-mn_2\right)+\frac{m^2-1}{12}\right]} \quad\textrm{for}\quad z\in D^{(0)}_{1/m} \,.
	\end{equation}
	Instead, if $z$ lies outside the diamond shaped domain, the $\Gamma$ functions will not trivialize, and we cannot find a simple expression for the limit using the specific modular property.
	
	As we mentioned in the beginning of this section, the limit~\eqref{eq:gen-cardy-Gamma} generalizes~\eqref{eq:pre-naive-cardy-lim-Gamma} in two ways.
	Firstly, it holds for finite $\mathrm{Im}(z)$.
	Secondly, it contains subleading corrections. 
	To leading order, it is easy to check that~\eqref{eq:gen-cardy-Gamma} reduces to~\eqref{eq:pre-naive-cardy-lim-Gamma}.
	One way to think about this result is that the modular property justifies the analytic continuation of the Fourier series in the first line of~\eqref{eq:pre-naive-cardy-lim-Gamma} to imaginary values of $z$, through  $B_3(z)$, even though the naive extension of the series is divergent.
	
	In the beginning of this section, recall that we were able to extend the generalized Cardy limit from the interval $z\in \left(-\frac{1}{m},0\right)$ to $z\in\mathbb{R}\setminus \tfrac{1}{m}\mathbb{Z}$, as indicated in~\eqref{eq:naive-gen-cardy-lim-Gamma}.
	The identity theorem for analytic functions then implies that~\eqref{eq:gen-cardy-Gamma} holds more generally, i.e.:
	\begin{equation}\label{eq:gen-cardy-Gamma-2}
		\lim_{\text{gen Cardy}} \Gamma(z;\tau,\sigma)=e^{-\frac{i\pi}{m}\left[Q\left([mz];m\tau+1-mn_1,m\sigma+1-mn_2\right)+\frac{m^2-1}{12}\right]} \,.
	\end{equation}
	The $Q$ polynomial is evaluated on the bracketed potential $[mz]$, to be defined momentarily, which has the crucial property that it is periodic under $z\to z+\tfrac{1}{m}$.
	In particular, this implies that in the generalized Cardy limit, $\Gamma(z;\tau,\sigma)$ develops the finer periodicity $z\to z+\tfrac{1}{m}$, as compared to its original periodicity under $z\to z+1$.
	In particular, this means that the initial domain of convergence $D^{(0)}_{1/m}$ is extended to an infinite set of diamond shaped domains $D^{(k)}_{1/m}$ which intersect the real axis in the intervals $\left(\frac{k-1}{m},\frac{k}{m}\right)$.
	
	Since the $Q$ polynomial only depends on $mz$, it is convenient to introduce a rescaled variable $\tilde{z}=mz$.
	In terms of this variable, the original domain of convergence, $D^{(0)}_{1/m}$ becomes of width one, and we denote it $D^{(0)}$.
	The full domain of convergence now consists of arbitrary integral horizontal shifts of $D^{(0)}$, which intersect the real axis in the intervals $\left(k-1,k\right)$.
	Let these intervals be denoted by $D^{(k)}$.
	Then, the bracket is defined similarly to the case of the ordinary modular property used in~\cite{Goldstein:2020yvj}:
	\begin{equation}\label{eq:bracket-defn}
		[\tilde{z}]\equiv \tilde{z}+n, \quad n\in \mathbb{Z} \quad \text{such that} \quad \tilde{z}+n\in D^{(0)} \,,
	\end{equation}
	Clearly, the bracket $[\tilde{z}]$ is only defined for $\tilde{z}\in D^{(k)}$ for some $k\in\mathbb{Z}$.
	If both $\tilde{z}\in D^{(k)}$ and $-\tilde{z}\in D^{(k')}$ for some $k,k'$, it has the properties:
	\begin{equation}
		[\tilde{z}+m]=[\tilde{z}] \,,\quad m\in \mathbb{Z} \,, \qquad [-\tilde{z}]=-[\tilde{z}]-1 \,,
	\end{equation}
	As in the case of the original modular property analyzed in~\cite{Goldstein:2020yvj} and opposed to bracket defined in~\cite{Benini:2018ywd}, the brackets do not satisfy:
	\begin{eqnarray}
		[\tilde{z}+\tilde{\tau}]=[\tilde{z}]+\tilde{\tau} \,, \qquad [\tilde{z}+\tilde{\sigma}]=[\tilde{z}]+\tilde{\sigma} \,,
	\end{eqnarray}
	where $\tilde{\tau}$ and $\tilde{\sigma}$ were defined in~\eqref{eq:tilded-tau-sigma-defn}.
	This is due to the fact that enough translations by $\tilde{\tau}$ (or $\tilde{\sigma}$) will take the point outside the diamond domain.
	The lack of this latter property will not be an issue for our purposes, since in the generalized Cardy limit~\eqref{eq:gen-cardy-limit}, $\tilde{\tau}$ and $\tilde{\sigma}$ are vanishing.
	
	Whenever $\mathrm{Im}(z)$ is close to the value~\eqref{eq:top-bottom-diamond}, $z$ itself will generically lie outside any diamond domain.
	In this case, since the bracket is not defined, the divergence of the generalized Cardy limit cannot be isolated inside the $Q$ function as in~\eqref{eq:gen-cardy-Gamma-2}.
	To avoid restrictions on $\tilde{z}$, we will consider the following regime in parameter space:
	\begin{align}\label{eq:im-phi-constraint}
		\begin{split}
			\left|\mathrm{Im}(\tilde{z})\right|\ll  \frac{\alpha_\tau\alpha_\sigma}{\alpha_\tau-\alpha_\sigma} \,.
		\end{split}
	\end{align}
	This limit zooms into the area around the intersection of the diamond with real axis.
	Again, for $\alpha_\tau\approx \alpha_\sigma$ close enough,~\eqref{eq:im-phi-constraint} is not very constraining.
	In this regime, $\tilde{z}\in D^{(k)}$ for some $k$ for generic values of $\tilde{z}$, and we can rewrite the domains in~\eqref{eq:tau-sigma-domains} as follows:
	\begin{align}\label{eq:approx-domains}
		\begin{split}
			\mathrm{Im}\left(\tilde{z}\right)>0&:\quad -1<\tilde{z}^{\text{s}+}<0 \,,\\
			\mathrm{Im}\left(\tilde{z}\right)<0&:\quad -1<\tilde{z}^{\text{s}+}<0 \,.
		\end{split}
	\end{align}
	where we defined $\tilde{z}^{\text{s}\pm}$ as the coordinate which measures the distance for fixed $\mathrm{Im}(\tilde{z})$ from the upper and lower right boundary of the diamond respectively:
	\begin{align}\label{eq:phi-normal}
		\begin{split}
			z^{\text{s}+}&=\mathrm{Re}(z)-\frac{1}{\alpha_\tau}\mathrm{Im}(z) \,,\\
			z^{\text{s}-}&=\mathrm{Re}(z)-\frac{1}{\alpha_\sigma}\mathrm{Im}(z) \,.
		\end{split}
	\end{align}
	These coordinates will play an important role in Section~\ref{ssec:gen-cardy-index}, where they are used to specify the various domains in which the brackets appearing in the index take on different values.

	\section{Revisiting the \texorpdfstring{$\mathcal{N}=4$}{N=4} superconformal index}
	\label{sec:revis-index}
	
	In this section, we will take the generalized Cardy limit of the full superconformal index of the $\mathcal{N}=4$ theory.
	Based on this, we conjecture an expression for the full index which suggests an interpretation as a sum over geometries on the gravitational side.
	Subsequently, we will analyze the associated free energy and the resulting entropy formula.
	Finally, we end the section with an interpretation of the entropy formula.
	Before getting there, we first examine the computation of the anomaly polynomial through the new modular property, as this will provide some useful hints on what to expect for the generalized Cardy limit.

	\subsection{Anomaly polynomial}\label{ssec:anom-pol-index}
	
	There exists an interesting relation between the modular property of the elliptic $\Gamma$ function, the anomaly polynomial, and the supersymmetric Casimir energy as discussed in~\cite{Bobev:2015kza,Nieri:2015yia,Brunner:2016nyk} and more recently in~\cite{Gadde:2020bov}.
	In this section, we wish to point out how the new modular property also yields expressions which can be interpreted as an anomaly polynomial.
	This is perhaps not surprising, since the new modular property is structurally very similar to the ordinary one. 
	However, we will point out some interesting differences that will also be important in the following sections.
	
	The basic relation between modularity and the anomaly polynomial can be obtained as follows.
	First, one replaces each of the elliptic $\Gamma$ functions that make up the perturbative part of the partition function with the right hand side of the modular property~\eqref{eq:ord-mod-prop-Gamma}.
	The sum of $Q$ polynomials one thus obtains represents a version of the anomaly polynomial, defined in Appendix~\ref{app:defs}.
	
	We now apply the exact same idea for the new modular property.
	We will use the expression for the index derived in~\cite{Goldstein:2020yvj} and reviewed in Section~\ref{ssec:review-index}.
	Focusing on the perturbative part of the index, we have:
	\begin{align}\label{eq:suN-index-pert-anomaly}
		\begin{split}
			Z^{(a_i)}_{\text{P}}&=\frac{(\Gamma(f_1)\Gamma(f_2)\Gamma(f_3))^{N-1}}{N!\,\Gamma(1)^{N-1}}\prod^{N-1}_{i< j}\frac{\prod^3_{b=1}\Gamma((f_{a_i}f_{a_j}^{-1})^{\pm}f_b)}{\Gamma((f_{a_i}f_{a_j}^{-1})^{\pm})}\prod^{N-1}_{i=1}\frac{\prod^3_{b=1}\Gamma(f_{a_i}^{\pm}f_b)}{\Gamma(f_{a_i}^{\pm})}\,.
		\end{split}
	\end{align}
	We remind the reader about the third comment below~\eqref{eq:suN-index-final} concerning the $\Gamma$ functions in the denominators,
	To include them in the perturbative part turns out to be natural for the anomaly polynomial computation, as we explained in~\cite{Goldstein:2020yvj} and to which we refer for more details.
	
	Thus, we replace the $\Gamma$ functions using the modular property~\eqref{eq:mod-prop-gamma-a} and collect the total $Q$ polynomial.
	Notice that each $Q$ polynomial only depends on:
	\begin{equation}\label{eq:defn-tilde-phi}
		\tilde{\phi}_{1,2}\equiv m\,\phi_{1,2}\,.
	\end{equation}
	There, we write it as a function of $\tilde{\phi}_{1,2}$ and furthermore suppress the dependence on $\tau$ and $\sigma$:
	\begin{align}\label{eq:Qtot-suN}
		\begin{split}
			Q^{(a_i)}_{\mathrm{tot}}(\tilde{\phi}_a)=& (N-1)\left(-Q_{\mathbf{m}}(0)+\sum^3_{c=1}Q_{\mathbf{m}}(\tilde{\phi}_c)\right)\\
			&+\sum_{\tilde{\phi}\in\lbrace \tilde{\phi}_{a_i},\tilde{\phi}_{a_{i}}-\tilde{\phi}_{a_{j}}\rbrace }\Bigg[ \sum^3_{b=1}\Big(Q_{\mathbf{m}}(\tilde{\phi}+\tilde{\phi}_b)+Q_{\mathbf{m}}(-\tilde{\phi}+\tilde{\phi}_b)\Big)\\
			&\qquad \qquad \qquad \qquad -Q_{\mathbf{m}}(\tilde{\phi})-Q_{\mathbf{m}}(-\tilde{\phi})\Bigg] \,,
		\end{split}
	\end{align}
	Recall that for the ordinary modular property, the analogous polynomial evaluates to the anomaly polynomial for $\phi_3=\tau+\sigma -\phi_1-\phi_2-1$, although without an \textit{a priori} justification for the integer $-1$ since in principle it could have been any (odd) integer.
	In particular, for this choice of $\phi_3$, the total $Q$ polynomial does not depend on $(a_i)$, i.e., the specific residue.
	Interpreting a residue as a vacuum, as instructed by the Higgs branch localization computation, this matches the physical expectation that the anomaly polynomial of a theory should not depend on the choice of vacuum~\cite{Gadde:2020bov}.
	
	It turns out that in this more general case, there is an analogue of this.
	In particular, suppose we define:\footnote{
		This expression for $\tilde{\phi}_3$ will be derived in Section~\ref{ssec:gen-cardy-index} when we consider the generalized Cardy limit.}
	\begin{equation}\label{eq:phi3-constraint-anomaly-pol}
		\tilde{\phi}_3=\tilde{\tau}+\tilde{\sigma}-\tilde{\phi}_1-\tilde{\phi}_2-1\,,
	\end{equation}
	where we recall:
	\begin{equation}\label{eq:redef-moduli}
		\tilde{\tau}\equiv m\tau+1-m n_1 \quad \textrm{and}\quad \tilde{\sigma}\equiv m\sigma+1-mn_2\,.
	\end{equation}
	For this value of $\tilde{\phi}_3$, one finds that $Q^{(a_i)}_{\mathrm{tot}}$ does not depend on the choice of $(a_i)$.
	In particular, one finds that the second and third line do not depend on the summation variable $\phi$.
	Plugging in the definition of $Q_{\mathbf{m}}$ and dropping the superscript $(a_i)$, we obtain:
	\begin{equation}\label{eq:anomaly-pol-abc}
		Q_{\mathrm{tot}}(\phi_{a})=\frac{(N^2-1)}{m}\left(\frac{\tilde{\phi}_1\tilde{\phi}_2\tilde{\phi}_3}{\tilde{\tau}\tilde{\sigma}}+\frac{m^2-1}{6}\right).
	\end{equation}
	Notice  this expression is exact, i.e., we have not taken any Cardy-like limit to obtain it.\footnote{
		For future reference, we note here that together with~\eqref{eq:normZ-to-pertZ}, this equation verifies our claim~\eqref{eq:mod-prop-gen-theory} for the $\mathcal{N}=4$ theory.}
	The function is very closely related to the supersymmetric Casimir energy and the anomaly polynomial~\cite{Assel:2014paa,Bobev:2015kza}.
	However, let us point out some differences:
	\begin{itemize}
		\item The polynomial $Q^{(a_i)}_{\mathrm{tot}}$ depends only on the rescaled chemical potentials $\tilde{\phi}_{1,2}$, $\tilde{\tau}$ and $\tilde{\sigma}$, the latter two of which are also shifted. 
		\item When written in terms of the original variables, the expression for $\tilde{\phi}_3$ in~\eqref{eq:phi3-constraint-anomaly-pol} reads:
		\begin{equation}\label{eq:phi3-constraint-anomaly-pol-2}
			\phi_3=\tau+\sigma-\phi_1-\phi_2+\frac{1}{m}-n_1-n_2 \,.
		\end{equation}
		Notice that this expression of $\phi_3$ is not related by integer shifts to the expression in the original case~\cite{Goldstein:2020yvj}.
		A naive interpretation of this equation as a constraint on $\phi_3$ might seem problematic, for the following reason.
		We can rewrite the index~\eqref{eq:trace-defn-index} in terms of $f_3$:
		\begin{equation}\label{eq:trace-defn-index-2}
			I_N=\mathrm{tr}_{\mathcal{H}}(-1)^F p^{J_1}q^{J_2}f_1^{Q_1}f_2^{Q_2}f_3^{Q_3}e^{-\beta \lbrace \mathcal{Q},\mathcal{Q}^\dagger\rbrace} \,.
		\end{equation}
		Now, the supercharge $\mathcal{Q}$ only anticommutes with the operator in the trace if $ \phi_3=\tau+\sigma-\phi_1-\phi_2+2k$ with $k\in\mathbb{Z}$.
		Naively, the above constraint therefore would imply the ``index'' is not protected.
		However, notice that the supercharge will anticommute with the operator in the trace if the Hilbert space $\mathcal{H}$ is reduced to a subspace $\mathcal{H}^m$ which consists of only those states whose charges are multiples of $m$.
		In the next subsection, we will see that such a projection on $\mathcal{H}$ is precisely what occurs in the generalized Cardy limit, thus explaining the peculiar constraint~\eqref{eq:phi3-constraint-anomaly-pol-2}.
		\item There is an additional constant term which was not present for the anomaly polynomial derived using the ordinary modular property.
		Clearly, this term is of subleading order in the Cardy limit, so it will not play an important role in the following.
		We do note here that the constant term in the generalized Cardy limit has been related to the partition function of $SU(N)_k$ Chern--Simons theory on $S^3$ and lens spaces~\cite{GonzalezLezcano:2020yeb,ArabiArdehali:2021nsx}.
	\end{itemize}
	In the next section, we will see how the generalized Cardy limit of the index leads to a derivation and explanation of~\eqref{eq:phi3-constraint-anomaly-pol} and~\eqref{eq:anomaly-pol-abc}.

	\subsection{Generalized Cardy limit of the index}\label{ssec:gen-cardy-index}
	
	To compute the generalized Cardy limit of the index, we again only focus on the perturbative part of the partition function.
	This is because the limit of the elliptic $\Gamma$ function, as captured by $Q_{\mathbf{m}}$, diverges as $\mathcal{O}\left(\frac{1}{\tilde{\tau}\tilde{\sigma}}\right)$.
	Instead, the limit of the vortex partition functions, which is captured by the $B$ polynomial defined in~\eqref{eq:result}, diverges as $\mathcal{O}\left(\frac{1}{\tilde{\tau}}\right)$ or $\mathcal{O}\left(\frac{1}{\tilde{\sigma}}\right)$.
	
	Thus, we need to consider the following function:
	\begin{align}\label{eq:suN-index-pert}
		\begin{split}
			Z_{\text{P}}&=\frac{(\Gamma(f_1)\Gamma(f_2)\Gamma(f_3))^{N-1}}{N!\,\Gamma(1)^{N-1}}\sum^\prime_{(a_i)}\prod^3_{b=1}\prod^{N-1}_{i< j}\Gamma((f_{a_i}f_{a_j}^{-1})^{\pm}f_b)\prod^{N-1}_{i=1}\Gamma((f_{a_i})^{\pm}f_b) \,,
		\end{split}
	\end{align}
	where we now canceled the elliptic $\Gamma$ functions in the denominator of~\eqref{eq:suN-index-pert-anomaly} against the same factors in the vortex partition functions, because they will not contribute at leading order.
	See the third comment below~\eqref{eq:suN-index-final} and~\cite{Goldstein:2020yvj} for more information.
	
	Now, we plug in the new modular property~\eqref{eq:mod-prop-gamma-a-2} for each of the elliptic $\Gamma$ functions $Z_{\text{P}}$ and take the generalized Cardy limit.
	Consider a single term in the sum labelled by some $(N-1)$ tuple $(a_i)$, which we call $Z_{\text{P}}^{(a_i)}$ as in~\eqref{eq:suN-index-pert-anomaly}.
	Its generalized Cardy limit is simply the application of~\eqref{eq:gen-cardy-Gamma-2} for each individual $\Gamma$ function and reads:
	\begin{equation}
		\lim_{\text{gen Cardy}} \log Z_{\text{P}}^{(a_i)}=-i\pi Q^{(a_i)}_{\text{tot}}\left([\tilde{\phi}_a]\right)+\mathcal{O}\left(\tfrac{1}{\tilde{\tau}}\right)+\mathcal{O}\left(\tfrac{1}{\tilde{\sigma}}\right) \,, 
	\end{equation}
	where the bracket is defined in~\eqref{eq:bracket-defn} and $\tilde{\phi}_{1,2}$ in~\eqref{eq:defn-tilde-phi}, we again suppress dependence on $\tau$ and $\sigma$, and:
	\begin{align}\label{eq:Qtot-suN-cardy}
		\begin{split}
			Q^{(a_i)}_{\text{tot}}&\left([\tilde{\phi}_a]\right)= (N-1)\left(-Q'_{\mathbf{m}}(0)+\sum^3_{b=1}Q'_{\mathbf{m}}([\tilde{\phi}_b])\right)\\
			&+\sum_{\tilde{\phi}\in\lbrace \tilde{\phi}_{a_i},\tilde{\phi}_{a_{i}}-\tilde{\phi}_{a_{j}}\rbrace} \sum^3_{b=1}Q'_{\mathbf{m}}\left([\tilde{\phi}+\tilde{\phi}_b]\right)+Q'_{\mathbf{m}}\left([-\tilde{\phi}+\tilde{\phi}_b]\right) \,.
		\end{split}
	\end{align}
	Notice that in terms of the brackets, $[\tilde{\phi_3}]$ is unambiguously defined:
	\begin{equation}
		[\tilde{\phi}_3]=[\tilde{\tau}+\tilde{\sigma}-\tilde{\phi}_1-\tilde{\phi}_2]=[-\tilde{\phi}_1-\tilde{\phi}_2] \,,
	\end{equation}
	where we have used $f_3=pq(f_1f_2)^{-1}$, as required for the cancellation of non-BPS states to the index~\eqref{eq:trace-defn-index-2}, and in the last equation we evaluated the generalized Cardy limit $\tilde{\tau}=\tilde{\sigma}=0$.
	
	To proceed, we evaluate all the brackets that appear in the arguments of the $Q_{\mathbf{m}}$ polynomial.
	Similar to our previous work, we focus on the part of $Q^{(a_i)}_{\text{tot}}$ which scales as $N^2$, i.e., the part of the summation of the second line of~\eqref{eq:Qtot-suN-cardy} for:
	\begin{equation}
		\phi\in \lbrace\tilde{\phi}_{a_{i}}-\tilde{\phi}_{a_{j}}\rbrace \,.
	\end{equation}
	Depending on the choice of residue $(a_i)$, one can have two types of contributions.
	There will be terms in the sum with $a_i=a_j$ and terms with $a_i\neq a_j$.
	For the terms with $a_i=a_j$, the sum over $b$ is most easily performed.
	We obtain:
	\begin{align}
		\begin{split}
			Q^{(a_i)}_{\text{tot}}&\approx 2N^2\sum^3_{b=1}Q'_{\mathbf{m}}\left([\tilde{\phi}_b]\right)=\\
			&\begin{cases}
				-\frac{2N^2}{m}\frac{[\tilde{\phi}_1][\tilde{\phi}_2][\tilde{\phi}_3]}{\tilde{\tau}\tilde{\sigma}}+\mathcal{O}(\tilde{\tau}^{-1})+\mathcal{O}(\tilde{\sigma}^{-1})+\mathcal{O}(N)  &\text{if}\quad [\tilde{\phi}_1]+[\tilde{\phi}_2]\in D^{(0)} \, , \\
				-\frac{2N^2}{m} \frac{[\tilde{\phi}_1]^\prime [\tilde{\phi}_2]^\prime[\tilde{\phi}_3]^\prime}{\tilde{\tau}\tilde{\sigma}}+\mathcal{O}(\tilde{\tau}^{-1})+\mathcal{O}(\tilde{\sigma}^{-1})+\mathcal{O}(N) &\text{if}\quad [\tilde{\phi}_1]+[\tilde{\phi}_2]\in D^{(-1)} \, ,
			\end{cases} 
		\end{split}
	\end{align}
	where we note that the answer does not depend on the value of $a_i$, we have defined:
	\begin{equation}\label{eq:phiPrimeDef}
		[\tilde{\phi}_a]^\prime \equiv [\tilde{\phi}_a]+1 \, ,
	\end{equation}
	and used the fact that there are two possible ways to evaluate the bracket $ [\tilde{\phi}_3]$:
	\begin{equation}\label{eq:ev-phi3-D0-D1}
		[\tilde{\phi}_3]=[-\tilde{\phi}_1-\tilde{\phi}_2]=\begin{cases}
			-[\tilde{\phi}_1]-[\tilde{\phi}_2]-1  &\text{if}\quad [\tilde{\phi}_1]+[\tilde{\phi}_2]\in D^{(0)} \, , \\
			-[\tilde{\phi}_1]-[\tilde{\phi}_2]-2 &\text{if}\quad [\tilde{\phi}_1]+[\tilde{\phi}_2]\in D^{(-1)} \, .
		\end{cases}
	\end{equation}
	If instead $[\phi_1]+[\phi_2]\notin D^{(0)}$ or $D^{(-1)}$ we cannot proceed, as explained in Section~\ref{ssec:gen-cardy-gamma}.
	Notice that here, for the first time, we see how the constraint~\eqref{eq:phi3-constraint-anomaly-pol} arises in the generalized Cardy limit.
	We will discuss the meaning of this in more detail in Section~\ref{ssec:lens}.
	
	When $a_i\neq a_j$, we have three possibilities: $(a_i,a_j)=(1,2),(1,3),(2,3)$.
	The analysis of the sums over $b$ for these values of $(a_i,a_j)$ is completely identical to the analysis in Section 3.2.2 of~\cite{Goldstein:2020yvj}.
	In particular, it turns out that for generic values $\tilde{\phi}_{1,2}^{\text{s}\pm}$ (see the end of Section~\ref{ssec:gen-cardy-gamma} for the definition of $\tilde{\phi}_{1,2}^{\text{s}\pm}$), the $Q$ polynomials for the pairs $(a_i,a_j)$ will not agree at leading order in the generalized Cardy limit.
	This implies that $Q^{(a_i)}_{\text{tot}}$ depends on the residue $(a_i)$, as opposed to the anomaly polynomial derived in Section~\ref{ssec:anom-pol-index}.
	In particular, to extract a useful Cardy limit of the full index, this would require some sort of resummation of the full residue sum in~\eqref{eq:suN-index-pert}.
	The difficulty in this resummation is that one needs to take into account subleading terms in the Cardy limit, since the summation depends on the relative phases of the residues.
	We have not been able to do this so far.
	
	However, the crucial observation in~\cite{Goldstein:2020yvj} is that there exists a special region in $\tilde{\phi}_{1,2}^{\text{s}\pm}$ space, where the residues take on a universal form, i.e., they become independent of $(a_i)$.
	The special regions correspond to the regions which include the unrefined points: 
	\begin{equation}
		[\tilde{\phi}]\equiv[\tilde{\phi}_1]=[\tilde{\phi}_2]=[\tilde{\phi}_3] \,.
	\end{equation}
	This equation can be solved in the two cases mentioned in~\eqref{eq:ev-phi3-D0-D1}:
	\begin{align}\label{eq:unref-points}
		\begin{split}
			[\tilde{\phi}]=-\tfrac{1}{3} \quad &\text{if}\quad [\tilde{\phi}_1]+[\tilde{\phi}_2]\in D^{(0)} \, ,\\
			[\tilde{\phi}]=-\tfrac{2}{3}\quad &\text{if}\quad [\tilde{\phi}_1]+[\tilde{\phi}_2]\in D^{(1)}\, .
		\end{split} 
	\end{align}
	Notice that these unrefined points are distinct by a factor $\frac{1}{m}$ from the unrefined point of the original index, reviewed in Section~\ref{ssec:review-index}, which would correspond to $[\phi]=-\tfrac{1}{3}$ or $[\phi]=-\tfrac{2}{3}$.
	
	\begin{figure}
		\centering
		\includegraphics[width=.4\textwidth]{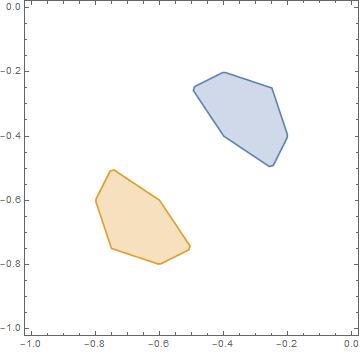}
		\caption{This figure shows the $-1<\tilde{\phi}_{1,2}^{\text{s}\pm}<0$ plane, where the $\tilde{\phi}_{1,2}^{\text{s}\pm}$ were defined in~\eqref{eq:phi-normal}.
			There are various domains in this region which correspond to regions where the brackets appearing in~\eqref{eq:defn-Qij} take on a fixed value.
			We only illustrated the relevant domains, blue and yellow, which correspond to the domains that contain the unrefined points $\phi_1=\phi_2=\phi_3\equiv \phi$ with $[m\phi]=-\tfrac{1}{3}$ and $[m\phi]=-\tfrac{2}{3}$, respectively.}
		\label{fig:large-n-domain}
	\end{figure}
	
	Let us define:
	\begin{equation}\label{eq:defn-Qij}
		Q_{ij}=\sum^3_{b=1}Q'_{\mathbf{m}}\left([\tilde{\phi}+\tilde{\phi}_b]\right)+Q'_{\mathbf{m}}\left([-\tilde{\phi}+\tilde{\phi}_b]\right)\,, \quad \tilde{\phi}=\tilde{\phi}_{a_i}-\tilde{\phi}_{a_j} \,.
	\end{equation}
	As explained in Section 3.2.2 of~\cite{Goldstein:2020yvj}, the various bracketed potentials appearing in this expression divide the $\tilde{\phi}_{1,2}^{\text{s}\pm}$ space into various domains where the set of brackets takes on a set of fixed values.
	The domains which contain the unrefined points~\eqref{eq:unref-points} are depicted in Figure~\ref{fig:large-n-domain}.
	In the blue domain, for which $[\tilde{\phi}_1]+[\tilde{\phi}_2]\in D^{(0)}$, we then have:
	\begin{align}
		\begin{split}
			Q_{12}&=-\frac{2}{m}\frac{[\tilde{\phi}_1][\tilde{\phi}_2][\tilde{\phi}_3]}{\tilde{\tau}\tilde{\sigma}}+\frac{([\tilde{\phi}_1]-[\tilde{\phi}_2])^2}{m\tilde{\tau}\tilde{\sigma}}+\mathcal{O}(\tau^{-1})+\mathcal{O}(\sigma^{-1})\,,\\
			Q_{13}&=-\frac{2}{m}\frac{[\tilde{\phi}_1][\tilde{\phi}_2][\tilde{\phi}_3]}{\tilde{\tau}\tilde{\sigma}}+\frac{([\tilde{\phi}_1]-[\tilde{\phi}_3])^2}{m\tilde{\tau}\tilde{\sigma}}+\mathcal{O}(\tau^{-1})+\mathcal{O}(\sigma^{-1})\,,\\
			Q_{23}&=-\frac{2}{m}\frac{[\tilde{\phi}_1][\tilde{\phi}_2][\tilde{\phi}_3]}{\tilde{\tau}\tilde{\sigma}}+\frac{([\tilde{\phi}_2]-[\tilde{\phi}_3])^2}{m\tilde{\tau}\tilde{\sigma}}+\mathcal{O}(\tau^{-1})+\mathcal{O}(\sigma^{-1})\,,
		\end{split}
	\end{align}
	where:
	\begin{equation}\label{eq:constraint-blue-dom}
		[\tilde{\phi}_3]=-[\tilde{\phi}_1]-[\tilde{\phi}_2]-1\,.
	\end{equation}
	This shows that, close enough to the unrefined point, the residues take on a universal form: $Q_{12}= Q_{23}= Q_{13}$.
	It is also clear from the full expression of the index in~\eqref{eq:suN-index-final} that subleading corrections, whether in $\tilde{\tau}$, $\tilde{\sigma}$ or $N$, will not spoil this universality. 
	Therefore, we may replace $Z_{\text{P}}^{(a_i)}$ with the full index $I_N$ in the generalized Cardy limit and write:\footnote{
		For a more detailed argument for the replacement, see the review of~\cite{Goldstein:2020yvj} in Section~\ref{ssec:review-index}.}
	\begin{equation}\label{eq:gen-cardy-limit-index-blue}
		\lim_{\text{gen Cardy}} \log I_N =-i\pi\frac{N^2}{m}\frac{[\tilde{\phi}_1][\tilde{\phi}_2][\tilde{\phi}_3]}{\tilde{\tau}\tilde{\sigma}}+\mathcal{O}\left(\frac{([\tilde{\phi}_a]-[\tilde{\phi}_b])^2}{m\tilde{\tau}\tilde{\sigma}}\right)+\mathcal{O}\left(\tfrac{1}{\tilde{\tau}}\right)+\mathcal{O}\left(\tfrac{1}{\tilde{\sigma}}\right)\,.
	\end{equation}
	Analogously, for the yellow region in Figure~\ref{fig:large-n-domain}, we can write:
	\begin{equation}\label{eq:gen-cardy-limit-index-yellow}
		\lim_{\text{gen Cardy}} \log I_N =-i\pi\frac{N^2}{m}\frac{[\tilde{\phi}_1]'[\tilde{\phi}_2]'[\tilde{\phi}_3]'}{\tilde{\tau}\tilde{\sigma}}+\mathcal{O}\left(\frac{([\tilde{\phi}_a]'-[\tilde{\phi}_b]')^2}{m\tilde{\tau}\tilde{\sigma}}\right)+\mathcal{O}\left(\tfrac{1}{\tilde{\tau}}\right)+\mathcal{O}\left(\tfrac{1}{\tilde{\sigma}}\right)\,,
	\end{equation}
	where now:
	\begin{equation}
		[\tilde{\phi}_3]'=-[\tilde{\phi}_1]'-[\tilde{\phi}_2]'+1\,,
	\end{equation}
	using the definition~\eqref{eq:phiPrimeDef} and that $[\tilde{\phi}_3]=-[\tilde{\phi}_1]-[\tilde{\phi}_2]-2$ in the yellow domain.  
	
	The expression~\eqref{eq:gen-cardy-limit-index-blue} shows how the generalized Cardy limit of the index reproduces the anomaly polynomial of the theory, discussed in Section~\ref{ssec:anom-pol-index}, close to the unrefined point and at large-$N$.
	In this case, however, the constraint~\eqref{eq:phi3-constraint-anomaly-pol} is derived rigorously.
	In particular, writing the limit~\eqref{eq:gen-cardy-limit-index-blue} in terms of the original chemical potentials, we have:
	\begin{equation}\label{eq:gen-cardy-index-or-var}
		\lim_{\text{gen Cardy}} \log I_N =-i\pi\frac{N^2}{m}\frac{[m\phi_1][m\phi_2]([m\phi_1]+[m\phi_2]+1)}{(m\tau+1-mn_1)(m\sigma+1-mn_2)}\,,
	\end{equation}
	where we see that the correct interpretation of the constraint~\eqref{eq:phi3-constraint-anomaly-pol-2} in the generalized Cardy limit is:
	\begin{equation}
		[\tilde{\phi_3}]=-[m\phi_1]-[m\phi_2]-1\,.
	\end{equation}
	Also, let us note that the generalized Cardy limit for $m>1$ exhibits a finer periodicity in the chemical potentials $\phi_a$ than the full index~\eqref{eq:trace-defn-index}: 
	\begin{equation}\label{newPeriodicity}
		\phi_{1,2}\to \phi_{1,2}+\frac{1}{m}\,.
	\end{equation}
	We will return to this observation in Section~\ref{ssec:lens}.
	Furthermore, at the end of Section~\ref{ssec:entropy}, we will discuss the relation of our result to the analysis of~\cite{Cabo-Bizet:2019eaf}.
	
	We close this section by making some comments about~\eqref{eq:gen-cardy-index-or-var}.
	First of all, we see that~\eqref{eq:gen-cardy-index-or-var} for any value of $m$ is periodic under:
	\begin{equation}\label{oldPeriodicity}
		\phi_{1,2}\to \phi_{1,2}+1\,.
	\end{equation}
	This is consistent with the fact that the trace definition of the index has the same periodicity, as discussed in Section~\ref{ssec:review-index}.
	This periodicity was already observed in the original $\tau,\sigma\to 0$ limit, performed in~\cite{Goldstein:2020yvj}.
	However, we now observe that the appearance of $n_{1,2}$ could also reflect the periodicity of the index under:
	\begin{equation}
		\tau\to\tau+1,\quad \sigma\to\sigma+1\, .
	\end{equation}
	The idea is that we think of (part of) the index as the following formal expression:\footnote{
		This sum may be divergent and then needs to be regularized.}
	\begin{equation}\label{eq:farey-tail}
		I_N=\sum_{\substack{m,n_1,n_2\in \mathbb{Z}\\ m\neq 0}}\exp\left(-i\pi \frac{N^2}{m}\frac{[m\phi_1][m\phi_2]([m\phi_1]+[m\phi_2]+1)}{(m\tau+1-m n_1)(m \sigma+1-m n_2)}\right)\,\times\, \alpha(\mathbf{m})\,,
	\end{equation}
	where $\alpha(\mathbf{m})$ denotes some unknown function. 
	Let us first discuss some motivations to write the index as such, after which we will turn to discuss some of its deficiencies.
	First of all, this expression is manifestly periodic in all chemical potentials, consistent with the trace definition of the index in~\eqref{eq:trace-defn-index}.
	Secondly, if we assume that $\alpha(\mathbf{m})$ regularizes the possibly divergent sum and that moreover in the generalized Cardy limit associated to $\mathbf{m}$ the dominating term in the sum has:
	\begin{equation}
		\alpha(\mathbf{m})=1+\text{corrections}\,,
	\end{equation}
	and the corrections are subleading, the expression reproduces the result~\eqref{eq:gen-cardy-index-or-var}.
	Thirdly, if we interpret the generalized Cardy limits in~\eqref{eq:gen-cardy-index-or-var} as providing the on-shell action of some Euclidean gravitational saddles,\footnote{See~\cite{Cabo-Bizet:2019eaf,Cabo-Bizet:2020nkr,ArabiArdehali:2021nsx} for further evidence for this interpretation, and~\cite{Aharony:2021zkr} for a recent explicit construction of the saddles.} which generalizes the case of ordinary AdS$_5$ black hole for $\mathbf{m}=(1,1,1)$ to arbitrary $\mathbf{m}$, the expression takes on a natural form from the gravitational perspective.
	Indeed, the sum would correspond to a sum over gravitational saddles, and the summand corresponds to $e^{-S_{\text{on-shell}}}$ with $S_{\text{on-shell}}$ the on-shell action of the solution.
	This third motivation is inspired by the familiar Farey tail expansion in two dimensions~\cite{Dijkgraaf:2000fq,Manschot:2007ha}, where the elliptic genus of a two-dimensional CFT is rewritten in a form that resembles the expected form of a Euclidean gravitational path integral.
	We will discuss this analogy extensively in Section~\ref{sec:disc}.
	
	The formula~\eqref{eq:farey-tail} only includes the saddles reached by the Cardy limits~\eqref{eq:gen-cardy-limit}.
	Let us now include the saddles from more general order three $SL(3,\mathbb{Z})$ elements, of the form $A=h\cdot S_{23}$ with $h\in H$, studied in Section~\ref{ssec:finite-order-elms}.
	(The $X_{\mathbf{m}}$ correspond to a special case of such elements.)
	The most general element of this form leads to a similar modular property as that of $X_{\mathbf{m}}$ (see Section~\ref{ssec:new-mod-prop}) but with $Q'_{\mathbf{m}}$ polynomial given by:\footnote{For a discussion of these more general elements, see Appendix~\ref{app:more-gen-order-3-elms}. For a (partial) derivation of the associated $Q$ polynomial, see Section~\ref{app:most-gen-mod-prop}.}
	\begin{equation}\label{eq:gen-Q-1}
		Q_{\text{gen}}(z;\tau,\sigma) =  Q(m z\,; m \tau+ n\,,  m(\sigma+k)+n^2) + \text{constant} \,,
	\end{equation}
	where $m$ and $n$ are subject to the constraint $m|n^3-1$, $k$ is a free integer, and we have not been able to determine the constant.
	For an element giving rise to~\eqref{eq:gen-Q-1}, there exists another element which leads to a modular property with $Q$ polynomial:
	\begin{equation}\label{eq:gen-Q-2}
		Q_{\text{gen}}(z;\tau,\sigma) =  Q(m z\,; m (\tau+k)+ n^2\,,  m\sigma+n) + \text{constant} \,.
	\end{equation}
	Since the constant is irrelevant at leading order in the generalized Cardy limit, the analysis in the beginning of this section applies equally well to these more general modular properties.
	In particular, the associated generalized Cardy limits for the modular properties are now:
	\begin{equation}
		(\tau\,,\sigma) \to \left(-\frac{n}{m}\,, -\frac{n^2}{m} -k\right)  \quad \text{and}\, \quad 	(\tau\,,\sigma) \to \left(-\frac{n^2}{m}-k\,, -\frac{n}{m} \right) \,. 
	\end{equation}
	The expressions for the index in the respective limits now read:
	\begin{align}\label{eq:gen-cardy-index-or-var-2}
		\begin{split}
			\lim_{\text{gen Cardy}} \log I_N &=-i\pi\frac{N^2}{m}\frac{[m\phi_1][m\phi_2]([m\phi_1]+[m\phi_2]+1)}{( m \tau+ n)(m(\sigma+k)+n^2)}\,,\\
			\lim_{\text{gen Cardy}} \log I_N &=-i\pi\frac{N^2}{m}\frac{[m\phi_1][m\phi_2]([m\phi_1]+[m\phi_2]+1)}{(m(\tau+k)+n^2)( m \sigma+ n)}\, .
		\end{split}
	\end{align}
	Even though each of these limits separately breaks the symmetry between $\tau$ and $\sigma$, the symmetry would be restored in a sum formula like~\eqref{eq:farey-tail}.
	After including this more general class of saddles, the sum in~\eqref{eq:farey-tail} is enlarged into
	\begin{equation}\label{eq:farey-tail2}
		I_N=\sum_{\widehat{\textbf{m}}}e^{-S^{\widehat{\mathbf{m}}}_{\text{on-shell}}}\,\times\, \alpha(\widehat{\mathbf{m}})\,,
	\end{equation}
	where $\widehat{\mathbf{m}}=(m,n,k)$ such that $m|n^3-1$ and $k\in \mathbb{Z}$ and $-S^{\widehat{\mathbf{m}}}_{\text{on-shell}}$ is given by~\eqref{eq:gen-cardy-index-or-var} or~\eqref{eq:gen-cardy-index-or-var-2}, depending on the type of $\widehat{\mathbf{m}}$.
	
	The formula~\eqref{eq:farey-tail2} includes all the phases associated to the modular properties studied in this paper.
	However, there are reasons to believe that this sum is still incomplete, and in particular should at least contain terms with the denominator of $S_{\text{on-shell}}$ equal to $(m\tau+n_1)(m\sigma+n_2)$ for arbitrary coprime $m$ and $n_i$~\cite{Cabo-Bizet:2019eaf,Cabo-Bizet:2020ewf,ArabiArdehali:2021nsx,Aharony:2021zkr}.
	To find the modular properties that allow for these more general Cardy limits, we will need to find relations in the group $G$ beyond the order three elements.
	We will suggest a different type of relation in Section~\ref{ssec:future}, which may allow one to take such more general limits and thus complete the sum.
	
	\newpage
	
	\subsection{Entropy}\label{ssec:entropy}
	
	In this section, we will compute the entropy from the free energy derived in the previous subsection by a Legendre transformation.
	The analysis will be similar to the case of the AdS$_5$ black hole studied in~\cite{Hosseini:2017mds,Cabo-Bizet:2018ehj,Choi:2018hmj,Benini:2018ywd}.
	We will focus on the free energy in the blue domain~\eqref{eq:gen-cardy-limit-index-blue}, which we can write as:
	\begin{equation}\label{eq:free-energy}
		F \equiv\lim_{\text{gen Cardy}} \log I_N = -i\pi\frac{N^2}{m}\frac{\tilde{\phi}_1\tilde{\phi}_2\tilde{\phi}_3}{\tilde{\tau}\tilde{\sigma}} \, .
	\end{equation}
	Here, $\tilde{\phi}_3=-\tilde{\phi}_1-\tilde{\phi}_2-1$ and the tilded chemical potentials are defined in~\eqref{eq:defn-tilde-phi} and~\eqref{eq:redef-moduli}.
	Furthermore, we have suppressed the brackets and only check that the saddle point values of the $\tilde{\phi}_{1,2,3}$ all lie in the domain $D^{(0)}$ at the end of the calculation.
	Finally, we have written the free energy in a fully refined manner, but caution the reader that our derivation of the expression is only valid parametrically close to the unrefined point. 
	
	The entropy associated to the generalized Cardy limit of the index is given by the Legendre transformation of $F$ for charges $J_{1,2}$ and $Q_{1,2,3}$:
	\begin{align}\label{eq:extrimi-modular}
		\begin{split}
			&S=\text{ext}_{\phi_a,\tau,\sigma,\Lambda}\left[F -2\pi i\left(\tau J_1+\sigma J_2+\phi_aQ_a+\Lambda\left(\tilde{\phi}_1+\tilde{\phi}_2+\tilde{\phi}_3-\tilde{\tau}-\tilde{\sigma}+1\right)\right)\right]\,,
		\end{split}
	\end{align}
	where the Lagrange multiplier $\Lambda$ enforces the constraint~\eqref{eq:constraint-blue-dom}.
	Notice that we reinstated $\tilde{\tau}$ and $\tilde{\sigma}$ in the constraint, even though they do not contribute to the constraint in the generalized Cardy limit, as we discussed in Section~\ref{ssec:gen-cardy-index}.
	This is done for completeness, but at the end of this section we will indeed see  that consistency of the saddle points with the generalized Cardy limit implies that their reinstatement is immaterial to the extremization to leading order.
	
	It will be convenient to rewrite the extremization in terms of tilded chemical potentials as:
	\begin{align}\label{eq:entropy-extremi-Xabc}
		\begin{split}
			S=\frac{1}{m}\,\text{ext}_{\tilde{\phi}_{1,2},\tilde{\tau},\tilde{\sigma},\tilde{\Lambda}}&\left[-i\pi N^2\frac{\tilde{\phi}_1\tilde{\phi}_2\tilde{\phi}_3}{\tilde{\tau}\tilde{\sigma}} -2\pi i\left(\tilde{\tau} \tilde{J}_1+\tilde{\sigma} \tilde{J}_2+\tilde{\phi}_a\tilde{Q}_a\right.\right.\\
			&\left.+\tilde{\Lambda}\left(\tilde{\phi}_1+\tilde{\phi}_2+\tilde{\phi}_3-\tilde{\tau}-\tilde{\sigma}+1\right)\right)\Bigg]\,,
		\end{split}
	\end{align}
	where we defined $\tilde{\Lambda}=m\Lambda$, simply pulled out a factor $\frac{1}{m}$, and introduced the shifted charges:
	\begin{equation}\label{eq:tilde-charges}
		\tilde{J}_{i}=J_{i}+\,C\quad \textrm{and}\quad \tilde{Q}_{a}=Q_{a}-\,C\,,
	\end{equation}
	with the constant $C$ given by:
	\begin{equation}
		C=(m n_1-1)J_1+(m n_2-1)J_2 \,.
	\end{equation}
	Notice that the arguments of the extremizations in~\eqref{eq:extrimi-modular} and~\eqref{eq:entropy-extremi-Xabc} are only equal on the saddle point.
	The final entropy from~\eqref{eq:entropy-extremi-Xabc} should be $C$-independent, which can be seen after integrating out $\tilde{\Lambda}$.
	The result can be written as the extremization over $C$-independent combinations of charges multiplying the chemical potentials and an additional constant $C$-dependent term:
	\begin{align}\label{eq:overall-extre-S}
		\begin{split}
			S=\frac{1}{m}\,\text{ext}_{\tilde{\phi}_{1,2},\tilde{\tau},\tilde{\sigma},\tilde{\Lambda}}&\left[-i\pi N^2\frac{\tilde{\phi}_1\tilde{\phi}_2\tilde{\phi}_3}{\tilde{\tau}\tilde{\sigma}} -2\pi i\left(\tilde{\tau} (\tilde{J}_1+\tilde{Q}_3)+\tilde{\sigma} (\tilde{J}_2+\tilde{Q}_3) \right. \right. \\
			& \left. 
			+\tilde{\phi}_1(\tilde{Q}_1-\tilde{Q}_3) +\tilde{\phi}_2(\tilde{Q}_2-\tilde{Q}_3)\right)  + 2\pi i \tilde{Q}_3 \Bigg]
		\end{split}
	\end{align}
	Clearly, the saddle point values of the chemical potentials will be $C$-independent. 
	Plugging these values back into the entropy function, we will acquire an entropy of the form $S=\frac{1}{m}(S_0 +2\pi i\tilde{Q}_3)$ with $S_0$ $C$-independent. 
	Imposing reality of the entropy requires that $2\pi\tilde{Q}_3=-\mathrm{Im}(S_0)$.
	This leads to an expression for the entropy purely in terms of the saddle point values of the chemical potentials, which are $C$-independent.\footnote{We thank Alejandro Cabo-Bizet for discussions regarding this point.} 
	For this reason, we will may remove all the tilded symbols in the charges $Q_a$ and $J_i$ in the rest of this section.  
	
	The rewriting in~\eqref{eq:entropy-extremi-Xabc}, while removing the tildes, makes the similarity to the extremization in the original case manifest~\cite{Hosseini:2017mds,Cabo-Bizet:2018ehj}.
	In particular, we can immediately borrow the saddle point values for the chemical potentials in terms of the charges from~\cite{Cabo-Bizet:2018ehj}:
	\begin{align}\label{eq:saddle-points-chem-pots}
		\begin{split}
			\tilde{\phi}_1^*&=-\Theta \hat{Q}_2\hat{Q}_3\,,\quad \tilde{\phi}_2^*=-\Theta \hat{Q}_1\hat{Q}_3\,,\quad \tilde{\phi}_3^*=-\Theta \hat{Q}_1\hat{Q}_2\,,\\
			\tilde{\tau}^*&=-\tfrac{1}{2}N^2\Theta \hat{J}_2\,,\quad \tilde{\sigma}^*=-\tfrac{1}{2}N^2\Theta \hat{J}_1\,,
		\end{split}
	\end{align}
	where we have defined:
	\begin{equation}\label{eq:JQhat}
		\hat{J}_i\equiv J_i-\tilde{\Lambda}^* \,, \qquad \hat{Q}_a\equiv Q_a+\tilde{\Lambda}^*\,,
	\end{equation}
	and
	\begin{equation}
		\Theta\equiv\frac{1}{(\hat{Q}_1\hat{Q}_2+\hat{Q}_2\hat{Q}_3+\hat{Q}_1\hat{Q}_3)-\frac{N^2}{2}(\hat{J}_1+\hat{J}_2)}\,.
	\end{equation}
	In~\eqref{eq:JQhat}, $\tilde{\Lambda}^*$ is the saddle point value of the Lagrange multiplier $\tilde{\Lambda}$, satisfying the cubic equation:
	\begin{equation}\label{eq:LambdaEq}
		\tilde{\Lambda}^3+p_2\tilde{\Lambda}^2+p_1\tilde{\Lambda}+p_0=0\,,
	\end{equation}
	where:
	\begin{align}\label{eq:pi}
		\begin{split}
			p_0&\equiv Q_1 Q_2 Q_3+\tfrac{1}{2}N^2 J_1 J_2\,,\\
			p_1&\equiv Q_1 Q_2+Q_2 Q_3+ Q_1 Q_3-\tfrac{1}{2}N^2(J_1+J_2)\,,\\
			p_2&\equiv Q_1+Q_2+Q_3+\tfrac{1}{2}N^2\,.
		\end{split}
	\end{align}
	Similar to the computation of the entropy for AdS$_5$ black hole, reality of the entropy selects one out of the three roots of~\eqref{eq:LambdaEq}.
	In particular, on the saddle the expression for the extremum simplifies significantly:
	\begin{equation}
		S=-\frac{2\pi i}{m}\tilde{\Lambda}^*\,.
	\end{equation}
	Hence, we see that for the entropy to be real, we need a purely imaginary positive root for $\tilde{\Lambda}^*$.
	Similar to the original case, this can only happen when:
	\begin{equation}\label{eq:tildepConstraint}
		p_0=p_1p_2 \,.
	\end{equation}
	which is equivalent to the non-linear charge constraint obeyed by the ordinary AdS$_5$ black hole~\cite{Gutowski:2004ez, Gutowski:2004yv,Chong:2005da,Chong:2005hr,Kunduri:2006ek}.\footnote{See also~\cite{Larsen:2019oll,Cabo-Bizet:2018ehj,Choi:2018hmj,Larsen:2021wnu} for more recent discussions.}
	When the charges obey~\eqref{eq:tildepConstraint}, the cubic equation~\eqref{eq:LambdaEq} can be rewritten as
	\begin{equation}
		(\tilde{\Lambda}^2+ p_1)(\tilde{\Lambda}+ p_2)=0 \,.
	\end{equation}
	The solution for which the entropy is real and positive is given by $\tilde{\Lambda}^*=i\sqrt{p_1}$.
	The entropy is then given by:\footnote{In v1 of this paper, we expressed the entropy in terms of the shifted charges~\eqref{eq:tilde-charges}. This incorrectly suggested a dependence of the entropy on $C$. As we explained below~\eqref{eq:overall-extre-S}, the entropy is independent of $C$. The present result for the entropy coincides with the entropy found for $(m,n)$ saddles in~\cite{Cabo-Bizet:2019eaf,Cabo-Bizet:2020nkr}. We compare our result in more detail at the end of this section.}
	
	\begin{tcolorbox}[ams align, colback=yellow!10!white]
		\begin{split}\label{eq:entropy-a}
			S&=\frac{2\pi}{m} \sqrt{p_1}=\frac{2\pi}{m}\sqrt{Q_1 Q_2+Q_2 Q_3+ Q_1Q_3-\tfrac{1}{2}N^2(J_1+J_2)}\,.
		\end{split}
	\end{tcolorbox}
	\noindent
	We will interpret this formula in Section~\ref{ssec:lens}.
	We also notice that the computation of the entropy for the more general modular properties, mentioned at the end of Section~\ref{ssec:gen-cardy-index} in~\eqref{eq:gen-cardy-index-or-var-2}, is completely equivalent to the above.
	In the remainder of this section, we check self-consistency of the analysis, and at the end comment on the relation to~\cite{Cabo-Bizet:2019eaf}.
	
	First of all, we derived the free energy in the generalized Cardy limit and close to the unrefined point:
	\begin{equation}
		\tilde{\tau},\tilde{\sigma}\to 0\,, \quad [\tilde{\phi}_{1,2,3}]\approx -\tfrac{1}{3}\,.
	\end{equation}
	The saddle point values of the chemical potentials in~\eqref{eq:saddle-points-chem-pots} should be consistent with this limit.
	It is not difficult to check that this can be achieved if the charges scale as follows~\cite{Choi:2018hmj}:
	\begin{equation}
		Q_a =\mathcal{O}\left(\mu^2\right)\,,\quad J_i =\mathcal{O}\left(\mu^3\right)\,,\quad \mu\to \infty\,.
	\end{equation}
	
	We now check that the values of $\tilde{\phi}_{1,2,3}$ lie inside the domain $D^{(0)}$, described in Section~\ref{ssec:gen-cardy-gamma}.
	Following~\cite{Benini:2018ywd}, we first use~\eqref{eq:saddle-points-chem-pots} to find:
	\begin{align}
		\begin{split}
			\frac{\tilde{\phi}_a^*}{\tilde{\tau}^*}&=-\frac{J_1-i\sqrt{p_1}}{Q_a+i\sqrt{p_1}}\,,\qquad
			\frac{\tilde{\phi}_a^*}{\tilde{\sigma}^*}=-\frac{J_2-i\sqrt{p_1}}{Q_a+i\sqrt{p_1}}\,,\qquad          \frac{\tilde{\sigma}^*}{\tilde{\tau}^*}=\frac{J_1-i\sqrt{p_1}}{J_2-i\sqrt{p_1}}\,.
		\end{split}
	\end{align}
	The imaginary parts of these expressions are given by:
	\begin{align}
		\begin{split}
			\mathrm{Im}\left(\frac{\tilde{\phi}_a^*}{\tilde{\tau}^*}\right)&=\frac{(J_1+Q_a)\sqrt{p_1}}{p_1+Q_a^2}\,,\qquad
			\mathrm{Im}\left( \frac{\tilde{\phi}_a^*}{\tilde{\sigma}^*}\right) =\frac{(J_2+Q_a)\sqrt{p_1}}{p_1+Q_a^2}\,,\\
			\mathrm{Im}\left( \frac{\tilde{\sigma}^*}{\tilde{\tau}^*}\right) &=\frac{(J_1-J_2)\sqrt{p_1}}{p_1+J_2^2}\,,\qquad \mathrm{Im}\left( \frac{\tilde{\tau}^*}{\tilde{\sigma}^*}\right) =\frac{(J_2-J_1)\sqrt{p_1}}{p_1+J_1^2}\,.
		\end{split}
	\end{align}
	Notice that the upper expressions are manifestly positive if $J_i+Q_a>0$.
	Furthermore, the lower two expressions vanish in the generalized Cardy limit.\footnote{
		This implies that $\alpha_\tau=\alpha_\sigma$ on the saddle point (see~\eqref{eq:gen-cardy-limit}), which is a divergent limit of the modular property. This divergence can be dealt in a similar way as is done in Theorem 5.2 in~\cite{Felder_2000}, as we discussed at the end of Section~\ref{ssec:new-mod-prop}. Keeping this in mind, the above analysis is self-consistent. At the moment, it is not clear how to make the saddle point analysis consistent with the generalized Cardy limit for $\alpha_\tau\neq\alpha_\sigma$.}
	We can then use the constraint to leading order to find:
	\begin{align}
		\begin{split}
			\mathrm{Im}\left(\frac{-1}{\tilde{\tau}^*}\right)&=\mathrm{Im}\left(\frac{\tilde{\phi}_1^*+\tilde{\phi}_2^*+\tilde{\phi}_3^*}{\tilde{\tau}^*}\right)\,,\qquad         \mathrm{Im}\left(\frac{-1}{\tilde{\sigma}^*}\right)=\mathrm{Im}\left(\frac{\tilde{\phi}_1^*+\tilde{\phi}_2^*+\tilde{\phi}_3^*}{\tilde{\sigma}^*}\right)\,.
		\end{split}
	\end{align}
	From the previous two equations we can conclude:
	\begin{align}
		\begin{split}
			\mathrm{Im}\left(\frac{-1}{\tilde{\tau}^*}\right)&>\mathrm{Im}\left(\frac{\tilde{\phi}_a^*}{\tilde{\tau}^*}\right)>0\,,\qquad 
			\mathrm{Im}\left(\frac{-1}{\tilde{\sigma}^*}\right)>\mathrm{Im}\left(\frac{\tilde{\phi}_a^*}{\tilde{\sigma}^*}\right)>0\,,
		\end{split}
	\end{align}
	which shows that $\tilde{\phi}^*_a$ lie inside the diamond domain~\eqref{eq:tau-sigma-domains}, which is what we wanted to show.
	For them to be parametrically close to the unrefined point, we should take:
	\begin{equation}
		Q_1\approx Q_2\approx Q_3\,.
	\end{equation}

	Finally, we wish to compare our result for the entropy to the work~\cite{Cabo-Bizet:2019eaf}.
	In that paper, the authors compute the superconformal index by a saddle point approximation in the large-$N$ limit. 
	Using an elliptic extension of the gauge integrand into the complex plane, they find saddle points labelled by two integers $(m,n)$.
	Such a saddle dominates their ensemble in the region around $\tau=-\frac{n}{m}$, and the associated free energy they find is given by:
	\begin{equation}
		S_{\text{eff}}=\frac{i\pi N^2}{27m}\frac{(2\tilde{\tau}-1)^3}{\tilde{\tau}^2}+i\pi N^2\varphi(m,n)\,, \qquad \tilde{\tau}=m\tau+n \,,
	\end{equation}
	with $\varphi(m,n)$ an unknown real constant independent of $\tau$ and $m+n=2\,\mod 3$.
	Up to conventions and the constant, this agrees precisely with the free energy we find in the generalized Cardy limit for $(m,n)=(m,1-m n_1)$, although we used an analytic extension of the gauge integrand to compute the index.
	Their solution for $m+n=1\,\mod 3$ agrees with our computation when we sit at the unrefined point in the yellow domain of Figure~\ref{fig:large-n-domain}.
	Finally, saddles for which $m+n=0\,\mod 3$ never dominate the ensemble. 
	For us, these constraints on the sum of $m$ and $n$ are not seen.
	In particular, $m+1-m n_1$ can take any value $\mod 3$ and for each generalized Cardy limit, we find two free energies: one in the blue domain and one in the yellow domain.
	
	In the subsequent analysis of~\cite{Cabo-Bizet:2019eaf}, the constraint $2\tau-2\phi=-1$ was integrated out such that their extremization becomes:\footnote{There is a factor of $2$ difference in the normalization of $R$-charges $Q$ from the original work~\cite{Cabo-Bizet:2019eaf}.}
	\begin{equation}\label{eq:elliptic-extremi-mn}
		S_{(m,n)}=\text{ext}_{\tau}\Big[-S_{\text{eff}}-2\pi i \tau (2 J+  2Q ) -2\pi i  Q\Big]\,,
	\end{equation}
	Notice that the extremization only depends on the combination $J+Q$.
	In our case, the extremization similarly only depends on $J_{1,2}+Q_3$ and $Q_{1,2}-Q_3$.
	In addition, the entropy is independent of the added constant $\varphi(m,n)$ for the same reason that our entropy does not depend on $C$, as described below~\eqref{eq:overall-extre-S}.
	
	To compare with our extremization problem~\eqref{eq:extrimi-modular}, we note that at the unrefined point $\phi\equiv \phi_1=\phi_2=\phi_3$ and $\tilde{\tau}=\tilde{\sigma}$ our constraint becomes:
	\begin{equation}
		3\phi-2\tau=\frac{1}{m}+2n_1.
	\end{equation}
	Notice that the $2n_1$ on the right hand side ensures that in the generalized Cardy limit, $\phi\in D^{(0)}_{1/m}$ as discussed in Section~\ref{ssec:gen-cardy-gamma}.
	Integrating out the constraint and taking $J\equiv J_1=J_2$ and $Q\equiv Q_1=Q_2=Q_3$, we find:
	\begin{equation}\label{eq:modular-extri-integratephi}
		S= \text{ext}_{\tau} \left[F -2\pi i \tau (2J+2Q) - \frac{2\pi i}{m} (1-2mn_1) Q  \right]\,,
	\end{equation}
	which is different from~\eqref{eq:elliptic-extremi-mn} in the last term due to different constraints on chemical potentials.
	Since the final entropy does not depend on this shift in the imaginary part we can conclude the final entropy will agree with the results from elliptic extension method~\cite{Cabo-Bizet:2019eaf}.

	\subsection{Interpretation}\label{ssec:lens}
	
	In this subsection, we will interpret the entropy formula derived in Section~\ref{ssec:entropy}.
	The formula is very similar to the entropy formula for ordinary AdS$_5$ black holes, up to a reduction by an overall factor $\frac{1}{m}$.
	We will give an interpretation of this reduction on both the field theory and the gravity sides.

	\paragraph{Field theory interpretation:}
	
	Recall from Section~\ref{ssec:gen-cardy-index} that the generalized Cardy limit of the index~\eqref{eq:gen-cardy-index-or-var} results in the finer periodicity~\eqref{newPeriodicity}. 
	This new periodicity is not satisfied by the full superconformal index, as can for example be seen from the trace definition of the index~\eqref{eq:trace-defn-index}, which we repeat in a slightly rewritten form:
	\begin{equation}\label{eq:trace-defn-index-3}
		I_N=\mathrm{tr}_{\mathcal{H}_{\text{BPS}}}e^{2\pi i \sigma J_1}e^{2\pi i \tau J_2}e^{2\pi i \phi_1 Q_1}e^{2\pi i \phi_2 Q_2}e^{2\pi i \phi_3 Q_3}\,.
	\end{equation}
	Here, $\mathcal{H}_{\text{BPS}}$ denotes the Hilbert space of BPS states on which the index~\eqref{eq:trace-defn-index-2} localizes, and we have used that $F=2Q_3$ such that now for $\phi_3=\tau+\sigma-\phi_1-\phi_2+2k+1$, $k\in\mathbb{Z}$ the supercharge anticommutes with the operator in the trace.
	As discussed in Section~\ref{ssec:review-index}, this index is only periodic under $\phi_{1,2}\to \phi_{1,2}+1$.
	Moreover, it depends on non-scaled variables $\phi_{1,2}$, $\tau$, and $\sigma$.
	Therefore, the new periodicity observed in~\eqref{eq:gen-cardy-index-or-var} suggests that in generalized Cardy limits~\eqref{eq:gen-cardy-limit} with fixed $m$ but arbitrary $n_{1,2}$, the Hilbert space is effectively projected onto a subsector $\mathcal{H}^m_{\text{BPS}}$ that only consists of states with charges 
	\begin{equation}\label{eq:subsector}
		\mathcal{H}^m_{\text{BPS}}:\qquad J_{i}=m\,J_{i}' \quad \textrm{and}\quad Q_{a}=m\,Q_{a}'\,,
	\end{equation}
	where $J_{i}', Q_{a}'\in \frac{1}{2}\mathbb{Z}$.
	Indeed, in the subsector $\mathcal{H}^m_{\text{BPS}}$, the index can be rewritten as follows:
	\begin{equation}\label{eq:trace-defn-index-4}
		I^{(m)}_N\equiv\mathrm{tr}_{\mathcal{H}^m_{\text{BPS}}}e^{2\pi i \tilde{\sigma} J_1'}e^{2\pi i \tilde{\tau} J_2'}e^{2\pi i \tilde{\phi}_1 Q_1'}e^{2\pi i \tilde{\phi}_2 Q_2'}e^{2\pi i \tilde{\phi}_3 Q_3'}\,,
	\end{equation}
	where we used that for any state:
	\begin{equation}
		e^{2\pi i \left((m n_1-1)(J_1'+Q_3')+(m n_2 -1)(J_2'+Q_3')\right)}=1\,.
	\end{equation}
	Note that~\eqref{eq:trace-defn-index-4} is periodic under the finer periodicity (\ref{newPeriodicity}): $\tilde{\phi}_{1,2}\to \tilde{\phi}_{1,2}+1$.
	Also, in this case we have:
	\begin{equation}
		\tilde{\phi}_3=\tilde{\tau}+\tilde{\sigma}-\tilde{\phi}_1-\tilde{\phi}_2+m\,(2k+1)\,.
	\end{equation}
	The index only depends on the $\mod 2$ value $m(2k+1)$:
	for even $m$, the states are counted with a plus sign while for odd $m$ with a minus sign.
	Taking $m(2k+1)=-1$ for odd $m$ and $m(2k+1)=-2$ for even $m$ reproduces the bracketed values of $\tilde{\phi}_3$ in~\eqref{eq:ev-phi3-D0-D1}.
	To conclude, we observe that if the generalized Cardy limit effectively projects the theory onto the subsector~\eqref{eq:subsector}, a finer periodicity~\eqref{newPeriodicity} emerges from the trace point of view that is consistent with the result~\eqref{eq:gen-cardy-index-or-var}.
	It would be interesting to understand if this projection of the Hilbert space can be related to observations in~\cite{ArabiArdehali:2021nsx}, where it is shown that in the generalized Cardy limit an effective Chern--Simons theory appears on a lens space.
	We will comment on this point more extensively at the end of this section.
	
	The reduced Hilbert space $\mathcal{H}^m_{\text{BPS}}$ helps us to understand the reduction of the entropy.
	In particular, since $\mathcal{H}^m_{\text{BPS}}$ only contains states with $Q_{1,2,3}=m\,Q_{1,2,3}'$, we expect less microstates and therefore a smaller entropy.
	Some intuition for the precise reduction may be obtained by comparing the entropy formula to the Hardy--Ramanujan formula.
	In particular, the Hardy--Ramanujan formula tells us that for large $n$, the number of partitions is given by $p(n)\sim e^{\sqrt{n}}$.
	If we want to compute the number of partitions where the parts are only allowed to multiples of $m$, we would find: $p(n)\sim e^{\sqrt{\frac{n}{m}}}$.\footnote{The number of unrestricted partitions of $n$ is given by the generating function $\sum_n p(n) q^n = \prod_{j=1}^\infty (1-q^j)^{-1}$. The number of partitions where all the parts are multiples of $m$ is given by the generating function $\sum_n p_m(n) q^n = \prod_{j=1}^\infty (1-q^{jm})^{-1}$.}
	
	The entropy formula~\eqref{eq:entropy-a} we derived in Section~\ref{ssec:entropy} is rather similar to the Hardy--Ramanujan formula, as is the entropy formula for AdS$_5$ black holes, except that the dependence on the analogue of $n$, i.e., the charges $Q_i$, is quadratic.\footnote{Since the index can be computed at weak coupling, one indeed expects the counting of states by the index for fixed charges to be related to some version of the counting of partitions.}
	We now see that if we replace the charges $Q_{1,2,3}$ in the ordinary AdS$_5$ black hole entropy formula with the $Q_{1,2,3}'$, as instructed by the analogy with Hardy--Ramanujan, we obtain:
	\begin{equation}
		S=2\pi \sqrt{Q'_1Q'_2+Q'_2Q'_3+Q'_1Q'_3}=\frac{2\pi}{m}\sqrt{Q_1Q_2+Q_2Q_3+Q_1Q_3}\,.
	\end{equation}
	Here, we have omitted the dependence on angular momenta since it is subleading in the generalized Cardy limit, as explained in Section~\ref{ssec:entropy}.
	Thus, we see how this analogy correctly predicts the entropy formula we find.
	
	Let us discuss also a more geometric interpretation of the reduction of the entropy by the factor $\frac{1}{m}$, still on the field theory side.
	For this, we first recall that one can think of the background geometry on which the field theory lives, $S^3\times S^1$, as the primary Hopf surface:
	$\mathbb{C}^2\setminus \lbrace (0,0)\rbrace/\mathbb{Z}$, where $\mathbb{Z}$ acts as:
	\begin{equation}
		(z,z')\sim (pz,qz'),
	\end{equation}
	and $p=e^{2\pi i \sigma}$ and $q=e^{2\pi i\tau}$, with $\mathrm{Im}(\tau)$ and $ \mathrm{Im}(\sigma)>0$.
	From this description it is clear that $\tau$ and $\sigma$ correspond to geometric parameters.
	In particular, we can ask what happens to the geometry in the generalized Cardy limit.
	To understand this, it is useful to write the identifications on $z=e^{2\pi i (\zeta_1+i\zeta_2)}$ and $z'=e^{2\pi i (\zeta'_1+i\zeta'_2)}$ as:
	\begin{align}
		\begin{split}
			\zeta_1&\sim \zeta_1+\mathrm{Re}(\sigma)\,,\quad \zeta_2\sim \zeta_2+\mathrm{Im}(\sigma)\,,\\
			\zeta'_1&\sim \zeta'_1+\mathrm{Re}(\tau)\,,\quad \zeta'_2\sim \zeta'_2+\mathrm{Im}(\tau)\,.
		\end{split}
	\end{align}
	Notice that we can think of the thermal circle as generated by $\zeta_2+ \zeta_2'$, which is non-contractible since $(z,z')=(0,0)$ is excluded.
	Similarly, we can think of the $\zeta_1$ and $\zeta_1'$ as the Euler angles on $S^3$, which contract at $z=0$ and $z'=0$ respectively~\cite{Assel:2014paa}.
	Since the generalized Cardy limit is defined as:
	\begin{equation}\label{eq:GCL2}
		\tau\to n_1-\tfrac{1}{m}\,, \quad \sigma\to n_2-\tfrac{1}{m}\,,
	\end{equation}
	we see that at the geometric level the Euler angles are quotiented by $\frac{1}{m}$ whereas the thermal circle pinches.\footnote{Similar conclusions recently appeared in~\cite{AharonyTalk,ArabiArdehali:2021nsx}.}
	This quotient is exactly the same as what one expects for a lens space $L(m,1)$.
	
	\paragraph{Aside on lens space partition functions:}
	
	Before turning to the gravitational interpretation of the entropy formula, let us point out some analogies and differences between $\mathcal{H}^m_{\text{BPS}}$, which appears in the generalized Cardy limit of the superconformal index, and the Hilbert space of a theory quantized on a lens space $L(m,1)$.
	The lens space $L(m,1)$ can be thought of as $S^3/\mathbb{Z}_m$, where $\mathbb{Z}_m$ quotients the Hopf fiber.
	Supersymmetric partition functions on such spaces were discussed in~\cite{Benini:2013yva,Razamat:2013jxa,Razamat:2013opa}.
	In the following, we collect some facts from~\cite{Razamat:2013opa}.
	
	Let us first consider a free chiral multiplet.
	Its lens space partition function can be obtained by projecting onto states invariant under $\mathbb{Z}_m$ action on a Hopf fiber.
	Since our supercharge $\mathcal{Q}$ has $j_2=0$ (see Section~\ref{ssec:review-index}), this projection can be achieved in the trace by adding a holonomy along the Hopf fiber associated to $j_2$.
	This simply projects onto the class of states with:
	\begin{equation}\label{eq:j2condition}
		j_2=\frac{1}{2}(J_1-J_2)=0\quad \mod m\,,    
	\end{equation} 
	which we will call the untwisted sector.
	On the other hand, there also exist twisted sectors.
	For a theory with only global symmetries, one can project onto a specific twisted sector by also adding a holonomy for the global symmetry along the Hopf fiber.
	In this case, one projects onto states with
	\begin{equation}
		j_2=\pm k \quad \mod m \qquad \textrm{for some}\quad k\in \{0,\ldots m-1\}\,.
	\end{equation}
	In contrast, for a gauge theory one has to sum over all twisted sectors.
	Schematically:
	\begin{equation}
		I_N=\sum_k \oint \mathcal{I}_k\,,
	\end{equation}
	where the sum is over all twisted sectors, the integral is over gauge holonomies along the temporal circle $S^1$, and the integrand $\mathcal{I}_k$ is the usual plethystic exponential of single particle indices but now with a gauge holonomy labelled by $k$ inserted along the Hopf fiber.
	
	We notice that there are some similarities between the lens space partition function and the projected Hilbert space $\mathcal{H}^m_{\text{BPS}}$.
	In particular, we see that the states in $\mathcal{H}^m_{\text{BPS}}$ satisfy the condition~\eqref{eq:j2condition}: $j_2=0\,\mod m$.
	However, there are some important differences:
	\begin{itemize}
		\item The constraint $j_2=0\, \mod m$ only constrains the difference of $J_1$ and $J_2$, while in $\mathcal{H}^m_{\text{BPS}}$ both $J_1$ and $J_2$ are $0\,\mod m$. 
		Moreover, the charges $Q_{1,2,3}$ are similarly constrained in $\mathcal{H}^m_{\text{BPS}}$. 
		Apparently, this cannot directly be understood from the lens space partition function.
		One way in which similar constraints on the $Q_{1,2,3}$ can arise is when there are lens like non-contractible cycles in the transverse space.
		In that case, one could add holonomies for these cycles with charges $Q_1-Q_2$, $Q_2-Q_3$, for which the supercharge $\mathcal{Q}$ has vanishing charge.
		This seems to be realized in the explicit orbifold solutions presented in~\cite{Aharony:2021zkr}.
		\item For a gauge theory such as the $\mathcal{N}=4$ theory, there should also be twisted sectors.
		At least naively, these are not seen in the generalized Cardy limit.
		\textit{A priori}, this may seem surprising, since for example in symmetric orbifolds in the two-dimensional context, it is usually the (maximally) twisted sector which provides the dominant contribution to the entropy.
		It would be very interesting to understand if and how such twisted sectors are encoded in our expression for the (generalized Cardy limit of) the superconformal index. 
		To study this, it may be helpful to compare to~\cite{ArabiArdehali:2021nsx}, where such twisted sectors are found in generalized Cardy limits. 
		A complicating factor in this comparison is the fact that in our approach we perform the gauge integral before analyzing the generalized Cardy limit, whereas they first take the generalized Cardy limit and then compute the gauge integral.
		This different order seems to be crucial in their identification of the twisted sectors.
	\end{itemize}

	\paragraph{Gravitational interpretation:}
	
	We will now present some ideas on how to interpret the entropy formula from the gravitational point of view.
	Again, we recapitulate the entropy formula for convenience:
	\begin{equation}\label{eq:new-entropy}
		S=\frac{2\pi}{m}\sqrt{Q_1Q_2+Q_2Q_3+Q_1Q_3}\,,
	\end{equation}
	where we recall from Section~\ref{ssec:entropy} that the generalized Cardy limit in the microcanonical ensemble reads:
	\begin{equation}\label{eq:Q-J-constraint}
		Q_a=\mathcal{O}(\mu^2)\,,\quad J_i=\mathcal{O}(\mu^3)\,, 
	\end{equation}
	for $\mu\to \infty$.
	Let us first emphasize that this formula is obtained from the index by Legendre transformation with respect to unshifted charges $Q_{1,2,3}$ and $J_{1,2}$.
	In particular, this means that it is obtained for states with the same energy as ordinary AdS$_5$ black holes with charges $Q_{1,2,3}$ and $J_{1,2}$.
	The entropy of the latter can be obtained by setting $m=1$ to find to leading order in the Cardy limit:
	\begin{equation}\label{eq:old-entropy}
		S=2\pi\sqrt{Q_1Q_2+Q_2Q_3+Q_1Q_3}\,.
	\end{equation}
	As already hinted at in the field theory interpretation, the reduction of the entropy by a factor $\frac{1}{m}$ could be interpreted in terms of a black hole with $L(m,1)$ horizon topology.\footnote{Another possible gravitational interpretation can be elicited from the giant graviton configuration appearing in the near extremal static black holes~\cite{Balasubramanian:2007bs}. In the absence of one $R$-charge, the central charge of the black hole is proportional to the intersection number between two species of giant gravitons. The form of the entropy in~\eqref{eq:old-entropy} indicates that this could potentially be described by intersections between unknown gravitational configurations. The $1/m$ factor can then be understood as reducing the degrees of freedom in each gravitational species by  $m$, thus lowering the intersections between these configurations by $1/m^2$, which results in the $1/m$ factor in the entropy.}
	Indeed, the generalized Cardy limit (\ref{eq:GCL2}) induces a $\mathbb{Z}_m$ quotient of $S^3$, which can be extended into the bulk. 
	However, we are not aware of any explicitly known solutions for black lenses with AdS$_5$ asymptotics, and therefore cannot compare the entropy formula with the actual horizon area of such a solution.\footnote{
		Solutions like this were announced as orbifold solutions in~\cite{AharonyTalk} and appeared in~\cite{Aharony:2021zkr} after v1 of the present paper appeared. Also, similar to our geometric analysis above, the asymptotics of such solutions were recently discussed in~\cite{ArabiArdehali:2021nsx}.}

	\section{Discussion}
	\label{sec:disc}

	In this work, we have derived a three-integer family of modular properties for the elliptic $\Gamma$ function.
	We applied these properties to the study of generalized Cardy-like limits of the superconformal index of the $\mathcal{N}=4$ super-Yang--Mills theory.
	A more detailed summary of our main results is given at the end of Section~\ref{sec:intro}.
	
	The elliptic $\Gamma$ function can be thought of as the superconformal index of a free chiral multiplet.
	The reason that we can use its modular properties to study the $\mathcal{N}=4$ theory is  that the index of the $\mathcal{N}=4$ theory consists of a collection of elliptic $\Gamma$ functions and $\theta$ functions, and only the $\Gamma$ functions are relevant to leading order in the generalized Cardy limits.
	However, on general grounds one expects that the \emph{normalized} superconformal indices of arbitrary $\mathcal{N}=1$ SCFTs, including the $\mathcal{N}=4$ theory, by themselves obey similar modular properties~\cite{Gadde:2020bov}. 
	In this section, we will discuss how these modular properties could constrain the ordinary superconformal index and whether they have a natural interpretation on the gravitational side, specifically in terms of the gravitational path integral.
	
	Let us start by recalling how the $SL(2,\mathbb{Z})$ modularity in AdS$_3$/CFT$_2$ is reflected on the gravitational side.
	In the two-dimensional CFT, the superconformal index $Z(z_{1,\ldots,r};\tau)$ (a.k.a.\ the elliptic genus) is invariant, up to a phase, under arbitrary transformations in the Jacobi group $J=SL(2,\mathbb{Z})\ltimes \mathbb{Z}^{2r}$:
	\begin{equation}\label{eq:sl2-reln-part-funct-disc}
		Z\left(z_i;\tau\right)=e^{i\phi_{g}\left(z_i;\tau\right)}
		Z\left(g^{-1}(z_i;\tau)\right), \quad g\in J \,.
	\end{equation}
	This property is crucial in defining a Farey tail transform of the elliptic genus~\cite{Dijkgraaf:2000fq,deBoer:2006vg,Manschot:2007ha}, which can be interpreted naturally in terms of the dual gravitational path integral.
	In particular, the expression for the elliptic genus obtained in~\cite{Manschot:2007ha} reflects the covariance~\eqref{eq:sl2-reln-part-funct-disc} by being written as an average over (part of) the modular group $SL(2,\mathbb{Z})$.
	The gravitational interpretation of this average is in terms of a sum over gravitational saddle points, corresponding to the $SL(2,\mathbb{Z})$ family of Euclidean BTZ black holes~\cite{Maldacena:1998bw}.
	From the gravitational path integral point of view, this family of saddles arises due to the fact that gravity can fill in any cycle of the boundary torus.
	
	We now want to ask whether the modular properties of four-dimensional normalized partition functions, discussed in~\cite{Gadde:2020bov} and this work, similarly have a natural interpretation on the gravitational side.
	More precisely: do the modular properties imply a covariance property like~\eqref{eq:sl2-reln-part-funct-disc} for the four-dimensional superconformal index?
	And could such a property be used to argue for the existence of an averaged expression for the index, which can subsequently be interpreted as a sum over saddles in the gravitational theory?
	
	Let us start by noting that our work has shown that, using modular properties of the elliptic $\Gamma$ functions, one can compute generalized Cardy limits for the full superconformal index.
	The resulting expressions match precisely with the on-shell actions of the supersymmetric AdS$_5$ black hole solution and certain orbifolds thereof.\footnote{
		Our prediction for the on-shell action is the same as obtained for the $(m,n)$ saddles in~\cite{Cabo-Bizet:2019eaf} --- see also the more recent discussion in~\cite{Cabo-Bizet:2020nkr,ArabiArdehali:2021nsx,Aharony:2021zkr}.}
	The modular properties of the elliptic $\Gamma$ function are generally labeled by \emph{relations} in $G=SL(3,\mathbb{Z})\ltimes \mathbb{Z}^{3r}$, in this work specifically a three-integer parameter family of order three elements $X_{\textbf{m}}^3=1$.
	Thus, the associated gravitational saddles can be labeled similarly. 
	This generalizes the observation of Gadde in~\cite{Gadde:2020bov} and our previous work~\cite{Goldstein:2020yvj}, which associates the order three element $Y\in SL(3,\mathbb{Z})$ to the ordinary AdS$_5$ black hole.
	This is somewhat reminiscent of AdS$_3$/CFT$_2$, although there Euclidean gravitational saddles are labeled by $SL(2,\mathbb{Z})$ \emph{elements}.
	However, it is not straightforward to push this analogy further, for the following conceptual and technical reasons.
	
	First of all, in AdS$_5$/CFT$_4$ either the $S^3$ or the $S^1$ factor from the boundary geometry can contract in the bulk, corresponding to thermal AdS or an AdS black hole, respectively~\cite{Witten:1998zw}.
	However, unlike the case for a two-dimensional boundary, there seems to be no natural combination of these cycles which can contract.
	The fact that the large diffeomorphism group of $S^3\times S^1$ is presumably small (or trivial) is a reflection of this fact.
	Indeed, the relevant modular group in four dimensions is the group of large diffeomorphisms of a three torus, which arises in a Heegaard-like decomposition of a class of four manifolds (see Section~\ref{ssec:sl3}).
	As such, it does not correspond to the large diffeomorphism group of the full four manifold, such as $S^3\times S^1$. 
	It is precisely for this reason that the four-dimensional (normalized) index is a rather different modular object than the two-dimensional elliptic genus.
	
	This brings us to two more technical issues.
	Firstly, it is really the ordinary, i.e., unnormalized, partition function $Z_g$ that features in the AdS/CFT correspondence.
	It is not immediately obvious whether modular properties of normalized partition functions have useful implications for the former.
	Secondly, the fact that the normalized partition functions are rather different modular objects than ordinary automorphic forms will in particular complicate the interpretation of their modular properties in terms of a covariance such as~\eqref{eq:sl2-reln-part-funct-disc}.
	Such a covariance is crucial to argue for an averaged expression of the index over a modular group, which could have a natural interpretation in terms of a gravitational path integral.
	
	In the following, we will first examine this second point in Section~\ref{ssec:mod-norm-Z}.
	In particular, we will discuss in detail how the modular property of normalized partition functions differs from the two-dimensional case, and how one can still interpret it as a type of covariance.
	However, it seems somewhat awkward to give this covariance property a gravitational interpretation.
	We then return to the first point in Section~\ref{ssec:mod-Z}, where we show that modularity of the normalized superconformal index still implies an interesting covariance of the ordinary superconformal index.
	This covariance turns out to allow for a more natural gravitational interpretation, which we will discuss.

	\subsection{Modularity of the normalized index}\label{ssec:mod-norm-Z}
	
	We recall from Sections~\ref{ssec:review-Gadde} and~\ref{ssec:sl3} that the modular property of $\hat{Z}^{a}_g$ under $G=SL(3,\mathbb{Z})\ltimes \mathbb{Z}^{3r}$ is as follows:
	\begin{equation}\label{eq:sl3-reln-part-funct-disc}
		\begin{aligned}
			\hat{Z}^{a}_{g_1 \cdot g_2}\left(\boldsymbol{\rho}\right)&=e^{i\phi_{g_1,g_2}\left(\boldsymbol{\rho}\right)}\hat{Z}^{a}_{g_1}\left(\boldsymbol{\rho}\right)\hat{Z}^{a}_{g_2}\left(g^{-1}_1\boldsymbol{\rho}\right) \,, \quad g_{1,2}\in G  \,.
		\end{aligned}
	\end{equation}
	In mathematical terms, $\hat{Z}^{a}_{g}$ is an element of the first group cohomology $H^{1}(G,N/M)$, as opposed to the elliptic genus which can be thought of as an element of $H^{0}(J,N/M)$~\cite{Gadde:2020bov}. 
	This property differs crucially from the two-dimensional case in that it involves three partition functions which are generically defined on non-diffeomorphic manifolds.
	Therefore, it cannot be interpreted straightforwardly as a covariance of a partition function defined on a single manifold.
	In addition, the property is labeled by a \emph{relation} in the modular group $G$, in this case:
	\begin{equation}\label{eq:relation}
		r=(g_1\cdot g_2)\cdot (g_2)^{-1}\cdot(g_1)^{-1}=1\,.
	\end{equation}
	This is opposed to the modular property for the elliptic genus, which is labeled by an \emph{element} in the modular group $J$.
	Notice that~\eqref{eq:sl3-reln-part-funct-disc} can be viewed as a factorization property of a normalized partition function, up to a phase, on some manifold $M_{g_1\cdot g_2}$ in terms of partition functions on $M_{g_1}$ and $M_{g_2}$.
	
	There is a close relation between the modular property~\eqref{eq:sl3-reln-part-funct-disc} and the holomorphic block factorization~\cite{Peelaers:2014ima,Yoshida:2014qwa,Nieri:2015yia}.
	Holomorphic blocks can be viewed as partition functions on the solid torus $D_2\times T^2$.
	An ordinary supersymmetric partition function on $M_g$ can sometimes\footnote{As discussed in~\cite{Gadde:2020bov}, holomorphic blocks provide a local trivialization of the \emph{normalized} partition function. Since the normalized partition functions are non-trivial elements of $H^{1}(G,N/M)$, holomorphic block factorization does not hold for all generators of $G$. This implies that also ordinary partition functions cannot be factorized in blocks for all generators.} be factorized into two such holomorphic blocks $\mathcal{B}^{a}_L(\boldsymbol{\rho})$ and $\mathcal{B}^{a}_R(g^{-1}\boldsymbol{\rho})$:
	\begin{equation}\label{eq:Zmg-hol-blocks}
		Z[M_g](\boldsymbol{\rho})\simeq \sum_a\mathcal{B}^{a}_L(\hat{\boldsymbol{\rho}})\,\mathcal{B}^{a}_R(g^{-1}\hat{\boldsymbol{\rho}}),
	\end{equation}
	where the sum runs over all Higgs branch vacua $|a\rangle $ of the mass deformed theory, the equality should be understood modulo a phase, and $\mathcal{B}^{a}_R(z;\tau,\sigma)=\mathcal{B}^{a}_L(z;-\tau,\sigma)$ are related by orientation reversal.
	The factorization is a reflection of the Heegaard-like decomposition of $M_g$, and the fact that the associated supersymmetric partition function only depends on the complex structures of $M_g$~\cite{Cecotti:2013mba}.
	
\begin{figure}
		\centering
		\includegraphics[width=.75\textwidth]{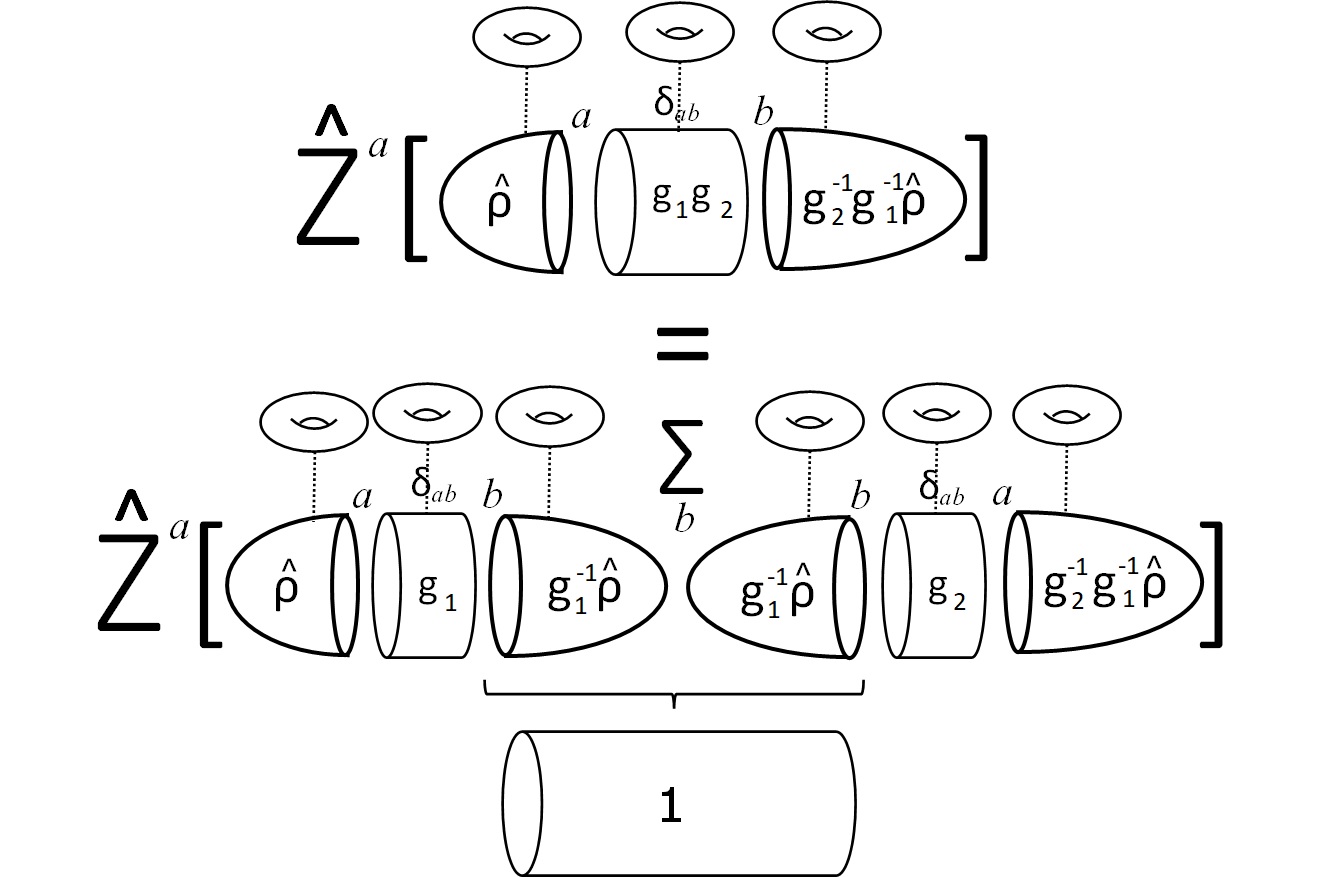}
		\caption{This figure depicts the covariance of the normalized partition function $\hat{Z}^{a}_{(\cdot)}$ under the split $M_{A}\to M_{A^{-1}}\cup M_{A^{-1}}$. Summation over the $a,b$ indices is \emph{not} implied, i.e., $\hat{Z}^a$ only corresponds to a single (normalized) term in~\eqref{eq:Zmg-hol-blocks} (see Section~\ref{ssec:review-Gadde}). In the \emph{top part}, the manifold $M_{A}$ with complex structure moduli $\boldsymbol{\rho}$ is represented as two solid three-tori with moduli $\hat{\boldsymbol{\rho}}$ and $A^{-1}\hat{\boldsymbol{\rho}}$, which are glued by the element $A$ (and orientation reversal).
			The hatted versus unhatted notation, introduced in Section~\ref{ssec:sl3}, serves to indicate the distinction between the $T^3$ moduli and the $M_A$ moduli. The relation between these moduli will be made explicit in Section~\ref{ssec:mod-Z}.
			In the \emph{lower part}, the splitting of $M_{A}\to M_{A^{-1}}\cup M_{A^{-1}}$ is illustrated by stretching the cylinder and factorizing it into two solid three-tori through a resolution of the identity: $A=A^{-1} 1 A^{-1}$~\cite{Gadde:2020bov}. The equality denoted should be understood $\mod M$. }
		\label{fig:split}
	\end{figure}

	Holomorphic block factorization underlies the derivation of the modular property for normalized partition functions, modulo the phase~\cite{Gadde:2020bov}.
	In Figure~\ref{fig:split}, we illustrate the derivation for an order three element $A^3=1$, such that $g_1=g_2=A^{-1}$ and $g_1g_2=A$.
	This figure serves to illustrate the geometric interpretation of~\eqref{eq:sl3-reln-part-funct-disc}: normalized partition functions $\hat{Z}^{a}_{(\cdot)}$ are covariant under the \emph{splitting} of a manifold $M_{g_1g_2}$ into the disjoint union $M_{g_1}\cup M_{g_2}$, with an appropriate mapping of the moduli.
	The non-trivial phase is a reflection of non-trivial Berry curvature on the space of parameters~\cite{Gadde:2020bov}, to which the partition functions are sensitive.
	Indeed, the property~\eqref{eq:sl3-reln-part-funct-disc} can be viewed as describing the Berry phase associated to the loop:
	\begin{equation}
		\boldsymbol{\rho}\to g_{1}^{-1}\boldsymbol{\rho}\to g_2^{-1}g_{1}^{-1}\boldsymbol{\rho}\to g_1g_2g_2^{-1}g_{1}^{-1}\boldsymbol{\rho}=\boldsymbol{\rho}\,.
	\end{equation}
	Notice that this loop is precisely captured by the relation~\eqref{eq:relation}.
	
	Admittedly, the above describes a rather different type of covariance than the one appearing in two dimensions, where the partition function on a single manifold is covariant under \emph{large diffeomorphisms} of the geometry.
	In particular, it seems awkward to interpret the splitting of the (boundary) manifold on the AdS side.
	However, here we recall that we are describing properties of normalized partition functions, which are not directly related to the gravitational description.
	Instead, we should understand the implications of modularity for the ordinary superconformal index.
	We will see in Section~\ref{ssec:mod-Z} that this understanding will lead to a more natural interpretation on the gravitational side.
	Before turning to this, let us analyze the modular properties of normalized partition functions in more detail.
	
	The property~\eqref{eq:sl3-reln-part-funct-disc} may seem closer related to the two-dimensional case when $g_1$ is an element of the subgroup $H=SL(2,\mathbb{Z})\ltimes \mathbb{Z}^{2+2r}\subset G$ (see Sections~\ref{ssec:sl3} and~\ref{ssec:partn-fn-sl3}).
	The partition functions evaluate trivially ($ \mod M$) on elements of $H$, such that~\eqref{eq:sl3-reln-part-funct-disc} becomes:
	\begin{equation}\label{eq:H-reln-part-funct-disc}
		\begin{aligned}
			\hat{Z}^a_{h \cdot g}\left(\boldsymbol{\rho}\right)&=e^{i\phi_{h,g}\left(\boldsymbol{\rho}\right)}\hat{Z}^a_{g}\left(h^{-1}\boldsymbol{\rho}\right) \,, \quad h\in H, \quad g\in G  \,.
		\end{aligned}
	\end{equation}
	We notice that this equation holds for the ordinary partition function as well, since the normalization factors on both sides are equal and therefore can be canceled.
	Even though this equation superficially resembles~\eqref{eq:sl2-reln-part-funct-disc}, an interpretation in terms of a similar covariance does not seem to go through.\footnote{This is a correction of an opposing statement in v1 of this paper.}
	To make the point, let us take for concreteness $g=S_{23}$ such that $M_{g}\cong S^3\times S^1$.
	The equation then states that the partition function on the manifold $M_{h\cdot S_{23}}$ with moduli $\boldsymbol{\rho}$ may be equivalently computed on the manifold $M_{S_{23}}$ with moduli $h^{-1}\boldsymbol{\rho}$ (up to a phase).
	However, even though these manifolds are both diffeomorphic to $S^3\times S^1$ for any $h\in H$, it seems that their moduli cannot be related by a large diffeomorphism.
	Indeed, the large diffeomorphism group of $S^3\times S^1$ is expected to be small or trivial, whereas $h\in H$ generate in particular a full $SL(2,\mathbb{Z})$. 
	Therefore, the left and right hand side should be treated as inequivalent partition functions, even though they are related computationally.\footnote{This fact will still turn out very useful in a different context to be described below.}
	For similar reasons, the partition functions $\hat{Z}_{h\cdot g}$ for distinct $h$ describe inequivalent partition functions.
	All of this is to say that~\eqref{eq:H-reln-part-funct-disc} cannot be interpreted like the covariance~\eqref{eq:sl2-reln-part-funct-disc}, and consequently cannot be used to argue for an expression of the (normalized) index in terms of some average over $H$.
	Indeed, viewed in the context of AdS/CFT, the partition functions would correspond to distinct boundary conditions on the AdS side.
	Therefore, also from the AdS side an averaged expression over $H$ is not expected since the gravitational path integral should sum only over geometries with the same boundary conditions.
	
	To end this section, we will describe a concrete family of examples of~\eqref{eq:sl3-reln-part-funct-disc}, which will help us to address the ordinary superconformal index in Section~\ref{ssec:mod-Z}.
	This is the three-integer family of order three elements $X_{\textbf{m}}$ studied in Section~\ref{ssec:finite-order-elms}, which we will denote by $A$ to avoid clutter.
	As described in Section~\ref{ssec:new-mod-prop}, applying the main modular property~\eqref{eq:sl3-reln-part-funct-disc} to the specific relation $A^3=1$ leads to:
	\begin{equation}\label{eq:gen-mod-prop-Xm-disc}
		\hat{Z}^a_{A}\left(\boldsymbol{\rho}\right)\hat{Z}^a_{A}\left(A^{-1}\boldsymbol{\rho}\right)\hat{Z}^a_{A}\left(A^{-2}\boldsymbol{\rho}\right)=e^{i\pi P_{\textbf{m}}\left(\boldsymbol{\rho}\right)},
	\end{equation}
	where $P_{\textbf{m}}$ is related to the anomaly polynomial of the theory.
	The advantages of evaluating on relations associated to order three elements are as follows.
	First of all, we see that this equation involves partition functions which are all defined on $M_A$, even though the distinct moduli mean that these partition functions are still inequivalent for the reasons discussed above.
	Furthermore, for the elements $X_{\textbf{m}}$ we can compute the Berry phase explicitly and, as already mentioned at the beginning of this section, these phases correspond to the on-shell actions of Euclidean gravitational solutions.
	In this case, the Berry phase is associated to the loop:
	\begin{equation}
		\boldsymbol{\rho}\to A\,\boldsymbol{\rho}\to \, A^2 \boldsymbol{\rho}\to  \boldsymbol{\rho}\,.
	\end{equation}
	To make the covariance manifest, we find it useful to rewrite~\eqref{eq:gen-mod-prop-Xm-disc}, using properties of $\hat{Z}^a_g$ described in Section~\ref{ssec:partn-fn-sl3}, as follows:
	\begin{equation}\label{eq:gen-mod-prop-Xm-disc-2}
		\frac{\hat{Z}^a_{A}\left(\boldsymbol{\rho}\right)}{\hat{Z}^a_{A^{-1}}\left(\boldsymbol{\rho}\right)\hat{Z}^a_{A^{-1}}\left(A\boldsymbol{\rho}\right)}=e^{i\pi P_{\textbf{m}}\left(\boldsymbol{\rho}\right)} \,.
	\end{equation}
	On the left hand side the numerator is the normalized partition function on the manifold $M_{A}$.
	Due to the order three property, we have $M_{A}\cong M_{A^{-2}}$.
	The denominator, then, reflects the factorization of the normalized partition function on $M_{A^{-2}}$, being the partition function on the disjoint union $M_{A^{-1}}\cup M_{A^{-1}}$.
	This equation was already illustrated in Figure~\ref{fig:split}.
	
	The property~\eqref{eq:gen-mod-prop-Xm-disc-2} may seem specific to the normalized partition function on the manifold $M_A$.
	However, we now show that it implies a family of covariances for the normalized superconformal index, i.e., $\hat{Z}^a_{S_{23}}$, by making use of~\eqref{eq:H-reln-part-funct-disc}.
	We first recall the explicit form of the order three elements $X_{\textbf{m}}=h\cdot S_{23}$ with $h\in H$ constrained such that $X_{\textbf{m}}^3=1$.
	Since $\hat{Z}_{h}=1$ for any $h\in H$, we find a useful rewriting of~\eqref{eq:gen-mod-prop-Xm-disc-2}:
	\begin{equation}\label{eq:gen-mod-prop-Xm-disc-3}
		\frac{\hat{Z}^a_{S_{23}}\left(S_{23}\boldsymbol{\rho}\right)}{\hat{Z}^a_{S_{23}^{-1}}\left(X_{\textbf{m}}\boldsymbol{\rho}\right)\,\hat{Z}^a_{S_{23}^{-1}}\left(X_{\textbf{m}}^{2}\boldsymbol{\rho}\right)}=e^{i\pi P_{\textbf{m}}\left(\boldsymbol{\rho}\right)} \,,
	\end{equation}
	where we redefined $\boldsymbol{\rho}\to X_{\textbf{m}}\boldsymbol{\rho}$ and used $P_{\textbf{m}}\left(X_{\textbf{m}}\boldsymbol{\rho}\right)=P_{\textbf{m}}\left(\boldsymbol{\rho}\right)$.
	This is the form of the modular property used in the main text~\eqref{eq:deriv-mod-prop-Xabc-2}.
	In particular, it is expressed purely in terms of normalized superconformal indices, i.e., partition functions on $M_{S_{23}}$.
	A disadvantage of this representation is that the geometric picture of Figure~\ref{fig:split} is somewhat obscured, which is due to the fact that we evaluated $\hat{Z}^a_{h}=1$.
	However, the key point is that the covariance of $\hat{Z}^a_{(\cdot)}$ as expressed in~\eqref{eq:gen-mod-prop-Xm-disc-2} can be understood as a family of covariances of the normalized superconformal index under the splitting $M_{S_{23}}\to M_{S_{23}^{-1}}\cup M_{S_{23}^{-1}}$ for \emph{any} $h\in H$ such that $X_{\textbf{m}}^3=1$.

	\subsection{A covariance property of the ordinary index}\label{ssec:mod-Z}
	
	As mentioned in the previous section, it is important to understand what~\eqref{eq:gen-mod-prop-Xm-disc-3} implies for the ordinary, i.e., unnormalized, superconformal index, since this would equal a gravitational path integral on the AdS side.
	To see this, we first use that the product of normalized superconformal indices appearing in~\eqref{eq:gen-mod-prop-Xm-disc-3} can be expressed in terms of a product of the perturbative part of ordinary superconformal indices:
	\begin{equation}\label{eq:normZ-to-pertZ}
		\hat{Z}^a_{S_{23}^{-1}}\left(\boldsymbol{\rho}\right) \hat{Z}^a_{S_{23}^{-1}}\left(X_{\textbf{m}}\boldsymbol{\rho}\right)\,\hat{Z}^a_{S_{23}^{-1}}\left(X_{\textbf{m}}^{2}\boldsymbol{\rho}\right)=Z_{\mathrm{P}}^{(a_i)}\left(\boldsymbol{\rho}\right)Z_{\mathrm{P}}^{(a_i)}\left(X_{\textbf{m}}\boldsymbol{\rho}\right)\,Z_{\mathrm{P}}^{(a_i)}\left(X_{\textbf{m}}^{2}\boldsymbol{\rho}\right)\,.
	\end{equation}
	Here, $Z_{\text{P}}^{(a_i)}$ is defined in~\eqref{eq:suN-index-pert-anomaly} (see also Section~\ref{ssec:review-index}).
	This identity was proven in~\cite{Gadde:2020bov} for $X_{\textbf{m}}=Y$ in the context of SQED, but a completely analogous proof applies to the $\mathcal{N}=4$ theory for general $X_{\textbf{m}}$.
	The proof uses that the vortex partition function  $Z^{(a_i),(k_i)}_{\mathrm{V}}(\phi_{a},\tau;\sigma)$, defined in~\eqref{eq:vortex-part-suN}, is invariant under $H$.
	In particular, one may easily verify invariance under $T_{31}$ and $S_{13}$, generating $SL(2,\mathbb{Z})\subset H$, using modular properties of $\theta(z;\sigma)$ and the fact that $f_{3}=pq(f_1f_2)^{-1}$.
	Furthermore, invariance under $T_{21}$ is trivial, while invariance under $T_{23}:\tau\to \tau +\sigma$ is shown by first solving for $f_3=pq(f_1f_2)^{-1}$ in the expression for $Z_{\mathrm{V}}$, and only then transform $q\to pq$.
	
	We now use~\eqref{eq:normZ-to-pertZ} to rewrite~\eqref{eq:gen-mod-prop-Xm-disc-3} once more:
	\begin{equation}\label{eq:gen-mod-prop-Xm-disc-4}
		Z_{\mathrm{P}}^{(a_i)}\left(\boldsymbol{\rho}\right)Z^{(a_i)}_{\mathrm{P}}\left(X_{\textbf{m}}\boldsymbol{\rho}\right)\,Z_{\mathrm{P}}^{(a_i)}\left(X_{\textbf{m}}^{2}\boldsymbol{\rho}\right)=e^{-i\pi P_{\textbf{m}}\left(\boldsymbol{\rho}\right)} \,.
	\end{equation}
	Note that the phase is independent of $(a_i)$, and indeed one can explicitly show that the product on the left hand side is independent of $(a_i)$ as well (see Section~\ref{ssec:anom-pol-index}).
	One might worry that this equation is ill-defined due to the fact that the explicit expression for $Z_{\text{P}}^{(a_i)}$ given in~\eqref{eq:suN-index-pert-anomaly} is vanishing.
	In the full index of the $\mathcal{N}=4$ theory, the simple zeros of $Z_{\text{P}}^{(a_i)}$ cancel against simple poles of $Z^{(a_i)}_{\mathrm{V}}$, making the whole index non-vanishing and well-defined (see the third comment below~\eqref{eq:suN-index-final}).
	The reason to not cancel these parts outright is to facilitate the proof of invariance of $Z^{(a_i)}_{\mathrm{V}}$ under $H$.
	Cancelling the poles and zeros first gives the same result, of course.
	
	Since the vortex partition functions cancel among themselves in~\eqref{eq:normZ-to-pertZ}, including the poles, we should check that the zeros of the $Z_{\text{P}}^{(a_i)}$ factors cancel from this equation as well. 
	And indeed, one can also show this by plugging in the explicit expression for $Z_{\text{P}}^{(a_i)}$ while ensuring that each of the $\Gamma$ functions has a convergent product expansion (see, e.g., the comments around~\eqref{eq:constraint-mod-prop}).
	Keeping in mind that~\eqref{eq:gen-mod-prop-Xm-disc-4} is to be read stripped from the simple zeros, we note that it is well-defined also in the case of the $\mathcal{N}=4$ theory.
	
	It should not be surprising that $Z^{(a_i)}_{\mathrm{P}}$ obeys a similar property as the normalized partition functions.
	Indeed, it consists of a product of elliptic $\Gamma$ functions, each of which is understood as a non-trivial element in $H^{1}(G,N/M)$.
	In fact,~\eqref{eq:gen-mod-prop-Xm-disc-4} is a direct consequence of the main modular property of the elliptic $\Gamma$ function, derived in Section~\ref{ssec:new-mod-prop}:
	\begin{equation}\label{eq:mod-prop-gamma-a-disc}
		\Gamma(z;\tau,\sigma) =e^{-i\pi Q'_{\mathbf{m}}\left(mz;\tau,\sigma\right)}\Gamma\left(\tfrac{z}{m\tau+n};\tfrac{\sigma-\tau}{m\tau+n},\tfrac{\tau-n_1}{m\tau+n}\right)\Gamma\left(\tfrac{z}{m\sigma+n};\tfrac{\tau-\sigma}{m\sigma+n},\tfrac{\sigma-n_1}{m\sigma+n}\right)\,,
	\end{equation}
	where we have specialized $n_1=n_2$ and written $n\equiv 1-mn_1$.
	Applying this equation to each elliptic $\Gamma$ function in $Z_{\text{P}}^{(a_i)}$ gives~\eqref{eq:gen-mod-prop-Xm-disc-4}, as we explicitly showed in Section~\ref{ssec:anom-pol-index}.
	
\begin{figure}
		\centering
		\includegraphics[width=1\textwidth]{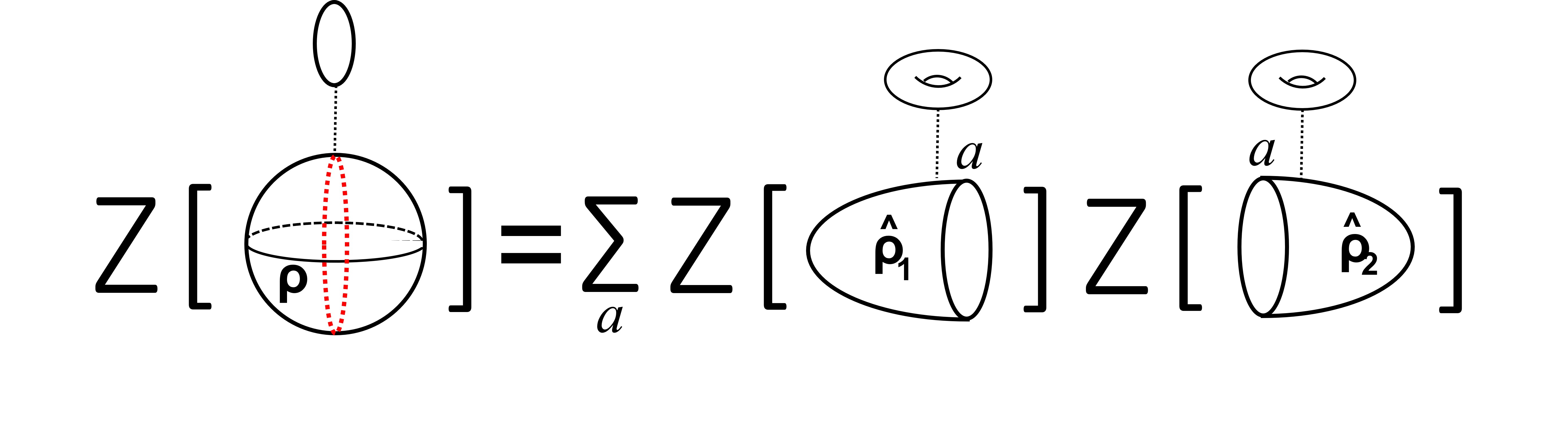}
		\caption{On the left side, a supersymmetric partition function on $S^3\times S^1$ geometry with moduli $\boldsymbol{\rho}$ is depicted. Up to a phase, the partition function on this geometry can be written as a sum of a product of two solid tori partition functions, i.e., holomorphic blocks, with moduli $\hat{\boldsymbol{\rho}}_1$ and $\hat{\boldsymbol{\rho}}_2$ defined in~\eqref{eq:rho1rho2}.	The red dotted line indicates how we may think of the geometry on the left as being split into the geometry on the right.}
		\label{fig:split2}
	\end{figure}

	Now, one could interpret~\eqref{eq:gen-mod-prop-Xm-disc-4} or~\eqref{eq:mod-prop-gamma-a-disc} similarly as in the previous section, in terms of a covariance under the splitting of a closed manifold $M_{S_{23}}$ into two other closed manifolds $M_{S_{23}}\cup M_{S_{23}}$.
	However, inspired by holomorphic block factorization~\cite{Nieri:2015yia}, we propose an alternative interpretation.
	First, let us recall that $\Gamma(z;\tau,\sigma)$ not only describes a supersymmetric partition function of the (anomaly-free) chiral multiplet on $S^3\times S^1$ with complex structure moduli $\tau$ and $\sigma$~\cite{Closset:2013sxa}, but can also be interpreted as the partition function on the solid torus $D_2\times T^2$ where the boundary $T^3$ has moduli $\tau$ and $\sigma$~\cite{Nieri:2015yia,Longhi:2019hdh}.
	As above, we denote these functions by $\mathcal{B}^a_{L,R}$.
	Then, we may read~\eqref{eq:mod-prop-gamma-a-disc} as a factorization, up to a phase, of the $S^3\times S^1$ partition function with moduli $\boldsymbol{\rho}$ into the two holomorphic blocks $\mathcal{B}^a_{L}(\hat{\boldsymbol{\rho}}_1)$ and $\mathcal{B}^a_{R}(\hat{\boldsymbol{\rho}}_2)$ where:
	\begin{align}\label{eq:rho1rho2}
		\begin{split}
			\hat{\boldsymbol{\rho}}_1&=S_{23}^{2}X_{\textbf{m}}\, \boldsymbol{\rho}=\left(\frac{z}{m\tau+n};\frac{\tau-\sigma}{m\tau+n},\frac{\tau-n_1}{m\tau+n}\right)\mod 1\\
			\hat{\boldsymbol{\rho}}_2&=S_{23}^{-1}X_{\textbf{m}}^{2}\,\boldsymbol{\rho}=\left(\frac{z}{m\sigma+n};\frac{\tau-\sigma}{m\sigma+n},\frac{\sigma-n_1}{m\sigma+n}\right)\mod 1\,,
		\end{split}
	\end{align}
	where we recall that $\mathcal{B}^a_{L,R}$ are related by orientation reversal.
	Here, the last entry is associated to the modulus of the non-contractible $T^2\subset D_2\times T^2$ while the middle entry corresponds to the modulus of the $T^2$ which contains the contractible cycle.
	This interpretation is depicted in Figure~\ref{fig:split2}, where for the chiral multiplet we note that the sum would only contain a single term as in~\eqref{eq:mod-prop-gamma-a-disc}.
	The somewhat complicated relation between the complex structure moduli of the solid tori and the $S^3\times S^1$ geometry can be traced to the fact that the gluing element, which identifies $\hat{\boldsymbol{\rho}}_1$ and $\hat{\boldsymbol{\rho}}_2$, involves all three cycles of the boundary $T^3$.
	This is also the case for the original holomorphic blocks~\cite{Nieri:2015yia}. 
	
	Thus, we see that~\eqref{eq:mod-prop-gamma-a-disc} provides a family of holomorphic blocks for the chiral multiplet superconformal index.
	In particular, the original example studied in~\cite{Nieri:2015yia} can be obtained by specializing to $m=n_1=1$ such that $n=0$.
	
	Let us pause here briefly to comment on this interpretation in the case when $\tau=\sigma$, a somewhat subtle limit of the modular property discussed at the end of Section~\ref{ssec:new-mod-prop}.
	As we saw there, the elliptic $\Gamma$ functions on the right hand side of the modular property become singular in this limit.
	Viewing these $\Gamma$ functions as holomorphic blocks, as above, we note that these divergences result geometrically from pinching cycles on the solid tori associated to the blocks.
	This can be seen for example from the complex structure parameters in~\eqref{eq:rho1rho2}, where one may note that one entry of $\hat{\boldsymbol{\rho}}_{1,2}$ obtains a vanishing imaginary part.
	On the other hand, it is clear that the left hand side of the modular property is completely well-defined for $\tau=\sigma$.
	Correspondingly, the $S^3\times S^1$ geometry is also completely regular for $\tau=\sigma$.
	It is therefore not surprising that the divergences cancel and one obtains the modular property associated to $\tau=\sigma$, which we repeat here for convenience:
	\begin{equation}\label{eq:spec-mod-prop-disc}
		\Gamma(z;\tau,\tau)=\frac{e^{-i\pi Q'_{\mathbf{m}}\left(mz;\tau,\tau\right)}}{\theta\left(\frac{z}{m\tau+1-mn};\frac{\tau -n}{m\tau+1-mn}\right)}\prod^{\infty}_{k=0}\left(\frac{\psi^{(m,k+1)}\left(\frac{-z-\frac{(k+1)}{m}}{m\tau+1-mn}\right)}{\psi^{(m,k)}\left(\frac{z-\frac{k}{m}}{m\tau+1-mn}\right)}\right)^m\,.
	\end{equation}
	If one follows the explicit computation (by setting $\sigma=\tau(1+\epsilon)$ and taking $\epsilon\to 0$), one finds that the left holomorphic block in \eqref{eq:mod-prop-gamma-a-disc} cancels entirely against a large part of the holomorphic block on the right. 
	The finite remainder exhibited in \eqref{eq:spec-mod-prop-disc} can thus be ascribed to a single holomorphic block.
	We postpone a detailed interpretation of this to future work.
	For now, let us roughly sketch what we suspect is happening.
	If we think of the $S^3$ as a torus fibration over an interval, we usually have in mind that the solid tori comprising the left and right halve of the interval are glued at the middle of the interval.
	The pinching limit $\tau=\sigma$, instead, should correspond to a limit of this fibration where one solid torus comprises essentially the entire interval, while the other ``solid torus'' sits at the end point.
	Both these solid tori have a pinched cycle in a sense: one has a cycle which blows up, while the other has a cycle which pinches.
	However, it is clear that the $S^3$ has not changed; we just changed our perspective of the torus fibration.
	We expect then that the finite remainder is due to the solid torus with blown up cycle, while a remainder of pinched torus has completely vanished.
	
	We end this section by stating a covariance property for the full, ordinary superconformal index.
	Recall from Section~\ref{ssec:review-index} that we can write the index as:
	\begin{equation}
		I_N=\sum_{(a_i)} Z^{(a_i)}_{\mathrm{P}}(f_a;\tau,\sigma)\, Z^{(a_i)}_{\mathrm{V}}(f_a,\sigma;\tau)\,Z^{(a_i)}_{\mathrm{V}}(f_a,\tau;\sigma).
	\end{equation}
	Here, the sum is over $N-1$ tuples $(a_1,\ldots,a_{N-1})$ with each entry taking three possible values $a_{i}=1,2,3$, and we have written:
	\begin{equation}
		Z^{(a_i)}_{\mathrm{V}}(f_a,\sigma;\tau)\equiv \sum_{(k_i)}Z^{(a_i),(k_i)}_{\mathrm{V}}(f_{a_i},\sigma;\tau)\,.
	\end{equation}
	Now, we plug in the holomorphic blocks for $Z^{(a_i)}_{\mathrm{P}}$, use the fact that the associated phase does not depend on $(a_i)$ and finally use invariance of $Z^{(a_i)}_{\mathrm{V}}$ under $H$.
	This allows us to write:
	\begin{align}
		\begin{split}
			I_N=e^{-i\pi P_{\textbf{m}}(\boldsymbol{\rho}) }\sum_{(a_i)}& \left[Z^{(a_i)}_{\mathrm{P}}\left(\tfrac{\sigma-\tau}{m\tau+n},\tfrac{\tau-n_1}{m\tau+n}\right)Z^{(a_i)}_{\mathrm{V}}\left(\tfrac{\tau-n_1}{m\tau+n}\right)\right]\\
			&\times \left[ Z^{(a_i)}_{\mathrm{P}}\left(\tfrac{\tau-\sigma}{m\sigma+n},\tfrac{\sigma-n_1}{m\sigma+n}\right) Z^{(a_i)}_{\mathrm{V}}\left(\tfrac{\sigma-n_1}{m\sigma+n}\right)\right]\,,
		\end{split}
	\end{align}
	where we have bracketed the two holomorphic blocks and suppressed the dependence on $f_a$.
	The total phase was computed explicitly in Section~\ref{ssec:anom-pol-index} and is given by:
	\begin{equation}\label{eq:anomaly-pol-abc-disc}
		P_{\textbf{m}}(\boldsymbol{\rho})\equiv Q_{\mathrm{tot}}(\phi_{a};\tau,\sigma)=\frac{(N^2-1)}{m}\left(\frac{\tilde{\phi}_1\tilde{\phi}_2\tilde{\phi}_3}{(m\tau+n)(m\sigma+n)}+\frac{m^2-1}{6}\right).
	\end{equation}
	Here, the $\tilde{\phi}_a$ were defined in~\eqref{eq:defn-tilde-phi} and~\eqref{eq:phi3-constraint-anomaly-pol}.
	
	This brings us finally to the four-dimensional analogy with the two-dimensional modular property~\eqref{eq:sl2-reln-part-funct-disc}.
	Instead of covariance under large diffeomorphisms of the background geometry, the covariance property in four dimensions relies on the various ways in which one can factorize the superconformal index.
	These factorizations are labeled (so far) by the order three elements $X_{\textbf{m}}$.\footnote{We note here that there may be more ways in which to factorize the index, and return to this point in Section~\ref{ssec:future}.}
	Similar to two dimensions, the covariance suggests an expression for the index where one averages over all possible factorizations. 
	We imagine the each summand in such an expression contains the associated phase and further corrections.
	As mentioned before, the associated phases correspond to the on-shell actions of Euclidean gravitational solutions.
	Therefore, such an expression has the form of a sum over Euclidean gravitational saddles.
	It will be very interesting to further exploit the covariance to also find subleading corrections, as achieved by the (modern) Farey tail~\cite{Dijkgraaf:2000fq,Manschot:2007ha}.
	Perhaps one should look separately at a Farey tail-like expression for $Z_{\mathrm{P}}$ and an ordinary one for the $Z_{\mathrm{V}}$, since the latter are described by ordinary automorphic forms.
	
	The covariance described above is also more appealing for the gravitational interpretation, since the factorization into holomorphic blocks can still be viewed as describing the partition function on a single closed boundary geometry, in this case $S^3\times S^1$.
	A concrete proposal for a gravitational interpretation of this factorization was presented in~\cite{Hosseini:2019iad}, where the holomorphic blocks of~\cite{Nieri:2015yia} were interpreted in AdS and were dubbed \emph{gravitational blocks}.
	The geometric interpretation of the gravitational blocks is analogous to holomorphic blocks (see Figure~\ref{fig:split2}).
	In particular, the blocks are defined on the north and south pole of the $S^3$ in the near-horizon geometry of the black hole solution respectively.
	The blocks are then glued with a specific identification on the attractor values of the supergravity fields.
	Our work suggests that the Euclidean saddles which generalize the black hole, discussed already in~\cite{Cabo-Bizet:2019eaf,Cabo-Bizet:2020nkr,ArabiArdehali:2021nsx} and now explicitly constructed in~\cite{Aharony:2021zkr}, can be thought of as a more general gluing of the gravitational blocks.
	We will return to a more explicit description in future work.
	
	\subsection{Future directions}\label{ssec:future}
	
	Let us briefly conclude the previous two sections, and then sketch some future directions.
	In Section~\ref{ssec:mod-norm-Z}, we have discussed the strict interpretation of modularity in four dimensions as a covariance property of \emph{normalized} partition functions under the splitting of a closed manifold $M_{g_1g_2}$ into two closed manifolds $M_{g_1}\cup M_{g_2}$.
	However interesting, this property is quite distinct from the automorphic property in two dimensions.
	Moreover, normalized partition functions represent a somewhat artificial object and in particular do not have a clear gravitational interpretation through AdS/CFT.
	We then observed in Section~\ref{ssec:mod-Z} that for specific relations in the group, corresponding to a family of order three elements, the modular property of the normalized partition functions implies a similar property for the perturbative part of the ordinary superconformal index.
	We reinterpreted the modular properties as a statement about the existence of a familiy of holomorphic block factorizations of the ordinary superconformal index, including the non-pertubartive vortex part, which generalizes the original holomorphic blocks discussed in~\cite{Nieri:2015yia}.
	Geometrically, this expresses the various ways in which the partition function on $S^3\times S^1$ can be factorized into partition functions on two solid three-tori, where the moduli of the solid three tori are related to the moduli of $S^3\times S^1$ by certain $SL(3,\mathbb{Z})$ relations.
	This type of covariance suggests an averaged expression for the superconformal index that would be the four-dimensional analogue of the (modern) Farey tail expression for the elliptic genus~\cite{Manschot:2007ha}.  
	On the gravitational side, this suggests an interpretation as a more general gluing of the so-called gravitational blocks of~\cite{Hosseini:2019iad}, which would give rise to the family of Euclidean saddles of~\cite{Aharony:2021zkr}, whose existence was already predicted by the superconformal index~\cite{Cabo-Bizet:2019eaf,Cabo-Bizet:2020nkr,ArabiArdehali:2021nsx} and this work.
	
	\newpage
	
	For future work, we note that there are reasons to believe that the family of modular properties described above is incomplete, as already discussed in Section~\ref{ssec:gen-cardy-index}.
	In particular, this family cannot be used to compute the Cardy limits for arbitrary rational values of $\tau$ and $\sigma$.\footnote{Such limits were obtained in~\cite{ArabiArdehali:2021nsx} using different methods. The fact that the resulting expression is very similar to the phases obtained from our modular properties suggests that there should exist more general modular properties which allow the more general Cardy limits.}
	Recall that our modular properties resulted from relations in $SL(3,\mathbb{Z})$ obeying the following criterion: when evaluating the chiral multiplet partition function on the relation, there should not be more than three elliptic $\Gamma$ functions involved.
	This ensures that the limit of the elliptic $\Gamma$ function can be expressed in terms of a simple phase. 
	Moreover, it is also crucial to the interpretation of the modular property in terms of holomorphic blocks.
	The criterion landed us on a specific family of order three elements $X_{\textbf{m}}$.
	However, let us note that there exist other relations in $G$, not corresponding to order three elements, which also result in modular properties obeying the same criterion.
	For example, the basic modular property of the elliptic $\Gamma$ function, associated to the relation $Y^3=1$, can also be derived from the relation $T_{13}T_{32}(T_{12}T_{32}T_{13})^{-1}=1$.
	The latter relation is clearly inequivalent to the order three relation $Y^3=1$, even though they give rise to the same modular property.
	Perhaps generalizations of this relation will give rise to more general relations among three elliptic $\Gamma$ functions, which would allow the computation of the limit of the $\Gamma$ function for arbitrary rational values of $\tau$ and $\sigma$ in terms of a simple phase.
	In particular, this would probably provide a complete set of saddles and their associated on-shell actions.
	It will be interesting if the summation over this set has a natural interpretation in terms of (some subgroup of) $SL(3,\mathbb{Z})$.
	
	We have seen that the phases $P_{\textbf{m}}(\boldsymbol{\rho})$, which capture (a version of) the anomaly polynomial of the $\mathcal{N}=4$ theory, can be related to the on-shell actions of the Euclidean gravitational saddles.
	It should also be possible to compute $P_{\textbf{m}}$ as the action of a five-dimensional Chern--Simons theory on a cobordism which interpolates between the manifolds $M_{X_{\textbf{m}}}$, $M_{X_{\textbf{m}}^{-1}}$ and $M_{X_{\textbf{m}}^{-1}}$~\cite{Gadde:2020bov}.
	It would be interesting to understand the connection of this five-dimensional Chern--Simons theory, if any, to the supergravity path integral.
	
	Finally, a different type of modular property between two four-dimensional lens space partition functions corresponding to different quotients of $S^3$ has been studied in~\cite{Shaghoulian:2016gol} in the context of conformal field theories without supersymmetry.
	Lens spaces are included in the manifolds $M_g$ and therefore carry $SL(3,\mathbb{Z})$ modular properties. 
	It would be interesting to study if the $SL(3,\mathbb{Z})$ modularity is related to the modularity reported by~\cite{Shaghoulian:2016gol}, which might provide a hint on how to generalize the $SL(3,\mathbb{Z})$ modularity to non-supersymmetric theories.

	\section*{Acknowledgements}
	We thank Kevin Goldstein for collaboration in the initial stages of this work.
	We are grateful to Ye Tian and Songyan Xie for helpful discussions.
	Finally, we thank Alejandro Cabo-Bizet and Sameer Murthy for patient explanations of their work~\cite{Cabo-Bizet:2019eaf} and discussions about the first version of this paper.
	VJ and SvL are supported by the Simons Foundation Mathematical and Physical Sciences Targeted Grants to Institutes, Award ID:509116.
	VJ is supported by the South African Research Chairs Initiative of the Department of Science and Technology and the National Research Foundation.  
	YL is supported by the UCAS program of special research associate and by the internal funds of the KITS. 
	WL is supported by NSFC No.\ 11875064, No.\ 11947302, and the Max-Planck Partnergruppen fund.

	\appendix

	\section{Definitions and properties of special functions}
	\label{app:defs}
	
	In this appendix, we collect definitions and properties of the special functions used in the main text.
	
	\subsection{The \texorpdfstring{$q$}{q}-Pochhammer symbol}
	
	The $q$-Pochhammer symbol is defined as
	\begin{equation}
		(a;q)_n := \prod_{j=0}^{n-1} (1-a q^j) \,, \qquad (a;q)_0 = 1 \,.
	\end{equation}
	This can be extended to an infinite product so that we may write $(a;q)_\infty$.
	Hence,
	\begin{equation}
		(a;q)_n = \frac{(a;q)_\infty}{(a q^n;q)_\infty} \,.
	\end{equation}
	This enables us to define the $q$-Pochhammer symbol for negative values of the subscript.
	Noting that
	\begin{equation}
		(a;q)_\infty^{-1} = \prod_{j=0}^\infty \frac{1}{1-a q^j} \,,
	\end{equation}
	the coefficient of $a^m q^n$ is the number of partitions of $n$ into at most $m$ parts.
	Enumeration problems of this type are at the heart of microstate counting~\cite{Balasubramanian:2005mg}, so it is no surprise that such expressions make an appearance in discussing the index.

	\subsection{The \texorpdfstring{$q$-$\theta$}{q-theta} function}
	The $q$-$\theta$ function, or modified Jacobi theta function $\theta(z;\tau)$, is defined as
	\begin{equation}
		\theta_q(x) \equiv \theta(z;\tau) = \exp \left(- \sum_{m=1}^\infty \frac{x^m+x^{-m}q^m}{m(1-q^m)} \right) = \prod_{n=0}^{\infty} (1-x q^n) (1-x^{-1}q^{n+1}) \,,
	\end{equation}
	where we have put $q=e^{2\pi i \tau}$ and $x=e^{2\pi i z}$.
	The summation formula is valid for $0 < \text{Im}(z) < \text{Im}(\tau)$, and the product formula is valid for $|q| < 1$. 
	Commonly used properties of the $\theta(z;\tau)$ function include the transformations under translation and reflection:
	\begin{align}
		\begin{split}
			\theta(z+m\tau+n; \tau) = (-x)^{-m} q^{ - \frac{m(m-1)}{2}} \theta(z;\tau) \,, \qquad
			\theta(-z;\tau) = \theta(z+\tau;\tau) \,.
		\end{split}
	\end{align}
	Moreover,
	\begin{equation}
		\theta(z;-\tau) = \frac{-x}{\theta(z;\tau)} \,.
	\end{equation}
	Under the $S$-transformation, the $q$-$\theta$ function enjoys the modular property:
	\begin{align}\label{eq:defn-B}
		\begin{split}
			\theta\left(\frac{z}{\tau};-\frac{1}{\tau} \right) &= e^{i\pi B(z,\tau)} \theta(z;\tau) \,,
			\\
			B(z,\tau)
			&= \frac{z^2}{\tau}+z\left(\frac{1}{\tau}-1\right)+\frac{1}{6}\left(\tau+\frac{1}{\tau}\right)-\frac{1}{2} \,.
		\end{split}
	\end{align}
	Apart from the $S$-transformation, we will need to determine what happens to $q$-$\theta$ functions under a general $SL(2,\mathbb{Z})$ transformation. 
	This will generalize the expression in~\eqref{eq:defn-B}. 
	We derive this formula following~\cite{mumford1983tata}.
	
	First of all, the Jacobi theta function is defined as
	\begin{eqnarray}
		\vartheta(z;\tau)
		&=& \prod_{n=0}^{\infty} (1-q^{ n+1}) (1+x q^{n+\frac{1}{2}}) (1+x^{-1}q^{n+\frac{1}{2}}) \,,
	\end{eqnarray}
	which is related to $q$-$\theta$ function by 
	\begin{equation}\label{eq:varthetatotheta}
		\vartheta(z;\tau) = e^{-\frac{\pi i \tau}{12}} \eta(\tau) \theta\left(z+ \frac{\tau+1}{2};\tau \right) \,.
	\end{equation}
	Here, $\eta(\tau)$ is the Dedekind $\eta$ function, which can be defined in terms of the $q$-Pochhammer symbol as $\eta(\tau)=q^{\frac{1}{24}}(q;q)_\infty$.
	It is shown in~\cite{mumford1983tata} that given $ab$ and $cd$ even, the $SL(2,\mathbb{Z})$ action on $\vartheta(z,\tau)$ is 
	\begin{equation}\label{eq:SL2Zmodularonvartheta}
		\vartheta\left(\frac{z}{c\tau+d};\frac{a\tau+b}{c\tau+d}\right) =\zeta \sqrt{c\tau+d}\ 
		\exp\left(\frac{\pi i c z^2}{c\tau+d} \right)	 \vartheta(z;\tau) \,.
	\end{equation}
	The phase $\zeta$ depends on whether $c$ is taken to be odd or even:
	\begin{equation}
		\zeta = \left\{ \begin{array}{ccc} e^{-\frac{\pi i c}{4}} \left( \frac{d}{c}\right)_J &\ \text{if}\ & c\in 2\mathbb{Z}+1 \,, \cr e^{\frac{\pi i (d-1)}{4}} \left( \frac{c}{|d|}\right)_J &\ \text{if}\ & c\in 2\mathbb{Z} \,, \end{array} \right.
	\end{equation}
	where the $(\cdot)_J$ is the Jacobi symbol.
	The general $SL(2,\mathbb{Z})$ action on the $\eta$ function is
	\begin{equation}\label{eq:etaSL2Z}
		\eta\left( \frac{a\tau+b}{c\tau+d} \right) = \epsilon(a,b,c,d)\ \sqrt{c \tau+d}\ \eta(\tau) \,,
	\end{equation}
	where 
	\begin{equation}
		\epsilon(a,b,c,d)= \left\{ \begin{array}{ccc} \exp\left(i\pi \left[\frac{a+d}{12 c}-s(d,c)-\frac{1}{4}\right] \right) &\ \text{for}\ & c\neq 0 \,, \cr \exp(i\pi b/12)&\ \text{for}\ & c=0 \,, \end{array} \right.
	\end{equation}
	with
	\begin{equation}
		s(d,c) = -\frac1c \sum_\omega \frac{1}{(1-\omega^d)(1-\omega)}+\frac14-\frac{1}{4c} \,, \qquad \omega^c = 1\ \text{and}\ \omega \ne 1 \,,
	\end{equation}
	the Dedekind sum of a pair of coprime integers.
	Then taking~\eqref{eq:varthetatotheta} and~\eqref{eq:etaSL2Z} into~\eqref{eq:SL2Zmodularonvartheta}, we obtain: 
	\begin{align}\label{eq:result}
		\theta\left(\frac{z}{ c\tau+d};\frac{a\tau+b}{c\tau+d} \right)  &= \theta(z;\tau) e^{i\pi B(z,\tau;a,b,c,d)} \,,
	\end{align}
	with:
	\begin{eqnarray}\nonumber
		B(z,\tau;a,b,c,d) &=& \frac{c z^2}{c\tau+d} + z\left(\frac{1}{c\tau+d}-1 \right) + \frac{c\tau^2}{6(c\tau+d)}\\ \label{eq:finalBpoly}
		&+& \frac{c (3 - 6 d) + 6 (-1 + d) d + b (-2 + 3 c^2 - 6 c d + 3 d^2)}{12(c\tau+d)} \\ \nonumber
		&+& \frac{(-2 a + 3 (-2 + a) c^2 - d - 6 (-1 + a) c d + 3 a d^2) \tau}{12(c\tau+d)} + \delta(a,b,c,d) \,,
	\end{eqnarray}
	where $\delta(a,b,c,d)$ is a constant that depends only on the $SL(2,\mathbb{Z})$ element: 
	\begin{equation}
		e^{i\pi  \delta(a,b,c,d)} = \frac{\epsilon(a,b,c,d)}{\zeta} \,.
	\end{equation}
	The result~\eqref{eq:result} is only valid when $ab$ and $cd$ are both even, because of the requirement that~\eqref{eq:SL2Zmodularonvartheta} holds. 
	Although it is possible to generalize to cases where one of the products is odd, in this paper we will be satisfied with using~\eqref{eq:result}.

	\subsection{The elliptic \texorpdfstring{$\Gamma$}{Gamma} function}

	The elliptic $\Gamma$ function can be defined as an infinite product when $\mathrm{Im}(\tau)>0$ and $\mathrm{Im}(\sigma)>0$ as follows:
	\begin{align}\label{eq:defn-ell-gamma-app}
		\begin{split}
			\Gamma(x)\equiv\Gamma(z;\tau,\sigma)=\prod^{\infty}_{m,n=0}\frac{1-x^{-1}p^{m+1}q^{n+1}}{1-x p^{m}q^{n}} \,,
		\end{split}
	\end{align}
	where $q=e^{2\pi i \tau}$, $p=e^{2\pi i \sigma}$, and $x=e^{2\pi i z}$.
	For $\mathrm{Im}(\tau)>0$, $\mathrm{Im}(\sigma)>0$ and $0<\mathrm{Im}(z)<\mathrm{Im}(\tau)+\mathrm{Im}(\sigma)$, the elliptic $\Gamma$ function can also be defined through the summation formula:
	\begin{equation}\label{eq:defn-ell-gamma-sum}
		\Gamma(z;\tau,\sigma)=\exp\left(\sum^{\infty}_{\ell=1}\frac{x^\ell-(x^{-1}p q)^\ell}{\ell(1-p^\ell)(1-q^\ell)}\right) \,.
	\end{equation}
	Basic properties that are manifest from these expressions include:
	\begin{align}
		\begin{split}
			\Gamma(z;\tau,\sigma)&=\Gamma(z;\sigma,\tau) \,, \\
			\Gamma(z+1;\tau,\sigma)&=\Gamma(z;\tau+1,\sigma)=\Gamma(z;\tau,\sigma+1)=\Gamma(z;\tau,\sigma) \,.
		\end{split}
	\end{align} 
	Furthermore, the elliptic $\Gamma$ function satisfies the shift properties:
	\begin{align}\label{eq:basic-shift-gamma-app}
		\begin{split}
			\Gamma(z+\tau;\tau,\sigma)&=\theta(z;\sigma)\Gamma(z;\tau,\sigma) \,,\\
			\Gamma(z+\sigma;\tau,\sigma)&=\theta(z;\tau)\Gamma(z;\tau,\sigma) \,.
		\end{split}
	\end{align}
	The product of two $\Gamma$ functions with reflected $z$ arguments simplifies:
	\begin{equation}\label{eq:gamma-theta-id-app}
		\Gamma(z;\tau,\sigma)\Gamma(-z;\tau,\sigma)=\frac{1}{\theta(z;\sigma)\theta(-z;\tau)}.
	\end{equation}
	
	Even though the definitions of the elliptic $\Gamma$ function in terms of the product formula only hold for positive imaginary parts of its arguments, it can also be extended to negative imaginary part $\mathrm{Im}(\tau)<0$ or $\mathrm{Im}(\sigma)<0$ via the summation formula~\eqref{eq:defn-ell-gamma-sum}.
	In particular, one has: 
	\begin{align}\label{eq:extend}
		\begin{split}
			& \Gamma(z;-\tau,\sigma) = \frac{1}{\Gamma(z+\tau;\tau,\sigma)} = \Gamma(\sigma-z;\tau,\sigma) \\
			& \Gamma(z;\tau,-\sigma) = \frac{1}{\Gamma(z+\sigma;\tau,\sigma)} = \Gamma(\tau-z;\tau,\sigma) 
		\end{split}
	\end{align}
	Thus, the elliptic $\Gamma$ function is defined for $\tau,\sigma \in \mathbb{C}-\mathbb{R}$.
	
	A modular property of the elliptic $\Gamma$ function was first derived  in~\cite{Felder_2000}:
	\begin{equation}
		\Gamma(z;\tau,\sigma)=e^{-i\pi Q(z;\tau,\sigma)}\frac{\Gamma\left(\frac{z}{\tau};\frac{\sigma}{\tau},-\frac{1}{\tau}\right)}{\Gamma\left(\frac{z-\tau}{\sigma};-\frac{\tau}{\sigma},-\frac{1}{\sigma}\right)} \,,
	\end{equation}
	whose $Q$ polynomial is defined as:
	\begin{align}\label{eq:defn-Q-FV1}
		\begin{split}
			Q(z;\tau,\sigma)&=\frac{z^3}{3\tau\sigma}-\frac{\tau+\sigma-1}{2\tau\sigma}z^2+\frac{\tau^2+\sigma^2+3\tau\sigma-3\tau-3\sigma+1}{6\tau\sigma}z\\
			&+\frac{1}{12}(\tau+\sigma-1)(\tau^{-1}+\sigma^{-1}-1) \,.
		\end{split}
	\end{align}
	Relations with Bernoulli polynomials may be found in~\cite{Nieri:2015yia}.

	\subsection{Anomaly polynomial}
	The anomaly polynomial encodes the 't Hooft anomalies of a supersymmetric gauge theory in its coefficients.
	We use the parameterization of the anomaly polynomial of a general $\mathcal{N}=1$ theory given in~\cite{Gadde:2020bov}:
	\begin{align}\label{eq:anomaly-polynomial}
		\begin{split}
			P(z_a;x_i)=\frac{1}{x_1x_2x_3}\left(k_{ijk}z_iz_jz_k+3k_{ijR}z_iz_jX+3k_{iRR}z_iX^2+k_{RRR}X^3-k_iz_i\tilde{X}-k_RX\tilde{X}\right) \,,
		\end{split}
	\end{align}
	where the various $k$ indicate the (mixed) R- and flavor symmetry anomalies, and:
	\begin{equation}
		X=\tfrac{1}{2}\sum^3_{i=1}x_i \,,\qquad \tilde{X}=\tfrac{1}{4}\sum^3_{i=1}x^2_i \,.
	\end{equation}

	\section{More general order three elements of \texorpdfstring{$SL(3,\mathbb{Z})$}{SL(3,Z)}}
	\label{app:more-gen-order-3-elms}
	
	In this appendix, we consider the general form of the matrix $A$ in~\eqref{eq:matrixA} and~\eqref{eq:a2-a1cubic-def-d}, namely
	\begin{equation}\label{eq:matrixAcubica1}
		A=	\left(
		\begin{array}{ccc}
			n	&\ m\ & 0 \\
			k	&\ -n\  & 1 \\
			d-n k	&\ -(m k +n^2)\ & 0
		\end{array}
		\right) \,,
	\end{equation}
	subject to the constraint
	\begin{equation}\label{eq:a2-a1cubic-def-d2}
		(1-n)(n^2+n+1)=d\, m, \quad d\in \mathbb{Z} \,.
	\end{equation}
	The matrix $A$ will reproduce $X_{\mathbf{m}}$ in~\eqref{eq:Xabc-matrix} once the parameters are specified as
	\begin{equation}\label{eq:specialconstraintXab}
		n =1-m n_1 \,,  \quad k=n_1(2-m n_1)-n_2 \,, \quad d = n_1(m^2n_1^2-3m n_1+3) \,.
	\end{equation}
	The integer $k$ is generically unconstrained.

	In the Cardy limit, the complex structure $\tau$ approaches the ratio $-n/m$, which cannot be any rational number due to the constraint $n^3+d m=1$. 
	In principle, $n$ can be any integer whereas the allowed values of $m$ are more restricted. 
	The constraint indicates that $n,m$ are coprime, i.e., $\gcd(m,n)=1$. 
	Then Euler's theorem, which generalizes Fermat's little theorem, states that
	\begin{equation}
		n^{\phi(m)} \equiv 1 \ (\mod m) \,,
	\end{equation}
	for a generic pair of coprime integers $(m,n)$, where $\phi(m)$ is the Euler totient function. 
	Integers $m$ coprime to $n$ thus form a finite group called the \emph{multiplicative group of integers modulo $m$}, which is usually denoted as $(\mathbb{Z}/m \mathbb{Z})^\times$. 
	The order of the group is exactly $\phi(m)$.  
	
	Integers $n$ satisfying the constraint $m|n^3-1$ form an order three subgroup of $(\mathbb{Z}/m \mathbb{Z})^\times$. 
	There are two different kinds of subgroup elements. 
	The identity element satisfies the order three criterion trivially, which corresponds to $n \equiv 1\ (\mod m)$.
	The corresponding $SL(3,\mathbb{Z})$ matrices are the $X_{\mathbf{m}}$ we have studied in Section~\ref{ssec:finite-order-elms}.
	More non-trivial group elements are those not of the identity type. 
	Lagrange's theorem states the order of a subgroup of a finite group  divides the order of the entire group, which in our case is $(\mathbb{Z}/m \mathbb{Z})^\times$. 
	Thus, when $n \neq 1- n_1 m$ for integer $n_1$, we conclude that $3|\phi(m)$.
	
	Let us focus on the non-trivial solutions of $3|\phi(m)$.
	The values of $m$ fall on two branches: multiples either (i) of $9$ or (ii) of a prime of the form $6\ell+1$~\cite{oeisA066498}.
	A few low values of  allowed $m$ are
	\begin{equation}\label{eq:setofa2}
		m=7,9,13,14,18,19,21,26,27,28,... \,.
	\end{equation}
	This set comprises \emph{almost all} the integers, i.e., the asymptotic density  of $m$ among integers approaches unity~\cite{dressler1975property}.
	The integer pairs $(m,n)$ can have non-identity type solutions when $m$ takes the values in~\eqref{eq:setofa2}. 
	A few non-trivial examples are $(m,n)=(7,2),(7,4),(9,4)$.
	
	The non-trivial subgroup satisfying $3|\phi(m)$ is the cyclic group $\mathbb{Z}_3$, which is due to the fact \emph{all the finite groups of the prime order are cyclic group}.
	The solutions of integer pairs $(m,n)$ are thus characterized by neat algebraic structures, even without referring to explicit solutions.
	We will not be able to characterize the complete set of solutions explicitly.
	In the following, with a number of simplifying assumptions, we will investigate special cases for which the constraint $n^3+d m=1$ can be parameterized by free integers. 
	
	We construct the free parameters of integer solutions by simplifying the constraint equations further. 
	The factorization of the cubic polynomial $1-n^3 =(1-n)(1+n+n^2) $ can help us reduce the cubic congruence equation to quadratic equations if we only focus on the $1+n+n^2$ part.
	For simplicity, we further restrict ourselves to the case where $\frac{m}{1-n} =r \in \mathbb{Z}$. 
	Then the problem reduces to solve
	\begin{equation}\label{eq:quadratic-congruence-equation}
		n^2+n+1 \equiv 0\ (\mod r) \,.
	\end{equation}
	One can also consider~\eqref{eq:quadratic-congruence-equation} to be a three variable Pell type equation in number theory:
	\begin{equation}\label{eq:Pellequation}
		(2n+1)^2 + (r-u)^2-(r+u)^2 =-3 \,.
	\end{equation}
	The basic idea behind solving a Pell equation is to find a special solution and apply recursive relations between solutions. 
	A simple solution to~\eqref{eq:Pellequation} is $ (n,r,u)=\pm (0,1,1)$.
	If $(n,r,u)$ is a solution to~\eqref{eq:Pellequation}, then $(n+2r,2+4n+4r+u,r)$ is also a solution. 
	This results in a solvable recursive sequence.
	The disadvantage of this procedure is that the solutions to~\eqref{eq:Pellequation} are always parameterized by powers of irrational numbers, even though they are integers in nature. 
	
	In the following, we will search for solutions to~\eqref{eq:quadratic-congruence-equation} by considering it as a congruence equation. 
	The solutions are classified by the congruence class of the integer $r$.
	Therefore, we parameterize $n$ and $m$ by 
	\begin{equation}
		n = 1-r \alpha -\alpha_i \,, \qquad m = r (r\alpha+\alpha_i) \,,
	\end{equation}
	where the $\alpha_i$s are the remainders such that $\alpha_i \in \{0,...,r-1\}$ with $n$ satisfying~\eqref{eq:quadratic-congruence-equation}. 
	Then the solutions of $\alpha_i$ are
	\begin{equation}
		\{ 	\alpha_i \} = \Big\{ \frac{r+1}{2}(3+\sqrt{-3}) \Big\}
	\end{equation}
	in the sense of  congruence class.
	
	To have solutions of $m,n$ in terms of free integers, we still need to write $\sqrt{-3}$ in $\alpha_i$ as an integer modulo $r$. 
	In the following we will study the case where $r$ is a prime of the form $6\ell+1$. The $m$ manifestly satisfies $3|\phi(m)$ as it falls into the branch (ii) mentioned above where non-trivial solutions are allowed.
	Fermat's little theorem states that any such integers are of the form 
	\begin{equation}\label{eq:fermattheorem-absquare}
		r= t^2+3s^2 \,,
	\end{equation}
	with one of $(t,s)$ being an odd integer. 
	Then the solutions to $x^2=-3\ (\mod r)$ are 
	\begin{equation}
		x\equiv t s^{-1} \  (\mod n)
	\end{equation}
	The parameterization problem of $(m,n)$ then reduces to finding $r$ such that $x$ is manifestly an integer. 
	Take $(t,s)=(2p,1)$ as an example. 
	We can find $(r,x)=(4p^2+3,2p)$. Then
	\begin{equation}
		\alpha_1 =2p^2+3 \   (\mod r) \,, \qquad \alpha_2=-2p^2-p\  (\mod r) \,.
	\end{equation}
	In terms of the parameters $p$ and $\alpha$, we can write $(m,n)$ as
	\begin{equation}
		m = (4p^2+3)^2 \alpha -(4p^2+3)(2p^2+p)  \,, \qquad n=-(4p^2+3)\alpha +2p^2+p+1 \,.
	\end{equation}
	
	What we have shown above supplies hints to generalize the integer pair $(m,n)$ to other cases. 
	Note that $r=4p^2+3$ is not generically a prime integer, but the manifest integer parameterization of $(m,n)$ is still valid for any $p$. 
	As generalizations of this, one can consider $r$ in~\eqref{eq:fermattheorem-absquare} to be multiples of prime of the form $6\ell+1$. Alternatively, one can also consider $m/(1-n)$ to be general rational numbers.

	\section{Derivation of \texorpdfstring{$Q$}{Q} polynomials}\label{app:deriv-Q}
	
	In this appendix, we will derive the phases associated to the modular properties used in the main text.
	For this, we will use the ``gauge fixed'' formalism of~\cite{Felder_2000}.
	We will first briefly review the relevant parts of~\cite{Felder_2000} before turning to the computation of the phase for the relations $Y^3=1$ and $X_{\mathbf{m}}^3=1$, respectively.
	The computations we describe are straightforward but also somewhat involved.
	We provide the details for the purpose of completeness.
	
	\subsection{General strategy}
	
	For convenience, we repeat some of the relevant formulae already mentioned in Section~\ref{sec:mod-prop-gamma}, to which we refer for more details.
	We will make frequent use of the main property of the normalized partition function of a free chiral multiplet, for which we omit the superscript $\hat{Z}^a_g=\hat{Z}_g$ as in Section~\ref{ssec:partn-fn-sl3}:
	\begin{equation}\label{eq:sl3-reln-part-funct-app}
		\hat{Z}_{g_1 \cdot g_2}\left(\textbf{x}\right)=\hat{Z}_{g_1}\left(\textbf{x}\right)\hat{Z}_{g_2}\left(g^{-1}_1\textbf{x}\right)\mod M \,.
	\end{equation}
	In~\cite{Felder_2000}, the functions $\phi^{k,l}_{i,j}$ are introduced to ``gauge fix'' the evaluation of $\hat{Z}_g$ of a chiral multiplet, i.e., the elliptic $\Gamma$ function, on the basic $SL(3,\mathbb{Z})$ relations:
	\begin{align}\label{eq:Felder-relations-app}
		\begin{split}
			&\hat{Z}_{T_{ij}}\left(\textbf{x}\right)\hat{Z}_{T_{kl}}\left(T_{ij}^{-1}\textbf{x}\right)=\phi^{k,l}_{i,j}\left(\textbf{x}\right)\hat{Z}_{T_{kl}}\left(\textbf{x}\right)\hat{Z}_{T_{ij}}\left(T_{kl}^{-1}\textbf{x}\right),\qquad i\neq l \,,\quad j\neq k \,,\\
			&\hat{Z}_{T_{ij}}\left(\textbf{x}\right)\hat{Z}_{T_{jk}}\left(T_{ij}^{-1}\textbf{x}\right)=\phi^{j,k}_{i,j}\left(\textbf{x}\right)\hat{Z}_{T_{ik}}\left(\textbf{x}\right)\hat{Z}_{T_{jk}}\left(T_{ik}^{-1}\textbf{x}\right)\hat{Z}_{T_{ij}}\left(T_{jk}^{-1}T_{ik}^{-1}\textbf{x}\right)  \,,\\
			&\hat{Z}_{S_{13}}\left(\textbf{x}\right)\hat{Z}_{S_{13}}\left(S_{13}^{-1}\textbf{x}\right)\hat{Z}_{S_{13}}\left(S_{13}^{-2}\textbf{x}\right)\hat{Z}_{S_{13}}\left(S_{13}^{-3}\textbf{x}\right)=1 \,.
		\end{split}
	\end{align}
	Here, $\phi^{k,l}_{i,j}(\textbf{x})=e^{i\pi L^{k,l}_{i,j}(\textbf{x})}$ and $\phi^{j,k}_{i,j}(\textbf{x})=e^{i\pi L^{j,k}_{i,j}(\textbf{x})}$ with:
	\begin{align}
		\begin{split}
			L^{3,2}_{1,2}(\textbf{x})=-L^{1,2}_{3,2}(\textbf{x})&=Q\left(\tfrac{Z-x_1}{x_1};\tfrac{x_2-x_3}{x_1},\tfrac{x_3-x_1}{x_1}\right)+Q\left(\tfrac{Z-x_1+x_3}{x_1-x_3};\tfrac{x_3}{x_1-x_3},\tfrac{x_2-x_1}{x_1-x_3}\right)\,,\\
			L^{1,2}_{3,1}(\textbf{x})&=Q\left(\tfrac{Z-x_1}{x_1};\tfrac{x_2-x_3}{x_1},\tfrac{x_3-x_1}{x_1}\right)\,,\\
			L^{3,2}_{1,3}(\textbf{x})&=-Q\left(\tfrac{Z-x_1+x_3}{x_1-x_3};\tfrac{x_2-x_1}{x_1-x_3},\tfrac{x_3}{x_1-x_3}\right)\,.
		\end{split}
	\end{align}
	All the other $L$ vanish, and the function $Q$ is defined in~\eqref{eq:defn-Q-FV1}.
	The combination of~\eqref{eq:sl3-reln-part-funct-app} and~\eqref{eq:Felder-relations-app} allows one to compute the modular properties exactly, i.e., beyond $\mod M$, as we will see in detail below.
	
	Also, recall from Section~\ref{ssec:partn-fn-sl3} that the only $T_{ij}$ such that $\hat{Z}_{T_{ij}}\neq 1$ are $T_{12}$ and $T_{32}$.
	Explicitly,
	\begin{align}
		\hat{Z}_{T_{12}}&=\Gamma\left(\tfrac{Z-x_2}{x_3};\tfrac{x_1-x_2}{x_3},-\tfrac{x_1}{x_3}\right)^{-1}\,,\\
		\hat{Z}_{T_{32}}&=\Gamma\left(\tfrac{Z}{x_1};\tfrac{x_2-x_3}{x_1},\tfrac{x_3}{x_1}\right)\,.
	\end{align}
	This collection of formulae is sufficient to derive the phase $f_r\left(\textbf{x}\right)$ for any relation $r=1$ in the group, where $r=e_1\cdots e_n$ is a reduced expression in terms of the generators $e_k\in \lbrace T_{ij}^{\pm 1}\rbrace $.
	Evaluating the partition function on this relation:
	\begin{equation}\label{eq:gen-r}
		\hat{Z}_{e_1}\left(\textbf{x}\right)\prod^{n-1}_{i=1} \hat{Z}_{e_{i+1}}\left(e_i^{-1}\cdots e_1^{-1} \textbf{x}\right) =e^{i\pi f_{r}\left(\textbf{x}\right)} \,.
	\end{equation}
	For notational convenience, in the following we will write:
	\begin{equation}\label{eq:notation}
		\hat{Z}_{e_1\cdots e_n}\left(\textbf{x}\right)\equiv \hat{Z}_{e_1}\left(\textbf{x}\right)\prod^{n-1}_{i=1} \hat{Z}_{e_{i+1}}\left(e_i^{-1}\cdots e_1^{-1} \textbf{x}\right).
	\end{equation}
	Even though this notation by~\eqref{eq:sl3-reln-part-funct-app} is correct $\mod M$, it is misleading in the gauge fixed formalism. 
	Indeed, when $r=e_1\cdots e_n=1$, $\hat{Z}_{e_1\cdots e_n}\left(\textbf{x}\right)=1$ exactly.
	The non-trivial phase in~\eqref{eq:gen-r} is only present for the product of partition functions labeled by a single group element $\hat{Z}_{e_i}$.
	Similarly, the basic relations written in this notation become:
	\begin{align}
		\begin{split}
			\hat{Z}_{T_{ij}T_{kl}}(\textbf{x})&=\phi^{k,l}_{i,j}(\textbf{x})\hat{Z}_{T_{kl}T_{ij}}(\textbf{x})\,,\qquad i\neq l\,,\quad j\neq k \,,\\
			\hat{Z}_{T_{ij}T_{jk}}(\textbf{x})&=\phi^{j,k}_{i,j}(\textbf{x})\hat{Z}_{T_{ik}T_{jk}T_{ij}}(\textbf{x})\,,\\
			\hat{Z}_{S_{13}^4}(\textbf{x})&=1\,.
		\end{split}
	\end{align}
	These equations are again misleading because the partition functions appearing on the left and right hand side are simply equal, i.e., strictly speaking there should not be a phase.
	The expressions only hold when we understand them as the notation defined in~\eqref{eq:notation}, in which case one finds the relations~\eqref{eq:Felder-relations-app}.
	
	To illustrate how one can use the basic $SL(3,\mathbb{Z})$ relations in~\eqref{eq:Felder-relations-app} to compute phases for relations in the group, let us consider first consider two simple examples:
	\begin{equation}
		\hat{Z}_{T_{32}T_{31}}\left(\textbf{x}\right)=\phi^{3,1}_{3,2}\left(\textbf{x}\right)\hat{Z}_{T_{31}T_{32}}\left(\textbf{x}\right) \,.
	\end{equation}
	This implies that
	\begin{align}
		\begin{split}
			\hat{Z}_{T_{32}}\left(\textbf{x}\right)&=\hat{Z}_{T_{32}}\left(T_{31}^{-1}\textbf{x}\right) \,,\\
			\Gamma\left(\tfrac{Z}{x_1};\tfrac{x_2-x_3}{x_1},\tfrac{x_3}{x_1}\right)&=\Gamma\left(\tfrac{Z}{x_1};\tfrac{x_2-x_3+x_1}{x_1},\tfrac{x_3-x_1}{x_1}\right)\,,
		\end{split}
	\end{align}
	where the left and right hand sides of the two expressions above are exactly the same.
	We have used $\phi^{3,1}_{3,2}\left(\textbf{x}\right)=1$ and the property~\eqref{eq:sl3-reln-part-funct-app} of $\hat{Z}_g$.
	Periodicity of the elliptic $\Gamma$ function can now be used to see that this equation is indeed correct.
	
	More non-trivially, we have:
	\begin{align}
		\begin{split}
			\hat{Z}_{T_{12}T_{32}}\left(\textbf{x}\right)&=\phi^{3,2}_{1,2}\left(\textbf{x}\right)\hat{Z}_{T_{32}T_{12}}\left(\textbf{x}\right)\\
			\Longleftrightarrow \qquad \hat{Z}_{T_{12}}\left(\textbf{x}\right)\hat{Z}_{T_{32}}\left(T_{12}^{-1}\textbf{x}\right)&=\phi^{3,2}_{1,2}\left(\textbf{x}\right)\hat{Z}_{T_{32}}\left(\textbf{x}\right)\hat{Z}_{T_{12}}\left(T_{32}^{-1}\textbf{x}\right)\\
			\Longleftrightarrow \qquad \frac{\Gamma\left(\frac{Z}{x_1-x_2};\frac{x_2-x_3}{x_1-x_2},\frac{x_3}{x_1-x_2}\right)}{\Gamma\left(\frac{Z-x_2}{x_3};\frac{x_1-x_2}{x_3},-\frac{x_1}{x_3}\right)}&=e^{i\pi L^{3,2}_{1,2}\left(\textbf{x}\right)}\frac{\Gamma\left(\frac{Z}{x_1};\frac{x_2-x_3}{x_1},\frac{x_3}{x_1}\right)}{\Gamma\left(\frac{Z-x_2}{x_3-x_2};\frac{x_1-x_2}{x_3-x_2},-\frac{x_1}{x_3-x_2}\right)}.
		\end{split}
	\end{align}
	We have not encountered this relation explicitly in the literature.
	One way to prove it is by adapting the proof of Theorem~4.1 in~\cite{Felder_2000} to this relation.
	In particular, one may check that the function
	\begin{equation}
		A\left(\textbf{x}\right)=\frac{\Gamma\left(\frac{Z}{x_1-x_2};\frac{x_2-x_3}{x_1-x_2},\frac{x_3}{x_1-x_2}\right)\Gamma\left(\frac{Z-x_2}{x_3-x_2};\frac{x_1-x_2}{x_3-x_2},-\frac{x_1}{x_3-x_2}\right)}{\Gamma\left(\frac{Z-x_2}{x_3};\frac{x_1-x_2}{x_3},-\frac{x_1}{x_3}\right)\Gamma\left(\frac{Z}{x_1};\frac{x_2-x_3}{x_1},\frac{x_3}{x_1}\right)}e^{-i\pi L^{3,2}_{1,2}\left(\textbf{x}\right)}
	\end{equation}
	is periodic under $Z\to Z+x_i$ for $i=1,2,3$.
	Since the function is then a triply periodic meromorphic function, this implies that it is equal to a constant.
	The constant itself can be fixed by evaluating the equation on a special value of $Z$ at which both sides simplify greatly.
	
	When we wish to compute the phases associated to relations that are more complicated than the basic relations, such as $Y^3=1$ and $X_{\mathbf{m}}^3=1$, we will make use of the following strategy.
	First, one proves the relation, e.g., $Y^3=1$, through a combination of the basic $SL(3,\mathbb{Z})$ relations.
	The total phase is then an accumulation of phases corresponding to the basic relations~\eqref{eq:Felder-relations-app}.
	Explicitly, we use:\footnote{Note that we are using the slightly misleading notation defined in and explained around~\eqref{eq:notation}.}
	\begin{align}\label{eq:Felder-relations-2}
		\begin{split}
			\hat{Z}_{g_1T_{ij}T_{kl}g_2}\left(\textbf{x}\right)&=\phi^{k,l}_{i,j}\left(g_1^{-1}\textbf{x}\right)\hat{Z}_{g_1T_{kl}T_{ij}g_2}\left(\textbf{x}\right)\,,\qquad i\neq l\,,\quad j\neq k \,,\\
			\hat{Z}_{g_1T_{ij}T_{jk}g_2}\left(\textbf{x}\right)&=\phi^{j,k}_{i,j}\left(g_1^{-1}\textbf{x}\right)\hat{Z}_{g_1T_{ik}T_{jk}T_{ij}g_2}\left(\textbf{x}\right) \,.
		\end{split}
	\end{align}
	Let us quickly prove the first relation:
	\begin{align}
		\begin{split}
			\hat{Z}_{g_1T_{ij}T_{kl}g_2}(\textbf{x})&=\hat{Z}_{g_1}(\textbf{x})\,\hat{Z}_{T_{ij}T_{kl}}(g_1^{-1}\textbf{x})\,\hat{Z}_{g_2}(T_{kl}^{-1}T_{ij}^{-1}g_1^{-1}\textbf{x})\\
			&=\hat{Z}_{g_1}(\textbf{x})\,\phi^{k,l}_{i,j}(g_1^{-1}\textbf{x})\,\hat{Z}_{T_{kl}T_{ij}}(g_1^{-1}\textbf{x})\,\hat{Z}_{g_2}(T_{ij}^{-1}T_{kl}^{-1}g_1^{-1}\textbf{x})\\
			&=\phi^{k,l}_{i,j}(g_1^{-1}\textbf{x})\,\hat{Z}_{g_1T_{kl}T_{ij}g_2}(\textbf{x}) \,.
		\end{split}
	\end{align}
	The second relation is established by means of a similar argument.
	Finally, in order to compute the phases for the relations $Y^3=1$ and $X_{\mathbf{m}}^3=1$, it will be convenient to collect similar formulae to~\eqref{eq:Felder-relations-app} for when some of the $T_{ij}$ are replaced with their inverse $T_{ij}^{-1}$:
	\begin{align}
		\begin{split}
			\hat{Z}_{T_{ij}T_{kl}^{-1}}(\textbf{x})&=(\psi_1)_{i,j}^{k,l}(\textbf{x})\hat{Z}_{T_{kl}^{-1}T_{ij}}(\textbf{x})\,,\qquad i\neq l\,,\quad j\neq k\,,\\
			\hat{Z}_{T_{ij}^{-1}T_{kl}}(\textbf{x})&=(\psi_2)_{i,j}^{k,l}(\textbf{x})\hat{Z}_{T_{kl}T_{ij}^{-1}}(\textbf{x})\,,\qquad i\neq l\,,\quad j\neq k\,,\\
			\hat{Z}_{T_{ij}^{-1}T_{kl}^{-1}}(\textbf{x})&=(\psi_3)_{i,j}^{k,l}(\textbf{x})\hat{Z}_{T_{kl}^{-1}T_{ij}^{-1}}(\textbf{x})\,,\qquad i\neq l\,,\quad j\neq k\,,
		\end{split}
	\end{align}
	where we have defined:
	\begin{align}\label{eq:psi-in-phi}
		\begin{split}
			(\psi_1)_{i,j}^{k,l}(\textbf{x})&=\frac{1}{\phi_{i,j}^{k,l}(\textbf{y})}\,,\quad \textbf{y}=T_{kl}\textbf{x}\,,\\
			(\psi_2)_{i,j}^{k,l}(\textbf{x})&=\frac{1}{\phi_{i,j}^{k,l}(\textbf{y})}\,,\quad \textbf{y}=T_{ij}\textbf{x}\,,\\
			(\psi_3)^{k,l}_{i,j}(\textbf{x})&=\phi^{k,l}_{i,j}(\textbf{y})\,, \quad \textbf{y}=T_{ij}T_{kl}\textbf{x}\,.
		\end{split}
	\end{align}
	Similarly,
	\begin{align}
		\begin{split}
			\hat{Z}_{T_{jk}T_{ij}}(\textbf{x})&=(\xi_1)^{i,j}_{j,k}(\textbf{x})\hat{Z}_{T_{ik}^{-1}T_{ij}T_{jk}}(\textbf{x})\,,\\
			\hat{Z}_{T_{ij}^{-1}T_{jk}^{-1}}(\textbf{x})&=(\xi_2)^{j,k}_{i,j}(\textbf{x})\hat{Z}_{T_{jk}^{-1}T_{ij}^{-1}T_{ik}}(\textbf{x})\,,\\
			\hat{Z}_{T_{jk}^{-1}T_{ij}^{-1}}(\textbf{x})&=(\xi_3)^{i,j}_{j,k}(\textbf{x})\hat{Z}_{T_{ij}^{-1}T_{jk}^{-1}T_{ik}^{-1}}(\textbf{x})\,,\\
			\hat{Z}_{T_{ij}T_{jk}^{-1}}(\textbf{x})&=(\xi_4)_{i,j}^{j,k}(\textbf{x})\hat{Z}_{T_{jk}^{-1}T_{ik}^{-1}T_{ij}}(\textbf{x})\,,\\
			\hat{Z}_{T_{ij}^{-1}T_{jk}}(\textbf{x})&=(\xi_5)_{i,j}^{j,k}(\textbf{x})\hat{Z}_{T_{jk}T_{ij}^{-1}T_{ik}^{-1}}(\textbf{x})\,,\\
			\hat{Z}_{T_{jk}T_{ij}^{-1}}(\textbf{x})&=(\xi_6)^{i,j}_{j,k}(\textbf{x})\hat{Z}_{T_{ij}^{-1}T_{ik}T_{jk}}(\textbf{x})\,,\\
			\hat{Z}_{T_{jk}^{-1}T_{ij}}(\textbf{x})&=(\xi_7)^{i,j}_{j,k}(\textbf{x})\hat{Z}_{T_{ik}T_{ij}T_{jk}^{-1}}(\textbf{x})\,,
		\end{split}
	\end{align}
	where
	\begin{align}\label{eq:xi-in-phi}
		\begin{split}
			(\xi_1)^{i,j}_{j,k}(\textbf{x})&=\frac{1}{\phi^{j,k}_{i,j}(\textbf{y})}\,, \quad \textbf{y}=T_{ik}\textbf{x}\,,\\
			(\xi_2)^{j,k}_{i,j}(\textbf{x})&=\phi^{j,k}_{i,j}(\textbf{y})\,, \quad \textbf{y}=T_{ij}T_{jk}\textbf{x}\,,\\
			(\xi_3)^{i,j}_{j,k}(\textbf{x})&=\frac{1}{\phi^{j,k}_{i,j}(\textbf{y})}\,, \quad \textbf{y}=T_{ij}T_{jk}\textbf{x}\,,\\
			(\xi_4)_{i,j}^{j,k}(\textbf{x})&=\frac{1}{\phi^{j,k}_{i,j}(\textbf{y})}\,, \quad \textbf{y}=T_{ik}T_{jk}\textbf{x}\,,\\
			(\xi_5)_{i,j}^{j,k}(\textbf{x})&=\frac{1}{\phi^{j,k}_{i,j}(\textbf{y})\phi^{j,k}_{i,k}(\textbf{y})}\,, \quad \textbf{y}=T_{ij}\textbf{x}\,,\\
			(\xi_6)^{i,j}_{j,k}(\textbf{x})&=\phi^{j,k}_{i,j}(\textbf{y})\,, \quad \textbf{y}=T_{ij}\textbf{x}\,,\\
			(\xi_7)^{i,j}_{j,k}(\textbf{x})&=\phi^{j,k}_{i,j}(\textbf{y})\phi^{j,k}_{i,k}(\textbf{y})\,, \quad \textbf{y}=T_{jk}\textbf{x}\,.
		\end{split}
	\end{align}
	With these preliminaries, we are now finally ready to compute the phases associated to the arbitrary relations in the group.
	
	\subsection{$Q$ polynomial for the order three element $Y$}
	\label{app:Y-phase}
	
	In this section, we will compute the phase associated to the relation $Y^3=1$, which played a central role in~\cite{Gadde:2020bov}.
	To do this, we will prove algebraically that $Y^3=1$, where $Y=S_{23}^{-1}S_{13}$, making use only of the basic $SL(3,\mathbb{Z})$ relations.
	The accumulated phase we acquire by repetitive use of~\eqref{eq:Felder-relations-2} will provide the total phase $f\left(\textbf{x}\right)$ such that:\footnote{In the following, we will use the slightly misleading notation defined in and explained around~\eqref{eq:notation}.}
	\begin{equation}
		\hat{Z}_{Y^3}\left(\textbf{x}\right)=e^{i\pi f\left(\textbf{x}\right)}\,.
	\end{equation}
	
	To prove $Y^3=1$ algebraically, we first notice that:
	\begin{equation}
		S_{23}^{-1}S_{13}=S_{13}S_{12}^{-1} \,,\quad S_{12}^{-1}S_{13}=S_{13}S_{23}\,,\quad S_{12}^{-1}S_{23}^{-1}=S_{23}^{-1}S_{13}\,.
	\end{equation}
	Using these relations, we have:
	\begin{align}
		\begin{split}
			S_{23}^{-1}S_{13}S_{23}^{-1}S_{13}S_{23}^{-1}S_{13}&=S_{13}S_{12}^{-1}S_{13}S_{12}^{-1}S_{23}^{-1}S_{13}\\
			&=S_{13}S_{13}S_{23}S_{12}^{-1}S_{23}^{-1}S_{13}\\
			&=S_{13}S_{13}S_{23}S_{23}^{-1}S_{13}S_{13}=1\,.
		\end{split}
	\end{align}
	To compute the accumulated phase of these manipulations, we need to compute the phases associated to:
	\begin{align}
		\begin{split}
			\hat{Z}_{S_{23}^{-1}S_{13}}\left(\textbf{x}\right)&=\chi_1\left(\textbf{x}\right)\hat{Z}_{S_{13}S_{12}^{-1}}\left(\textbf{x}\right)\,,\\ \hat{Z}_{S_{12}^{-1}S_{13}}\left(\textbf{x}\right)&=\chi_2\left(\textbf{x}\right)\hat{Z}_{S_{13}S_{23}}\left(\textbf{x}\right)\,,\\ \hat{Z}_{S_{12}^{-1}S_{23}^{-1}}\left(\textbf{x}\right)&=\chi_3\left(\textbf{x}\right)\hat{Z}_{S_{23}^{-1}S_{13}}\left(\textbf{x}\right)\,.
		\end{split}
	\end{align}
	For the last step, the phase is trivial since:
	\begin{equation}
		\hat{Z}_{S_{23}S_{23}^{-1}}=\hat{Z}_{S_{13}^{4}}=1 \,.
	\end{equation}
	To compute the total phase, we therefore first determine the phases $\chi_1$, $\chi_2$ and $\chi_3$.
	Recall that:
	\begin{equation}
		S_{23}^{-1}=T_{23}^{-1}T_{32}T_{23}^{-1}\,,\quad S_{13}=T_{13}T_{31}^{-1}T_{13}\,,\quad S_{12}^{-1}=T_{21}T_{12}^{-1}T_{21}\,.
	\end{equation}
	Thus, we find:
	\begin{align}
		\begin{split}
			\hat{Z}_{T_{23}^{-1}T_{32}T_{23}^{-1}T_{13}T_{31}^{-1}T_{13}}\left(\textbf{x}\right)&=\frac{\hat{Z}_{T_{13}T_{31}^{-1}T_{13}T_{21}T_{12}^{-1}T_{21}}\left(\textbf{x}\right)}{\phi^{3,2}_{1,3}\left(T_{12}T_{23}\textbf{x}\right)\,\phi^{1,2}_{3,1}\left(T_{31}T_{12}T_{23}T_{13}^{-1}\textbf{x}\right)}\,,
		\end{split}
	\end{align}
	where we made repetitive use of~\eqref{eq:Felder-relations-2}.
	Similarly, we have:
	\begin{align}
		\begin{split}
			\hat{Z}_{T_{21}T_{12}^{-1}T_{21}T_{13}T_{31}^{-1}T_{13}}\left(\textbf{x}\right)
			&=\frac{\phi^{3,2}_{1,3}\left(T_{32}T_{12}T_{23}^{-1}T_{31}T_{13}^{-1}\textbf{x}\right)\,\phi^{3,2}_{1,2}\left(T_{32}T_{12}T_{23}^{-1}T_{31}T_{13}^{-1}\textbf{x}\right)}{\phi^{1,2}_{3,1}\left(T_{31}T_{12}T_{21}^{-1}T_{23}^{-1}T_{13}^{-1}\textbf{x}\right)}\\
			&\times\hat{Z}_{T_{13}T_{31}^{-1}T_{13}T_{23}T_{32}^{-1}T_{23}}\left(\textbf{x}\right)\,.
		\end{split}
	\end{align}
	Finally, we have:
	\begin{align}
		\begin{split}
			\hat{Z}_{T_{21}T_{12}^{-1}T_{21}T_{23}^{-1}T_{32}T_{23}^{-1}}\left(\textbf{x}\right)&=\phi^{3,2}_{1,3}\left(T_{12}T_{21}^{-1}T_{23}\textbf{x}\right)
			\hat{Z}_{T_{23}^{-1}T_{32}T_{23}^{-1}T_{31}^{-1}T_{13}T_{31}^{-1}}\left(\textbf{x}\right) \,.
		\end{split}
	\end{align}
	This provides us with explicit expressions for the phases $\chi_{1,2,3}$:
	\begin{align}
		\begin{split}
			\chi_1\left(\textbf{x}\right)&=\frac{1}{\phi^{3,2}_{1,3}\left(T_{12}T_{23}\textbf{x}\right)\,\phi^{1,2}_{3,1}\left(T_{31}T_{12}T_{23}T_{13}^{-1}\textbf{x}\right)}=1 \,,\\
			\chi_2\left(\textbf{x}\right)&= \frac{\phi^{3,2}_{1,3}\left(T_{32}T_{12}T_{23}^{-1}T_{31}T_{13}^{-1}\textbf{x}\right)\,\phi^{3,2}_{1,2}\left(T_{32}T_{12}T_{23}^{-1}T_{31}T_{13}^{-1}\textbf{x}\right)}{\phi^{1,2}_{3,1}\left(T_{31}T_{12}T_{21}^{-1}T_{23}^{-1}T_{13}^{-1}\textbf{x}\right)}=1\,,\\
			\chi_3\left(\textbf{x}\right)&=\phi^{3,2}_{1,3}\left(T_{12}T_{21}^{-1}T_{23}\textbf{x}\right)=e^{-i\pi Q\left(\frac{Z-x_2}{x_2};\frac{x_1}{x_2},\frac{x_3}{x_2}\right)} \,,
		\end{split}
	\end{align}
	from where we see that only $\chi_3$ is non-trivial.
	
	The total accumulated phase can now be computed:\footnote{Note that we are using the slightly misleading notation defined in and explained around~\eqref{eq:notation}.}
	\begin{align}
		\begin{split}
			\hat{Z}_{Y^3}\left(\textbf{x}\right)&=\hat{Z}_{(S_{23}^{-1}S_{13})^3}\left(\textbf{x}\right)=\hat{Z}_{S_{13}S_{12}^{-1}S_{13}S_{12}^{-1}S_{23}^{-1}S_{13}}\left(\textbf{x}\right)=\hat{Z}_{S_{13}S_{13}S_{23}S_{12}^{-1}S_{23}^{-1}S_{13}}\left(\textbf{x}\right)\\
			&=\chi_3\left(S_{23}^{-1}S_{13}^{-2}\textbf{x}\right)\,\hat{Z}_{S_{13}S_{13}S_{23}S_{23}^{-1}S_{13}S_{13}}\left(\textbf{x}\right)=\chi_3\left(S_{23}^{-1}S_{13}^{-2}\textbf{x}\right)\,.
		\end{split}
	\end{align}
	Thus, we find:
	\begin{align}
		\begin{split}
			\hat{Z}_{Y}\left(\textbf{x}\right)\hat{Z}_{Y}\left(Y^{-1}\textbf{x}\right)\hat{Z}_{Y}\left(Y^{-2}\textbf{x}\right) &=e^{-i\pi Q\left(\frac{Z-x_2}{x_2};\frac{x_1}{x_2},\frac{x_3}{x_2}\right)}\,.
		\end{split}
	\end{align}
	The function on the right hand side is symmetric under any exchange of the $x_{1,2,3}$.
	In particular, if we exchange $x_2$ and $x_1$ and identify $z=\frac{Z}{x_1}$, $\tau=\frac{x_2}{x_1}$, and $\sigma=\frac{x_3}{x_1}$, we can evaluate this property explicitly as:
	\begin{equation}\label{eq:pre-Y-mod-prop}
		\Gamma\left(z;\tau,\sigma\right)\Gamma\left(\tfrac{z}{\sigma};\tfrac{\tau}{\sigma},\tfrac{1}{\sigma}\right)\Gamma\left(\tfrac{z}{\tau};\tfrac{\sigma}{\tau},\tfrac{1}{\tau}\right)=e^{-i\pi Q\left(z-1;\tau,\sigma\right)}\,,
	\end{equation}
	where we used that $\hat{Z}_{S_{23}}=\Gamma(z+\sigma;\tau,\sigma)$ (see~\eqref{eq:sci-from-z-t32}) and properties of the elliptic $\Gamma$ function.
	One may easily verify that this expression matches with the modular property written in~\cite{Gadde:2020bov}.
	Finally, we can rewrite this expression as 
	\begin{equation}\label{eq:ord-mod-prop-Gamma-app}
		\Gamma(z;\tau,\sigma)=e^{-i\pi Q(z;\tau,\sigma)}\frac{\Gamma\left(\frac{z}{\tau};\frac{\sigma}{\tau},-\frac{1}{\tau}\right)}{\Gamma\left(\frac{z-\tau}{\sigma};-\frac{\tau}{\sigma},-\frac{1}{\sigma}\right)}\,,
	\end{equation}
	which is the alternative form of the modular property that we employed in~\cite{Goldstein:2020yvj}.

	\subsection{$Q$ polynomial for the order three element $X_{\mathbf{m}}$}\label{app:Xabc-phase}
	
	We would now like to perform a similar analysis for the element:
	\begin{align}
		\begin{split}
			X_{\mathbf{m}}&=\begin{pmatrix}
				\ 1-mn_1\ &\ m\ &\ 0\ \\
				\ (2-mn_1)n_1-n_2\ &\ mn_1-1\ &\ 1\ \\
				\ (1-mn_1)n_2+n_1\ &\ mn_2-1\ &\ 0\ 
			\end{pmatrix}\,,
		\end{split}
	\end{align}
	which may be written in terms of generators as:
	\begin{equation}
		X_{\mathbf{m}}=T_{23}T_{21}^{n_1-n_2}T_{31}^{n_2}T_{13}^{-m}T_{31}^{n_1}T_{21}^{-n_2}S_{23}\,.
	\end{equation}
	This element is also of order three.
	Thus we would like to compute:\footnote{In the following, we are using the slightly misleading notation defined in and explained around~\eqref{eq:notation}.}
	\begin{equation}
		\hat{Z}_{X_{\mathbf{m}}^3}\left(\textbf{x}\right)=e^{i\pi f\left(\textbf{x}\right)}\,.
	\end{equation}
	Following the same strategy as in the $Y^3=1$ case, we first prove algebraically that this element has order three, using only to the fundamental relations of $SL(3,\mathbb{Z})$.
	Specifically, we will first show that:
	\begin{equation}
		X_{\mathbf{m}}^2=S_{23}^{3}T_{21}^{n_2}T_{31}^{-n_1}T_{13}^{m}T_{31}^{-n_2}T_{21}^{n_2-n_1}T_{23}^{-1}=X_{\mathbf{m}}^{-1}\,.
	\end{equation}
	To show this, it is useful to record the following identities and their associated phases:
	\begin{align}
		\begin{split}
			T_{12}^mS_{23}&=S_{23}T_{13}^m\,,\qquad T_{13}^{-m}S_{23}=S_{23}T_{12}^{m}\,,\\
			T_{21}^{-n_1}S_{23}&=S_{23}T_{31}^{-n_1}\,,\qquad T_{31}^{n_1}S_{23}=S_{23}T_{21}^{-n_1}\,,
		\end{split}
	\end{align}
	where the phases are given by:
	\begin{align}
		\begin{split}
			\hat{Z}_{T_{12}^mS_{23}}\left(\textbf{x}\right)&=\prod_{k=0}^{m-1}\left[(\psi_1)_{1,2}^{3,2}\left(T_{12}^{-k}T_{23}^{-1}T_{13}^{-m}\textbf{x}\right)(\xi_4)_{1,3}^{3,2}\left(T_{13}^{-k}T_{23}^{-1}\textbf{x}\right)\right]\hat{Z}_{S_{23}T_{13}^a}\left(\textbf{x}\right)\,,\\
			\hat{Z}_{T_{13}^{-m}S_{23}}\left(\textbf{x}\right)&=\prod_{k=0}^{m-1}\left[(\xi_2)_{1,3}^{3,2}\left(T_{13}^{k}T_{23}^{-1}\textbf{x}\right)\right]\hat{Z}_{S_{23}T_{12}^{m}}\left(\textbf{x}\right)\,,\\
			\hat{Z}_{T_{21}^{-n_1}S_{23}}\left(\textbf{x}\right)&=\hat{Z}_{S_{23}T_{31}^{-n_1}}\left(\textbf{x}\right)\,,\\
			\hat{Z}_{T_{31}^{n_1}S_{23}}\left(\textbf{x}\right)&=\hat{Z}_{S_{23}T_{21}^{-n_1}}\left(\textbf{x}\right)\,.
		\end{split}
	\end{align}
	Here, the $\psi_i$ and $\xi_i$ were defined in~\eqref{eq:psi-in-phi} and~\eqref{eq:xi-in-phi}, respectively.
	
	The first step is to commute the $S_{23}T_{23}$ in the middle to the left:
	\begin{align}
		\begin{split}
			X_{\mathbf{m}}^2&=T_{23}T_{21}^{n_1-n_2}T_{31}^{n_2}T_{13}^{-m}T_{31}^{n_1}T_{21}^{-n_2}S_{23}T_{23}T_{21}^{n_1-n_2}T_{31}^{n_2}T_{13}^{-m}T_{31}^{n_1}T_{21}^{-n_2}S_{23}\\
			&=T_{23}S_{23}T_{23}T_{31}^{n_1-n_2}T_{21}^{-n_1}T_{12}^{m}T_{31}^{n_1}T_{21}^{-n_2}S_{23}=:M_1 \,.
		\end{split}
	\end{align}
	The only non-trivial phase occurs when commuting $T_{13}^{-m}$ and $S_{23}$, and yields:
	\begin{equation}
		\hat{Z}_{X_{\mathbf{m}}^2}\left(\textbf{x}\right)=\prod_{k=0}^{m-1}\left[(\xi_2)_{1,3}^{3,2}\left(T_{13}^{k}T_{23}^{-1}T_{31}^{-n_2}T_{21}^{n_2-n_1}T_{23}^{-1}\textbf{x}\right)\right]\hat{Z}_{M_1}\left(\textbf{x}\right)\,.
	\end{equation}
	Since $S_{23}=T_{23}T^{-1}_{32}T_{23}=T^{-1}_{32}T_{23}T^{-1}_{32}$ we have:
	\begin{equation}
		T_{23}S_{23}T_{23}=T_{23}T^{-1}_{32}T_{23}T^{-1}_{32}T_{23}=S_{23}^2T_{32}\,.
	\end{equation}
	Notice that we only inserted the definition of $S_{23}$ and an insertion of the identity.
	Therefore, this step will not acquire any phase.
	Thus, we arrive at:
	\begin{equation}
		X_{\mathbf{m}}^2=S_{23}^2T_{32}T_{31}^{n_1-n_2}T_{21}^{-n_1}T_{12}^{m}T_{31}^{n_1}T_{21}^{-n_2}S_{23}=:M_2\,.
	\end{equation}
	Now we continue to commute through the $S_{23}$ on the right all the way to the left.
	One finds:
	\begin{align}
		\begin{split}
			X_{\mathbf{m}}^2&=S_{23}^2T_{32}S_{23}T_{21}^{n_2-n_1}T_{31}^{-n_1}T_{13}^{m}T_{21}^{-n_1}T_{31}^{-n_2}\\
			&=S_{23}^3T_{23}^{-1}T_{21}^{n_2-n_1}T_{31}^{-n_1}T_{13}^{m}T_{21}^{-n_1}T_{31}^{-n_2}=:M_3\,,
		\end{split}
	\end{align}
	where in the last line we used $T_{32}S_{23}=T_{32}T^{-1}_{32}T_{23}T^{-1}_{32}=S_{23}T_{23}^{-1}$.
	The only non-trivial phase in this manipulation is when commuting $T_{12}^{m}$ and $S_{23}$, and is given by:
	\begin{align}
		\begin{split}
			\hat{Z}_{M_2}\left(\textbf{x}\right)&=\prod_{k=0}^{m-1}\left[(\psi_1)_{1,2}^{3,2}\left(T_{12}^{-k}T_{23}^{-1}T_{13}^{-k}T_{21}^{n_1}T_{31}^{n_2-n_1}T_{32}^{-1}S_{23}^2\textbf{x}\right)\right]\\
			&\times\prod_{k=0}^{m-1}\left[(\xi_4)_{1,3}^{3,2}\left(T_{13}^{-k}T_{23}^{-1}T_{21}^{n_1}T_{31}^{n_2-n_1}T_{32}^{-1}S_{23}^2\textbf{x}\right)\right] \hat{Z}_{M_3}\left(\textbf{x}\right).
		\end{split}
	\end{align}
	Now one can commute $T_{23}^{-1}$ to the right to finally obtain:
	\begin{equation}
		X_{\mathbf{m}}^2=S_{23}^{3}T_{21}^{n_2}T_{31}^{-n_1}T_{13}^{m}T_{31}^{-n_2}T_{21}^{n_2-n_1}T_{23}^{-1}\,,
	\end{equation}
	which is what we wanted to show.
	There is no non-trivial phase associated to these last moves.
	Thus, we see that
	\begin{align}
		\begin{split}
			\hat{Z}_{X_{\mathbf{m}}^3}\left(\textbf{x}\right)&=\prod_{k=0}^{m-1}\left[(\xi_2)_{1,3}^{3,2}\left(T_{13}^{k}T_{23}^{-1}T_{31}^{-n_2}T_{21}^{n_2-n_1}T_{23}^{-1}\textbf{x}\right)(\psi_1)_{1,2}^{3,2}\left(T_{12}^{-k}T_{23}^{-1}T_{13}^{-m}T_{21}^{n_1}T_{31}^{n_2-n_1}T_{32}^{-1}S_{23}^2\textbf{x}\right)\right]\\
			&\times\prod_{k=0}^{m-1}\left[(\xi_4)_{1,3}^{3,2}\left(T_{13}^{-k}T_{23}^{-1}T_{21}^{n_1}T_{31}^{n_2-n_1}T_{32}^{-1}S_{23}^2\textbf{x}\right)\right]\hat{Z}_{X_{\mathbf{m}}^{-1}X_{\mathbf{m}}}\left(\textbf{x}\right)\,.
		\end{split}
	\end{align}
	Notice that:
	\begin{equation}
		\hat{Z}_{X_{\mathbf{m}}^{-1}X_{\mathbf{m}}}\left(\textbf{x}\right)=\hat{Z}_{S_{23}^4}\left(\textbf{x}\right)\,.
	\end{equation}
	There is still a non-trivial phase associated with $\hat{Z}_{S_{23}^4}(\textbf{x})$, which we will now compute.
	First, note that:
	\begin{equation}
		\hat{Z}_{S_{12}^{-1}S_{13}^4S_{12}}\left(\textbf{x}\right)=1\,.
	\end{equation}
	This is due to the basic relations $\hat{Z}_{S_{13}^4}=1$ and $\hat{Z}_{T_{ij}^{-1}T_{ij}}=1$.
	Now, we wish to compute the phase associated to the relation $S_{23}=S_{12}^{-1}S_{13}S_{12}$.
	Since $S_{12}=T_{12}T_{21}^{-1}T_{12}$ and $S_{13}=T_{13}T_{31}^{-1}T_{13}$, one may derive the following two identities:
	\begin{align}
		\begin{split}
			\hat{Z}_{S_{12}^{-1}T_{13}S_{12}}\left(\textbf{x}\right)&=\hat{Z}_{T_{23}}\left(\textbf{x}\right)\,,\\
			\hat{Z}_{S_{12}^{-1}T_{31}^{-1}S_{12}}\left(\textbf{x}\right)&=\left(\xi_5\right)^{1,2}_{3,1}\left(S_{12}\textbf{x}\right)\,\left(\psi_2\right)^{1,2}_{3,2}\left(T_{12}\textbf{x}\right)\,\hat{Z}_{T_{32}^{-1}}\left(\textbf{x}\right)\,.
		\end{split}
	\end{align}
	These relations imply:
	\begin{align}
		\begin{split}
			\hat{Z}_{S_{12}^{-1}S_{13}S_{12}}\left(\textbf{x}\right)&=\left(\xi_5\right)^{1,2}_{3,1}\left(S_{12}T_{23}^{-1}\textbf{x}\right)\,\left(\psi_2\right)^{1,2}_{3,2}\left(T_{12}T_{23}^{-1}\textbf{x}\right)\,\hat{Z}_{S_{23}}\left(\textbf{x}\right)\,.
		\end{split}
	\end{align}
	Let us define the associated phase as:
	\begin{equation}
		\lambda\left(\textbf{x}\right)=\left(\xi_5\right)^{1,2}_{3,1}\left(S_{12}T_{23}^{-1}\textbf{x}\right)\,\left(\psi_2\right)^{1,2}_{3,2}\left(T_{12}T_{23}^{-1}\textbf{x}\right)\,.
	\end{equation}
	We then have:
	\begin{align}
		1=\hat{Z}_{S_{12}^{-1}S_{13}^4S_{12}}\left(\textbf{x}\right)&=\lambda\left(\textbf{x}\right)\,\lambda\left(S_{23}^{-1}\textbf{x}\right)\,\lambda\left(S_{23}^{-2}\textbf{x}\right)\,\lambda\left(S_{23}^{-3}\textbf{x}\right)\,\hat{Z}_{S_{23}^4}\left(\textbf{x}\right)\,.
	\end{align}
	From this, we conclude that
	\begin{align}
		\hat{Z}_{S_{23}^4}\left(\textbf{x}\right)=e^{i\pi \frac{2Z+x_1}{x_1}}\,.
	\end{align}
	This then leads to the final answer:
	\begin{align}
		\begin{split}
			\hat{Z}_{X_{\mathbf{m}}^3}\left(\textbf{x}\right)&=\prod_{k=0}^{m-1}\left[(\xi_2)_{1,3}^{3,2}\left(T_{13}^{k}T_{23}^{-1}T_{31}^{-n_2}T_{21}^{n_2-n_1}T_{23}^{-1}\textbf{x}\right)(\psi_1)_{1,2}^{3,2}\left(T_{12}^{-k}T_{23}^{-1}T_{13}^{-m}T_{21}^{n_1}T_{31}^{n_2-n_1}T_{32}^{-1}S_{23}^2\textbf{x}\right)\right]\\
			&\times\prod_{k=0}^{m-1}\left[(\xi_4)_{1,3}^{3,2}\left(T_{13}^{-k}T_{23}^{-1}T_{21}^{n_1}T_{31}^{n_2-n_1}T_{32}^{-1}S_{23}^2\textbf{x}\right)\right]e^{i\pi \frac{2Z+x_1}{x_1}} \,.
		\end{split}
	\end{align}
	Plugging in the definitions of the functions, we find that the phase can written as:\footnote{We recall that we are using the slightly misleading notation defined in and explained around~\eqref{eq:notation}.}
	\begin{equation}\label{eq:pre-mod-prop-Xabc-app}
		\hat{Z}_{X_{\mathbf{m}}^3}\left(\textbf{x}\right)=e^{\frac{i\pi}{m} \left[ Q\left(\frac{mZ-x_1}{x_1};\frac{mx_2+(1-mn_1)x_1}{x_1},\frac{mx_3+(1-mn_2)x_1}{x_1}\right)+\frac{(m+1)(m+3)}{4}\right]}=:e^{\frac{i\pi}{m}Q_{\mathbf{m}}\left(mz;\tau,\sigma\right)}\,.
	\end{equation}
	Notice that this function is invariant under the exchanges:
	\begin{align}\label{eq:sym-Qabc}
		\begin{split}
			x_1&\quad\leftrightarrow\quad mx_2+(1-mn_1)x_1\,,\\ x_1&\quad\leftrightarrow\quad mx_3+(1-mn_2)x_1\,,\\
			x_2+(1-mn_1)x_1&\quad\leftrightarrow\quad mx_3+(1-mn_2)x_1\,.
		\end{split}
	\end{align}
	Also, we note that for $m=n_1=n_2=1$, this property reduces to the property~\eqref{eq:pre-Y-mod-prop}, as it should since $X_{(1,1,1)}=Y^{-1}$.
	
	Evaluating the relation~\eqref{eq:pre-mod-prop-Xabc-app} explicitly, we find:
	\begin{align}\label{eq:pre-mod-prop-gamma-a-app}
		\begin{split}
			\frac{\Gamma\left(\tfrac{z+\sigma-n_2}{m\sigma+1-m\, n_2};\tfrac{\tau-\sigma+n_2-n_1}{m\sigma+1-m\,n_2},\tfrac{\sigma-n_2}{m\sigma+1-m\,n_2}\right)}{\Gamma\left(\tfrac{z+2\tau-\sigma+n_2-2n_1}{m\tau+1-m\,n_1};\tfrac{\tau-\sigma+n_2-n_1}{m\tau+1-m\,n_1},\tfrac{\tau-n_1}{m\tau+1-m\,n_1}\right)\Gamma(z;\tau,\sigma)}=e^{\frac{i\pi}{m}Q_{\mathbf{m}}\left(mz;\tau,\sigma\right)}\,,
		\end{split}
	\end{align}
	where we made the usual identifications $z=\frac{Z}{x_1}$, $\tau=\frac{x_2}{x_1}$ and $\sigma=\frac{x_3}{x_1}$.
	We have performed a number of consistency checks on this formula.
	For example, one may verify that both the left and right hand sides have the symmetries listed in~\eqref{eq:sym-Qabc}.
	Moreover, the left and right hand sides transform in the same way under the shifts $z\to z+1$, $z\to z+\tau$, and $z\to z+\sigma$.
	This implies that the product of the left hand side with the inverse of the right hand side is a triply periodic meromorphic function, and hence a constant.
	
	In the main text, we use a slightly different version of~\eqref{eq:pre-mod-prop-gamma-a-app}, which is the direct generalization of the property~\eqref{eq:ord-mod-prop-Gamma-app}.
	One can derive it using properties of the $\theta$ and elliptic $\Gamma$ functions collected in Appendix~\ref{app:defs}.
	This property is given by:
	\begin{align}\label{eq:mod-prop-gamma-a-app}
		\begin{split}
			\Gamma(z;\tau,\sigma)&=e^{-\frac{i\pi}{m}Q'_{\mathbf{m}}\left(mz;\tau,\sigma\right)}\frac{\Gamma\left(\tfrac{z}{m\sigma+1-mn_2};\tfrac{\tau-\sigma+n_2-n_1}{m\sigma+1-mn_2},\tfrac{\sigma-n_2}{m\sigma+1-mn_2}\right)}{\Gamma\left(\tfrac{z+\tau-\sigma+n_2-n_1}{m\tau+1-mn_1};\tfrac{\tau-\sigma+n_2-n_1}{m\tau+1-mn_1},\tfrac{\tau-n_1}{m\tau+1-mn_1}\right)} \,,
		\end{split}
	\end{align}
	where
	\begin{equation}
		Q'_{\mathbf{m}}\left(mz;\tau,\sigma\right)=Q\left(mz;m\tau+1-mn_1,m\sigma+1-mn_2\right)+\tfrac{m^2-1}{12}\,.
	\end{equation}

	\subsection{$Q$ polynomial for more general order three element of $SL(3,\mathbb{Z})$}\label{app:most-gen-mod-prop}
	In this section, we derive the modular property of the elliptic $\Gamma$ function that corresponds to more general order three elements of $SL(3,\mathbb{Z})$, given by the matrix~\eqref{eq:matrixAcubica1} under the constraint~\eqref{eq:a2-a1cubic-def-d2}. 
	The matrix $A$ can be decomposed as 
	\begin{equation}\label{eq:Adecomposition}
		A = T_{31}^{-k} \Lambda T_{23}^{n} T_{21}^k S_{23} \,,
	\end{equation}
	where the $\Lambda$ matrix is given by an element in $SL(2,\mathbb{Z})$:
	\begin{equation}\label{eq:SL2ZLambda}
		\Lambda=	\left(
		\begin{array}{ccc}
			\ n	&\ 0\ & -m\ \\
			\ 0	&\ 1\  & 0\ \\
			\ d	&\ 0\ & n^2\ 
		\end{array}
		\right) \,, \qquad n^3 + d m = 1 \,.
	\end{equation}

	The corresponding modularity property for $A$ would be
	\begin{equation}
		1=	Z_{S_{23}} ( S_{23} A^{-1}\boldsymbol{\rho}) 	Z_{S_{23}} (S_{23}  A^{-2} \boldsymbol{\rho}) Z_{S_{23}} (S_{23}   \boldsymbol{\rho}) \  \mod M \,.
	\end{equation}
	The generalization  of $e^{-i \pi \frac{Q}{m}}$ in~\eqref{eq:mod-prop-gamma-a} would be
	\begin{align}\label{eq:def-Q}
		\begin{split}
			\mathcal{A}(z;\tau,\sigma)\equiv
			\frac{\Gamma\left(\tfrac{z}{ m (\sigma +k) +n^2}, \tfrac{\tau-n^2(\sigma+k )+n d }{m (\sigma +k) +n^2}, \tfrac{n (\sigma +k) -d}{m (\sigma +k) +n^2} \right)}{\Gamma(z,\tau,\sigma) \Gamma\left(\tfrac{z+n\tau-(\sigma+k)}{m \tau+n}, \tfrac{ n^2 \tau-d}{m \tau +n}, \tfrac{n\tau-(\sigma+k)}{m \tau +n} \right) }
		\end{split} \,.
	\end{align}
	
	We can see explicitly that $k$ works as the integer shift of $\sigma$ modulus. Due to the periodicity under integer shift, $k$ would not introduce any new physical effects. 
	We therefore set $k=0$ without loss of generality below. 
	To avoid overloading notation, we will use the following expressions for the moduli:
	\begin{align}\label{def:tildehattausigma}
		\tilde{z} &=\frac{z}{m \,\tau+ n}\,, & \quad \tilde{\tau} &= \frac{n^2 \,\tau-d}{m \,\tau+n}\,, & \quad \tilde{\sigma} &= \frac{n \,\tau-\sigma}{m \,\tau+ n} \,, \\
		\hat{z} &= \frac{z}{ m\,\sigma + n^2}\,, & \quad \hat{\tau} &= \frac{\tau-n^2\, \sigma + n \,d }{ m\,\sigma+ n^2}\,, & \quad \hat{\sigma} &= \frac{n \,\sigma -d }{ m\,\sigma + n^2} \,.
	\end{align}
	To calculate $	\mathcal{A}(z;\tau,\sigma)$, we are going to follow the method of Theorem~4.1 from~\cite{Felder_2000}, namely to consider ratios like $	\mathcal{A}(z+\tau;\tau,\sigma)/ 	\mathcal{A}(z;\tau,\sigma)$.
	Due to the quasi-periodicity of elliptic $\Gamma$ functions, this ratio will be reduced to ratios between $q$-theta functions, which is calculable. 
	Without loss of generality, we will show explicitly
	how to calculate $\tau$-shift ratio $	\mathcal{A}(z+\tau;\tau,\sigma)/ 	\mathcal{A}(z;\tau,\sigma)$. Other ratios like $\sigma$-shift ratio and $1$-shift ratio can be calculated similarly. 
	
	Although we cannot solve the constraint analytically, it is still useful to for calculating the $\tau$ shift. 
	Note the following facts:
	\begin{align}\label{eq:decompositionoftau-shift}
		\begin{split}
			\tau &= n (n^2\tau- d) +d(m \tau +n) \,, \\
			\tau &=(\tau-n^2 \sigma+n d  ) + n (n \sigma -d )   \,.
		\end{split}
	\end{align}
	Then 
	\begin{align}
		\begin{split}
			& \Gamma\left(\tfrac{z+\tau}{m \,\sigma +n^2}, \tfrac{\tau-n^2 \sigma+n d }{m\, \sigma +n^2}, \tfrac{n\sigma -d}{m \,\sigma +n^2} \right) \\
			&\quad= \Gamma(\hat{z}+\hat{\tau}+n \hat{\sigma},\hat{\tau},\hat{\sigma}) =  \theta(\hat{z}+n \hat{\sigma},\hat{\sigma} )
			\Gamma(\hat{z},\hat{\tau} ,\hat{\sigma} )  \prod_{l=0}^{n-1} \theta(\hat{z } +l\hat{\sigma},\hat{\tau}) \,, \\
			& \Gamma\left(\tfrac{z+\tau+n \tau-\sigma}{m\, \tau + n}, \tfrac{ n^2 \tau-d}{m \,\tau + n}, \tfrac{n \tau-\sigma}{m \,\tau + n} \right) \\
			&\quad=  \Gamma(\tilde{z} +\tilde{\sigma}+n \tilde{\tau},\tilde{\tau},\tilde{\sigma}) = \theta(\tilde{z}+n\tilde{\tau},\tilde{\tau}) \Gamma(\tilde{z},\tilde{\tau},\tilde{\sigma}) 
			\prod_{l=0}^{n-1} \theta(\tilde{z}+l \tilde{\tau},\tilde{\sigma}) \,.
		\end{split}
	\end{align}
	The $\tau$-shift ratio is
	\begin{align}
		\begin{split}
			\frac{	\mathcal{A}(z+\tau;\tau,\sigma)}{	\mathcal{A}(z;\tau,\sigma)} = \frac{\theta(\hat{z},\hat{\sigma} )}{\theta(z,\sigma)}
			e^{2\pi i (\tilde{z}-\hat{z})n +\pi i (\tilde{\tau}-\hat{\sigma}) n(n-1)}
			\prod_{l=0}^{n-1} \frac{\theta(\hat{z } +l\hat{\sigma},\hat{\tau})}{\theta(\tilde{z}+ l \tilde{\tau},\tilde{\sigma})} \,.
		\end{split}
	\end{align}
	To simplify this expression, we need to use the formula~\eqref{eq:finalBpoly} which explicit shows the ratio between two $q$-theta functions related by general $SL(2,\mathbb{Z})$ transformation.\footnote{
		For~\eqref{eq:finalBpoly}  being valid to use, we need to require $n d$ and $ m n^2$ to be even simultaneously. 
		For example, $n$ to be even are the sets of solution which actually covers majority portion of solutions to the constraint.
	} 
	Note that
	\begin{equation}
		(\hat{z},\hat{\sigma})  = 	\Lambda_S \cdot 
		(z,\sigma)  \,, \qquad \Lambda_S\equiv \left(
		\begin{array}{cc}
			n &  -d\\
			m	& n^2
		\end{array}
		\right)
	\end{equation}
	which tells us
	\begin{equation}
		\frac{\theta(\hat{z},\hat{\sigma} )}{\theta(z,\sigma)} = \exp\Big[i \pi B(z,\sigma;n,-d,m,n^2) \Big] \,,
	\end{equation}
	where we use the $B$ polynomial calculated in~\eqref{eq:result}. Similarly 
	\begin{eqnarray}\label{eq:infiniteproducttheta-2}
		\frac{	\theta(\hat{z}+l\, \hat{\sigma},\hat{\tau})}{
			\theta(\tilde{z}+l\,\tilde{\tau},\tilde{\sigma})} &=& \frac{	\theta(\hat{z}+l\,\hat{\sigma},\hat{\tau})}{\theta(\tilde{z}+l\,\tilde{\tau} -l n\,\tilde{\sigma},\tilde{\sigma})}
		\frac{\theta(\tilde{z}+l\,\tilde{\tau} -l n \tilde{\sigma},\tilde{\sigma})}{\theta(\tilde{z}+l\,\tilde{\tau},\tilde{\sigma})} \\ \nonumber
		&= & \frac{	\theta(\hat{z}+l\,\hat{\sigma},\hat{\tau})}{\theta(\tilde{z}+l\,\tilde{\tau} -l n \tilde{\sigma},\tilde{\sigma})}
		\exp \Big[i \pi l n +2\pi i l n (\tilde{z}+l\,\tilde{\tau}) -\pi i \tilde{\sigma}ln(ln+1)\Big] \,.
	\end{eqnarray}
	The moduli $(\hat{z}+l\hat{\sigma},\hat{\tau})$ and $(\tilde{z}+l\tilde{\tau},\tilde{\sigma})$ are again related by $\Lambda_S$ transformation. Therefore, 
	\begin{align}
		\begin{split}
			\frac{	\theta(\hat{z}+l\hat{\sigma},\hat{\tau})}{\theta(\tilde{z}+l\tilde{\tau} -l n\tilde{\sigma},\tilde{\sigma})}  = \exp[-i \pi B(\hat{z}+l\hat{\sigma},\hat{\tau};n,-d,m,n^2)] \,.
		\end{split}
	\end{align}
	Collecting all of these results, we have 
	\begin{eqnarray}\label{eq:taushiftinA}
		\frac{	\mathcal{A}(z+\tau;\tau,\sigma)}{	\mathcal{A}(z;\tau,\sigma)} 
		&=&\exp\Big[i \pi B(z,\sigma;n,-d,m,n^2)+2\pi i (\tilde{z}-\hat{z}) n +\pi i (\tilde{\tau}-\hat{\sigma}) n(n-1) \Big] 
		\\ \nonumber
		&\times &
		\prod_{l=0}^{n-1} \exp \Big[i \pi l n +2\pi i l n (\tilde{z}+l\tilde{\tau}) -\pi i \tilde{\sigma}l n(l n+1)-i \pi B(\hat{z}+l\hat{\sigma},\hat{\tau};n,-d,m,n^2)\Big]\,.
	\end{eqnarray}
	Since we should think of~\eqref{eq:taushiftinA} as exponentials of the difference between two polynomials, we can identify $\mA$ with the following $Q$ polynomial after some admittedly cumbersome manipulations:\footnote{
		The $Q$ polynomial for nonzero $k$ is just to replace $\sigma$ in~\eqref{eq:generalQpoly} by $\sigma+k$.}
	\begin{equation}\label{eq:generalQpoly}
		\log 	\mathcal{A} (z;\tau,\sigma) =  \frac{i\pi}{m }Q(m z; m \tau +n , m \sigma +n^2) + \text{constant} \,.
	\end{equation}
	The equation~\eqref{eq:generalQpoly} is the $Q$ polynomial for most general $SL(3,\mathbb{Z})$ transformation, as a generalization of~\eqref{eq:abc-Q-poly}.
	The constant term would be hard to be written concisely due to several reasons. First of all, there is an ambiguity between $Q$ polynomial and  log of $ \mathcal{A}$ ratios since the difference up to an even integer is trivial due to the exponential structure. The constant term in the $Q$ polynomial also contains the Dedekind sum, which is hard to simplify generically.

	\bibliographystyle{JHEP}
	\bibliography{bib-bh}
	
\end{document}